\documentclass[useAMS,usenatbib]{mn2e}  
\usepackage {graphicx,subfig,aas_macros}
\newcommand{\equ}[1]{eq.~(\ref{eq:#1})}

\newcommand{\se}[1]{\S\ref{sec:#1}}
\newcommand{\fig}[1]{Fig.~\ref{fig:#1}}

\newcommand{\Fig}[1]{Figure~\ref{fig:#1}}

\newcommand{\tab}[1]{Table~\ref{tab:#1}}
\newcommand{\be}{\begin{equation}}
\newcommand{\ee}{\end{equation}}
\newcommand{\bea}{\begin{eqnarray}}
\newcommand{\eea}{\end{eqnarray}}

\newcommand{\no}{\noindent}

\newcommand{\msun}{{\rm M}_\odot}

\newcommand{\ifm}[1]{\relax\ifmmode#1\else$\mathsurround=0pt #1$\fi}
\newcommand{\kms}{\ifmmode\,{\rm km}\,{\rm s}^{-1}\else km$\,$s$^{-1}$\fi}
\newcommand{\hmpc}{\,\ifm{h^{-1}}{\rm Mpc}}

\newcommand{\Mpc}{\,{\rm Mpc}}
\newcommand{\kpc}{\,{\rm kpc}}
\newcommand{\pc}{\,{\rm pc}}
\newcommand{\Gyr}{\,{\rm Gyr}}

\newcommand{\Myr}{\,{\rm Myr}}

\newcommand{\ltsima}{$\; \buildrel < \over \sim \;$}
\newcommand{\lsim}{\lower.5ex\hbox{\ltsima}}
\newcommand{\gtsima}{$\; \buildrel > \over \sim \;$}
\newcommand{\gsim}{\lower.5ex\hbox{\gtsima}}

\def\la{\langle}

\def\omm{\Omega_{\rm m}}
\def\oml{\Omega_{\Lambda}}
\def\omb{\Omega_{\rm b}}
\def\sy{\,M_\odot\, {\rm yr}^{-1}}

\def\cmc{\,{\rm cm}^{-3}}

\def\Mv{M_{\rm v}}
\def\Rv{R_{\rm v}}
\def\Vv{V_{\rm v}}

\def\Md{M_{\rm d}}
\def\Mc{M_{\rm c}}

\def\Rd{R_{\rm d}}
\def\Hd{H_{\rm d}}

\def\vcm{\vec v_{\rm cm}}
\def\jd{\vec j_{\rm d}}

\def\td{t_{\rm d}}
\def\tm{t_{\rm mig}}

\def\Sigs{\Sigma_{*}}

\def\Fn{F_{\rm N}}
\def\Fw{F_{\rm W}}
\def\delrho{{\delta}_{\rm {\rho}}}
\def\delmin{{\delta}_{\rm {\rho}}^{\rm min}}

\def\rd0{r_{{\rm d}0}}
\def\Vdi0{V_{{\rm d}0}}

\def\insitu{{\it in-situ\ }}
\def\Insitu{{\it In-situ\ }}
\def\exsitu{{\it ex-situ\ }}
\def\Exsitu{{\it Ex-situ\ }}
\def\bulg{{\it bulge\ }}

\def\Vd{V_{\rm d}}

\newcommand{\Halpha}{H${\alpha}$}

\usepackage{color}

\begin{document} 

\large 

\title[Giant Clumps in High-$z$ Galaxies]
{The Population of Giant Clumps in Simulated High-$z$ Galaxies:
      In-situ and Ex-situ, Migration and Survival}

\author[Mandelker et al.] 
{\parbox[t]{\textwidth} 
{ 
Nir Mandelker$^1$, 
Avishai Dekel$^1$, 
Daniel Ceverino$^2$, 
Dylan Tweed$^1$, 
Christopher E. Moody$^3$, 
Joel Primack$^3$
} 
\\ \\  
$^1$Racah Institute of Physics, The Hebrew University, Jerusalem 91904, Israel\\ 
$^2$Grupo de Astrofisica, Universidad Autonoma de Madrid, Madrid E-28049,
Spain\\
$^3$Department of Physics, University of California, Santa Cruz, CA 95064, USA} 
\date{} 
 
\pagerange{\pageref{firstpage}--\pageref{lastpage}} \pubyear{0000} 
 
\maketitle 
 
\label{firstpage} 
 
\begin{abstract} 
We study the properties of giant clumps and their radial gradients in high-$z$ 
disc galaxies using AMR cosmological simulations. Our sample consists of 
770 snapshots in the redshift range $z=4-1$ from 29 galaxies that at $z=2$
span the stellar mass range $(0.2-3)\times 10^{11}\msun$. Extended gas discs 
exist in 83\% of the snapshots. Clumps are identified by gas density in 3D 
and their stellar and dark matter components are considered thereafter. 
While most of the overdensities are diffuse and elongated, 91\% of their mass 
and 83\% of their star-fromation rate (SFR) are in compact round clumps. Nearly 
all galaxies have a central, massive bulge clump, while 70\% of the discs show 
off-center clumps, 3-4 per galaxy. The fraction of clumpy discs peaks at 
intermediate disc masses. Clumps are divided based on dark-matter content 
into \insitu and \textit{ex-situ}, originating from violent disc instability 
(VDI) and minor mergers respectively. 60\% of the discs are in a VDI phase showing 
off-center \insitu clumps, which contribute 1-7\% of the disc mass and 5-45\% of 
its SFR. The \insitu clumps constitute 75\% of the off-center clumps in terms of 
number and SFR but only half the mass, each clump containing on average 1\% of the 
disc mass and 6\% of its SFR. They have young stellar ages, $100-400\Myr$, 
and high specific SFR (sSFR), $1-10\Gyr^{-1}$. They exhibit gradients resulting from 
inward clump migration, where the inner clumps are somewhat more massive and older, 
with lower gas fraction and sSFR and higher metallicity. Similar observed gradients 
indicate that clumps survive outflows. The \exsitu clumps have stellar ages 
$0.5-3\Gyr$ and sSFR $\sim 0.1-2 \Gyr^{-1}$, and they exhibit weaker gradients. 
Massive clumps of old stars at large radii are likely \exsitu mergers, 
though half of them share the disc rotation.
\end{abstract} 
 

\section{Introduction} 
\label{sec:intro} 

The typical massive star-forming galaxies (SFGs) at high redshift 
are different from their counterparts at $z=0$. While low-redshift 
discs like the Milky Way form stars quiescently with typical 
star-formation rates (SFRs) of a few $\sy$ \citep[e.g.][]{Brinchmann04}, 
the SFR in typical SFGs at $z\sim 2$ is on the order of $100\sy$ 
\citep{Genzel06,Forster06,Elmegreen07,Genzel08,Stark08}. Many of 
the massive SFGs have been spectroscopically confirmed to be rotating 
discs, with baryonic masses of $\sim 10^{11} \msun$ within radii of 
$\sim 10 \kpc$ \citep{Genzel06,Shapiro08,Forster09}. Their peak rotation 
velocities are $V\sim 150-250 \kms$, and their one-dimensional velocity 
dispersions are $\sigma\sim 20-80 \kms$, namely $V/{\sigma} \sim 2-7$, 
as opposed to $10-20$ in today's spiral galaxies \citep{Elmegreen05b,
Genzel06,Forster06,Forster09,Cresci09}. Their gas fractions, estimated 
from CO measurements, are in the range $0.2-0.8$ \citep{Tacconi08,
Tacconi10}, much higher than the fractions of $0.05-0.1$ in today's 
discs \citep{Saintonge11}. This is driven by the intense inflow of 
cold gas in narrow streams along the filaments of the cosmic web 
\citep[e.g.,][]{Dekel09}. Such a high gas fraction, combined with 
the high disc surface density that is imposed by the high density 
of the Universe at high redshift, lead to gravitational disc instability 
\citep{toomre64}. Under such conditions, the instability involves giant 
clumps and it operates on short, orbital timescales \citep[][hereafter 
DSC09]{DSC}. It is therefore termed ``violent" disk instability (VDI), 
as opposed to the slow, "secular" instability in today's discs. The 
large fraction of SFGs among $z\sim2$ galaxies \citep[e.g.][]{Elmegreen07,
Tacconi08} suggests that this phase is long lived, on the order of one 
to a few Gyr, comparable to the age of the Universe at $z\sim2$. 

The massive high-$z$ SFG discs tend to be broken into several giant 
clumps, each typically $\sim 1\kpc$ in diameter and $\sim 10^9 \msun$ 
in mass, where a large fraction of the star formation occurs \citep{Elmegreen05b,
Forster06,Genzel08,Forster11b,Guo12,wisnioski12}. Therefore, these 
discs were at some point referred to as "chain" or "clump-cluster" 
galaxies, depending on their orientation relative to the line of sight 
\citep{Cowie95,Bergh96,Elmegreen04b,Elmegreen04a,Elmegreen05a}. 
The clumps are observed in both rest-frame UV and rest-frame optical 
emission \citep{Genzel08,Forster09,Forster11b}, and they do not appear 
to be a bandshift artifact, in the sense that images of low redshift 
galaxies would not appear as clumpy if observed at high redshift with 
limited resolution and low S/N \citep{Elmegreen09}. Scaled down versions 
of these giant clumps, with radii of only a few hundred $\pc$ but similarly 
high SFR surface densities, are observed in strongly lensed 
$\sim 10^{10} \msun$ disc galaxies at simillar redshifts \citep{Jones10}.

The gravitational fragmentation of a gas-rich turbulent disc has been 
addressed in idealized simulations of isolated galaxies \citep{Noguchi99,
Gammie01,Immeli04a,Immeli04b,Bournaud07,Bournaud09,Elmegreen08,Hopkins12}  
as well as in a cosmological context, either analytically \citep[DSC09;][]
{Cacciato12,Genel12b}, or via cosmological simulations \citep{Agertz09,
CDB,Ceverino12,Genel12a}. \citet{toomre64} showed that a rotating disc 
becomes unstable to local gravitational collapse once the surface density 
of gas and ``cold" stars (${\Sigma}$) becomes large enough for self-gravity 
to overcome the stabilizing effects of pressure (represented here by a velocity 
dispersion, ${\sigma}$) and centrifugal forces (represented by the disc angular 
velocity, ${\Omega}$). This happens when the Toomre parameter 
$Q \propto \sigma\Omega/\Sigma$ becomes smaller than a critical value of order 
unity, $Q_{\rm c}$. The characteristic scale of fragmentation grows with the 
gas fraction of the system, which explains why the $z\sim 2$ giant clumps are 
so much larger than the low-redshift giant molecular clouds (GMCs). 

During a VDI phase, the disc can maintain a self-regulated, marginally 
unstable steady state, with a high velocity dispersion that keeps 
$Q \sim Q_{\rm c}$ \citep[DSC09;][]{krumholz_burkert10,Cacciato12,
Genel12b,Forbes12,Forbes13}. During this phase, clumps of a few 
percent of the disc mass and $\sim 10\%$ of its radius are formed. 
The perturbed disc induces angular momentum outflow and mass inflow 
towards the center, partly due to clump migration caused by torques, 
dynamical friction, and clump-clump interactions within the disc. 
In the high-$z$ gas-rich discs, the timescale for these proccesses 
is fast - comparable to the orbital time at the disc edge (DSC09 and 
references therein). The gravitational energy gained by the inflow 
in the disc, possibly along with clumpy accretion and feedback from 
stars and supernovae drive turbulence which keeps the disc in a 
marginally unstable state \citep[DSC09;][]{krumholz_burkert10,Cacciato12,
Genel12b}. This inflow can contribute to the growth of a central bulge 
(\citealp{Bournaud07}; DSC09) and may fuel AGN \citep{Bournaud11,Bournaud12}, 
in a way that may depend on the SFR in the disc and whether the clumps 
survive intact until they coalesce at the center. While the growing 
bulge and the transformation of gas to stars tend to stabilize the disc 
\citep{Martig09,Cacciato12}, the continuous input of gas from the cosmic 
streams ensures the high gas surface density needed to maintain VDI for 
cosmological times, till $z \sim 1$ \citep{DSC,Cacciato12,Forbes13}.

This simple theoretical framework, where most of the observed clumps 
are formed {\it in situ} in the discs by gravitational instability, 
has been confirmed in high-resolution zoom-in cosmological simulations
\citep{Agertz09,CDB,Ceverino12}. These studies used AMR hydrodynamics 
to zoom in on halos of mass $\Mv \sim 5 {\times} 10^{11} \msun$ at 
$z \sim 2.3$ and revealed discs of a few times $10^{10} \msun$ with 
bulges of comparable mass. These discs remain in a marginally unstable 
state for $\sim 1 \Gyr$ while continuously forming giant clumps of 
masses $10^8-10^9 \msun$. 

In these simulations, which employ thermal energy driven feedback 
from stellar winds and supernovae, the clumps remain intact for $1-2$ 
disc orbital times as they migrate towards the galactic center where 
they coalesce with the bulge. In previous work \citep{Ceverino12} we 
studied the internal kinematics of the clumps and found that they are 
in approximate Jeans equilibrium, supported mostly by rotation and 
partly by pressure against further gravitational collapse. In addition 
to the \insitu clumps, we found a few \exsitu clumps, which seem to 
have joined the disc as minor mergers. These clumps exhibited old stellar 
ages and low gas fractions, while appearing quite similar to the \insitu 
clumps in gas and stellar maps of the disc. From a limited sample of 
$\sim 70$ \insitu clumps, we estimated the mean stellar age of the clumps 
to be $\sim 150\Myr$ and found an apparent trend where clumps seem younger 
and more gas rich near the disc edge, consistent with early indications by 
observations. This prelimenary analysis is expanded upon in this work 
(\se{clump_grad}).

Our main goal in this paper is to analyze the properties of the clumps 
in a much larger suite of cosmological simulations. Our motivation is 
to verify the hypothesis of VDI-generated clumps, and to try to distinguish 
between the population of clumps formed \insitu by VDI and the clumps 
that formed \exsitu as galaxies and came in as minor mergers.

Our second goal is to address whether the \insitu 
clumps survive stellar-feedback-driven outflows and remain intact 
during their migration to the bulge \citep{KrumholzDekel,DK13}, or 
whether they disrupt on dynamical timescales by enhanced feedback 
\citep{murray10}. SPH cosmological simulations by \citet{Genel12a} 
which employ an enhanced-outflow version of the phenomenological model 
of \citet{OppenDave06,OppenDave08}, pushed the effect of outflows to 
the extreme and caused clumps to disrupt on a dynamical timescale of 
$\sim 50\Myr$. Simillar results were obtained in isoltaed disc simulations 
by \citet{Hopkins12}, though these simulations also overestimate the 
strength of radiation pressure (see \se{survival}). This has consequences 
on the disc structure and the bulge growth.

The simulations studied in this paper employ the same recipees 
for star-formation and feedback as in \citet{Ceverino12}. Thus, 
the clumps survive intact for orbital timescales of $\sim 250 \Myr$, 
typically caught in what seems to be internal dynamical equilibrium while 
they accrete mass during their migration to the center. This may or may 
not be more realistic than the scenario where clumps disrupt in a dynamical 
time, and reality likely lies somewhere in between. Our aim here is to study 
the long-lived clumps produced in these simulations, and come up with 
observable properties that may help distinguish between the two extreme 
scenarios. These include radial variation of clump properties within the 
disc in comparison with radial gradients in the diffuse inter-clump component.
Future work incorporating simulations with stronger feedback and 
outflows (N. Mandelker et al. in preparation) will help sharpen these 
distinguishing features and determine the effect of feedback on VDI.

Using a suite of 29 galaxies in the redshift range $1<z<4$, we identify  
clumps as overdensities in the three-dimensional gas distribution in 
the disc, and then measure the clump stellar and dark matter components. 
By identifying the clumps in the gas component, we trace the perturbations 
that lead to clump formation, and also capture young clumps in their early 
stages of collapse and star formation. The gas distribution also serves as 
a proxy for the distribution of star formation and young stars, which is 
traced, for example, by H$_\alpha$ observations. By considering the associated 
stars, we facilitate a comparison to observed stellar clumps. We assemble in 
this way a statistical sample of nearly 2000 compact, spherical clumps. 
Within this sample, we distinguish between \insitu and \exsitu clumps 
and quantify their respective contributions to the total clump population 
as well as to the total disc mass and SFR. We collect statistics on the 
masses, SFRs, ages, gas fractions and metallicities of each clump population 
and use these to predict observables for distinguishing between \insitu and 
\exsitu clumps. 

This paper is organized as follows: 
In \se{sim} we briefly present the suite of simulations, introduce 
the galaxy sample and describe our method for defining the galactic 
disc and identifying the clumps. More details are provided in three 
apendices: \se{art} elaborates on the simulation method, \se{disc-frame} 
describes the disc definition and \se{clump_finding} outlines our clump 
finding algortim. \se{clump_classification} describes our method for 
classifying the clumps into compact-spherical clumps and diffuse-elongated 
ones, as well as the distinction between \textit{in-situ}, \exsitu and \bulg clumps. 
In \se{clumpy_discs} we discuss the statistics of the discs that host 
off-center clumps, and in particular those discs undergoing VDI. 
In \se{clump_prop} we address the distributions of various properties of 
the clumps. In \se{clump_grad} we discuss radial variations of these 
properties within the disc. In \se{discussion} we discuss our results, 
particularly their implication for clump survival, and compare to 
preliminary observations. In \se{conc} we present our conclusions. 

\section{Analyzing the Simulations} 
\label{sec:sim} 
 
\subsection{The Cosmological Simulations}
\label{art_brief}

We use zoom-in hydro cosmological simulations of 29 galaxies 
whose virial masses at $z=2$ are in the range 
$2\times 10^{11}\le \Mv \le 3{\times}10^{12} \msun$. 
All the simulations were evolved to redshifts $z \lsim 2$ 
and several of them were evolved to redshift $z \sim 1$, 
with an AMR maximum resolution of $35-70\pc$ at all times. 
They utilize the ART code \citep{Kravtsov97,Kravtsov03}, 
which accurately follows the evolution of a gravitating N-body 
system and the Eulerian gas dynamics using an adaptive mesh. 
Beyond gravity and hydrodynamics, the code incorporates at the 
subgrid level many of the physical processes relevant for galaxy 
formation. These include gas cooling by atomic hydrogen and helium, 
metal and molecular hydrogen cooling, photoionization heating by a 
UV background with partial self-shielding, star formation, stellar 
mass loss, metal enrichment of the ISM and feedback from stellar 
winds and supernovae, implemented as local injections of thermal 
energy. Further details concerning the simulation method are provided 
in an appendix, \se{art}, as well as in \citet{Ceverino09}, \citet{CDB}, 
and \citet{Dekel13}.

The dark-matter halos were drawn from N-body simulations of 
the $\Lambda$CDM comsology with the WMAP5 parameters (\se{art}), 
in a comoving cosmological box. They were selected to have a 
virial mass in a desired range at a target redshift $z=1$ (a 
few at $z=0$). The only other selection criterion was that 
they show no ongoing major merger at that target time, which 
eliminates less than $10\%$ of the halos. Five of the galaxies, 
MW01-MW04 and SFG1 (see \se{art}), have been studied in some detail 
in \citet{CDB,Ceverino12}, where they are referred to, respectively, 
as galaxies B, C, A, D and E, ordered by virial mass at $z=2$. 
Further information about the simulated halos, including their 
virial properties at $z = 2$, target halo mass and final redshift 
can be found in \se{art}.

\subsection{The Galaxy Sample}

We focus on the redshift range $1\le z \le4$ where we 
have a total of 772 snapshots among the 29 simulations. 
We note that the outputs were equally spaced in expansion 
factor $a=(1+z)^{-1}$, so they become denser in time at 
lower redshifts. The time spacing between consecutive 
snapshots is on the order of $100-200 \Myr$, which 
is roughly half an orbital time at the disc edge or 
$10-20$ clump dynamical times. This means that the clumps 
can evolve significantly from snapshot to snapshot, even if 
they do not disrupt or complete their migration. Therefore, 
while consecutive snapshots from the same simulation are not 
truly independent of each other, we treat them as such in our 
analysis, noting that several clumps in fact survive for 2-3 
snapshots.

\begin{figure*} 
\includegraphics[width =0.95 \textwidth]{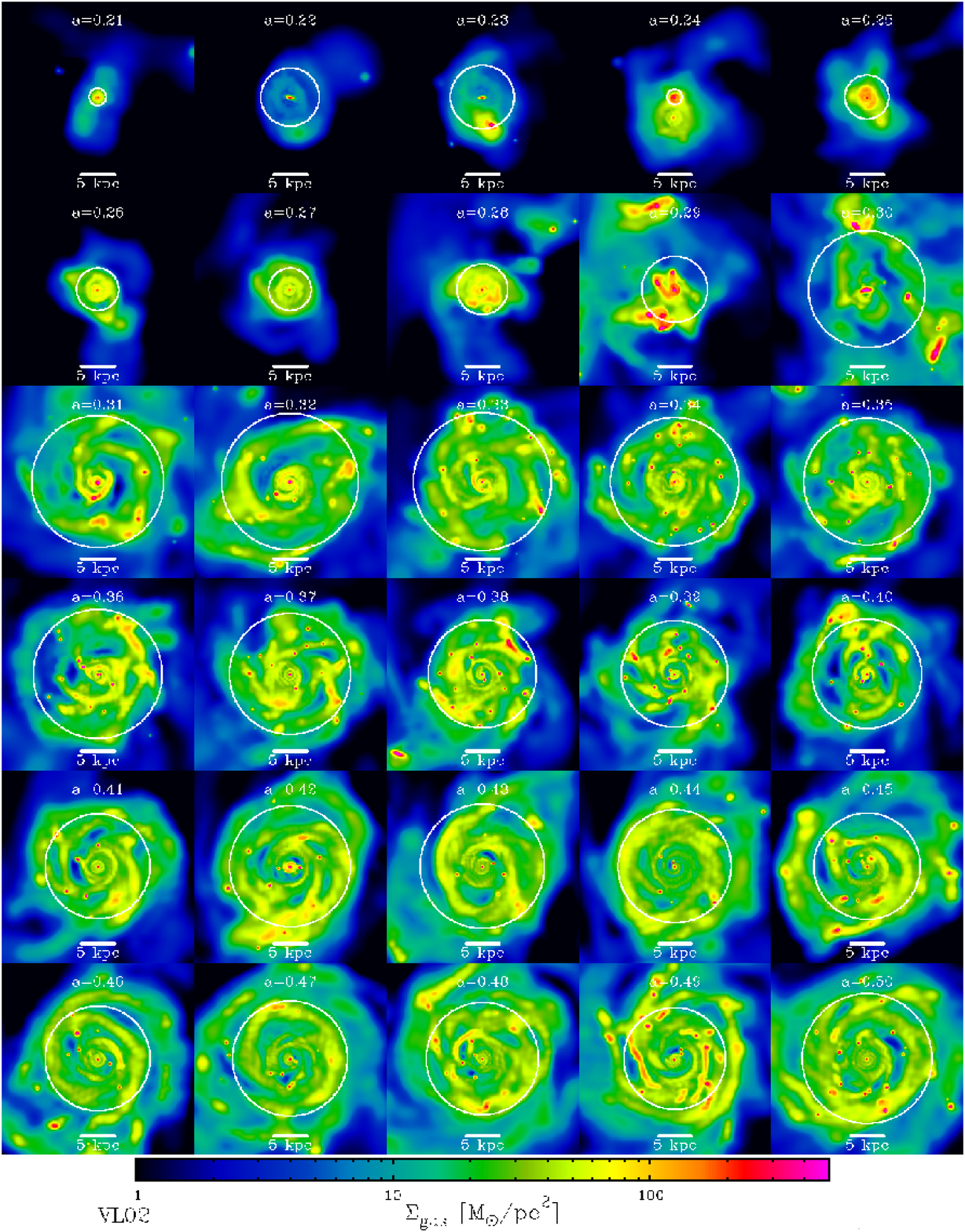}
\caption{
Evolution of gas density in galaxy VL02 between $z=4$ and $z=1$, 
face on views. The panels correspond to those shown in edge on 
projection in \fig{vl02_edge}, and the \textbf{x} axis is the same. 
The expansion factor $a$ is marked at the top of each panel. Color 
represents gas surface density according to the color bar. The disc 
radius $\Rd$ is marked by a white circle. The box shown is $\pm 15\kpc$ 
in each of the three directions. During several Gyr, the gas in this 
galaxy is an extended disc showing large-scale perturbations and giant 
compact clumps.
}
\label{fig:vl02_face}
\end{figure*}

\begin{figure*} 
\includegraphics[width =0.95 \textwidth]{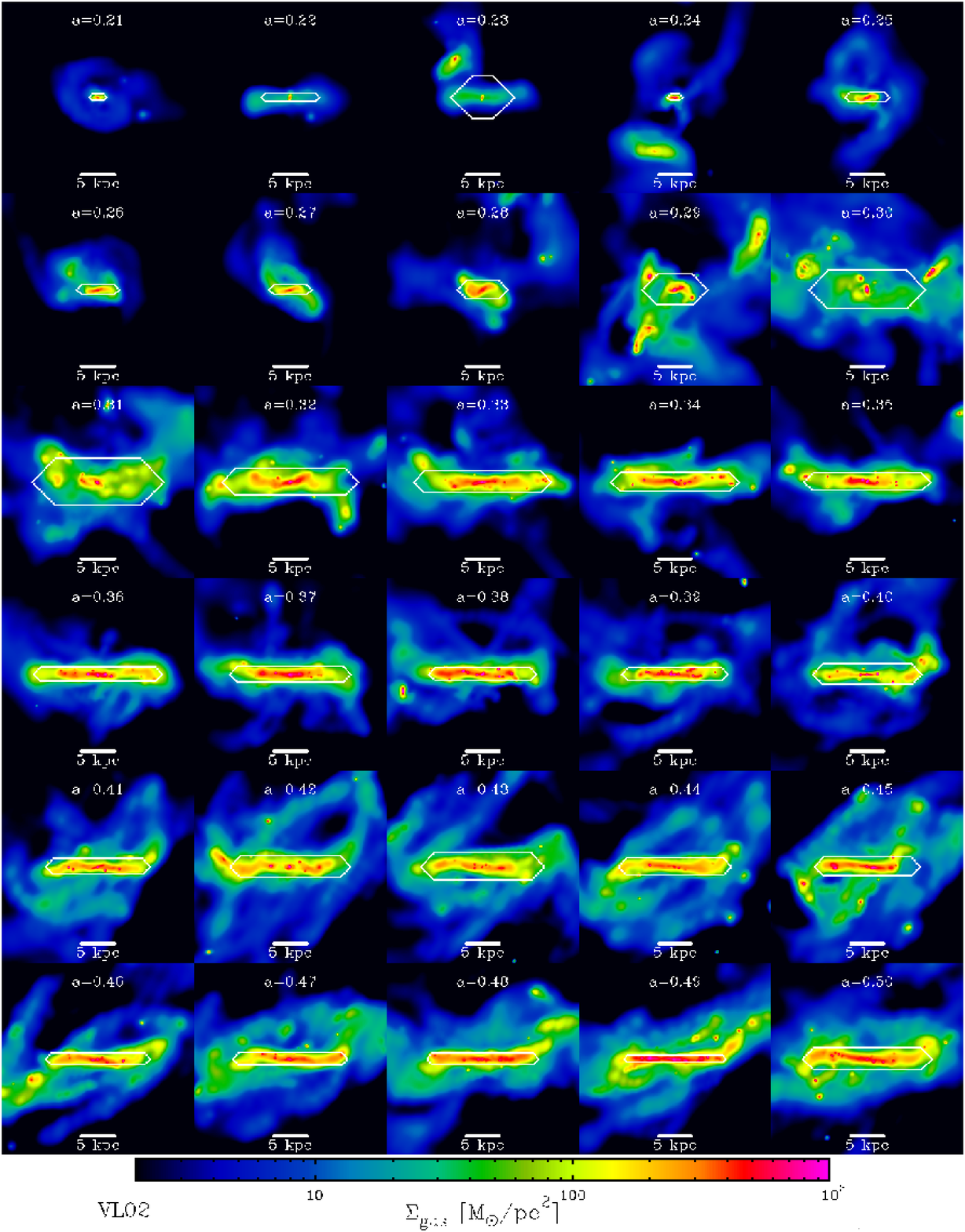}
\caption{
Evolution of gas density in galaxy VL02 between $z=4$ and $z=1$, 
edge on views. The panels correspond to those shown in face on 
projection in \fig{vl02_face}, and the \textbf{x} axis is the same. 
The expansion factor $a$ is marked at the top of each panel. Color 
represents gas surface density according to the color bar. The disc 
boundaries are marked by a white line. The box shown is $\pm 15\kpc$ 
in each of the three directions. The disc is perturbed, often showing 
extended arms above and below the midplane.
}
\label{fig:vl02_edge}
\end{figure*}

Our purpose is to study giant clumps formed in extended discs 
undergoing VDI, so we begin by identifying the disc component 
of each galaxy, modelling the disc as a cylinder with radius 
$\Rd$ and half thickness $\Hd$. The disc 
plane and dimensions are determined iteratively. The disc axis, 
$\hat z$, is defined by the angular momentum of cold gas 
($T<1.5{\times}10^4 {\rm K}$), which on average accounts for 
$\sim 97 \%$ of the total gas mass in the disc. The radius $\Rd$ 
is chosen to contain $85 \%$ of the cold gas mass in the 
galactic midplane out to $0.15$ of the virial radius $\Rv$, and 
the height $\Hd$ is defined so that $85\%$ of the cold gas 
mass in a thicker cylinder, where both radius and height equal 
$\Rd$, lies within $\pm \Hd$ of the midplane. Further details can 
be found in \se{disc-frame}.

All the gas contained within the final cylinder is considered part 
of the disc. An additional kinematic selection is 
applied to the stellar particles in order to seperate the disc stars 
from the bulge stars. A star particle is assigned to the disc only if 
the $z$-component of its angular momentum $j_z$ is higher than a fraction 
$f_J$ of the maximum angular momentum for the same orbital energy, 
$j_{max}=m|v|r$. Here, $m$ is the mass of the star particle, $|v|$ is 
the magnitude of its velocity and $r$ is its cylindrical radial distance 
from the galactic center. We adopt as default $f_J=0.7$ \citep{CDB}. 

The disc SFR is obtained crudely by $M_*(<\Delta t)/\Delta t$, where 
$M_*(<\Delta t)$ is the mass in disc stars younger than $\Delta t$. 
We adopt $\Delta t = 30 \Myr$, which is long enough to ensure good 
statistics and short enough to be comparable to the dynamical time 
and to keep the error of ignoring stellar mass loss small. The absolute 
values of the SFR as such should not be trusted to better than a factor 
of two, also given that an uncertainty at such a level is intrinsic to our 
current simulations (see below), but the ratio of SFRs between discs and 
between clumps should be more reliable. 

In order to get a sense of the time evolution of one of our clumpy 
galaxies, \fig{vl02_face} and \fig{vl02_edge} show a sequence of 30 
snapshots of VL02 in the range $0.20\le a \le 0.50$ ($4\ge z \ge 1$), 
projected both face on and edge on. From $a=0.32$ and on, we see an 
extended, clumpy disc with average dimensions $\Rd=9.2\kpc$ ($\sim 7\%$ 
of the average virial radius) and $\Hd=1.4\kpc$. During this phase, 
there are on average 8 off-center, compact, \insitu clumps in the disc. 

\begin{figure*}
\begin {center}
\includegraphics[width =0.90 \textwidth]{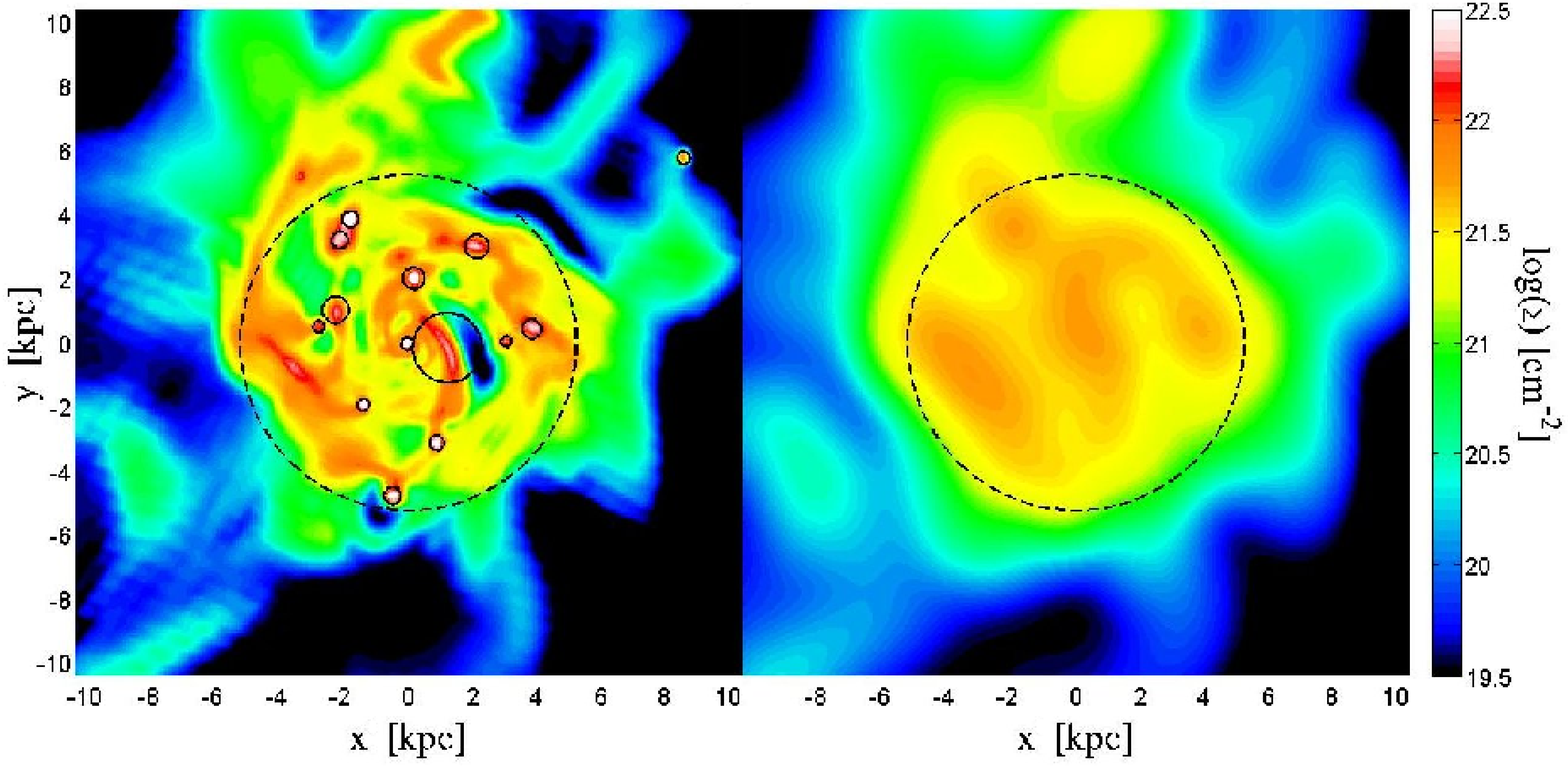}
\end {center}
\caption{
Clump identification. 
Shown is the gas surface density in a face-on view of MW3 at $a=0.30$ 
($z\simeq 2.33$). The disc radius, $\Rd=5.3\kpc$, is marked by a dashed 
circle. The scale of the image is $\pm 2\Rd$ and the integration depth 
is $\pm 2\Hd$, where $\Hd=1.1\kpc$. 
{\bf Right:} The large-scale surface density of the background disc as 
obtained by broad Gaussian filtering of the 3D density prior to projection.
{\bf Left:} The small-scale density as obtained by smoothing with a narrow 
Gaussian filter, comparable to the simulation resolution scale. The identified 
clumps are marked by circles with radii equal to the clumps' 3D radii. The 
clumps identified by the automatic algorithm coincide with the visual identification.
}
\label{fig:disc_smoothing} 
\end{figure*}

Since discs udergoing VDI are typically extended and gas rich, we pick up 
for analysis the discs having $\Rd \ge R_{\rm min}={\rm max}(2\kpc,\:0.03\Rv)$. 
This sample of discs comprises $\sim 83\%$ of the snapshots. 
At high redshifts, $2.5\le z\le 4$, there are 222 discs spanning 
$\Md \simeq 8{\times}10^{8}-7{\times}10^{10} \msun$, with a median 
mass of $1{\times}10^{10} \msun$. 
At intermediate redshifts, $1.5\le z < 2.5$, there are 270 discs 
with $\Md \simeq 4{\times}10^{9}-1{\times}10^{11} \msun$ and a 
median value of $4{\times}10^{10} \msun$. 
At low redshifts, $1\le z < 1.5$, there are 149 discs with 
$\Md \simeq 1{\times}10^{10}-1{\times}10^{11} \msun$ and a 
median of $6{\times}10^{10} \msun$. 
Our discs thus span a similar mass range to typical observed 
SFGs at similar redshifts \citep{Forster09,Forster11a,Guo12,Wuyts12}.

Previous studies using a subsample of these galaxies 
\citep{CDB,Ceverino12} have shown that they are fairly 
consistent with certain observed properties of galaxies 
in similar ranges of mass and redshift, such as the 
relation between SFR and stellar mass and the Tully-Fisher 
relation. 

However, a significant caveat to the simulations used here 
relates to the implementation of feedback, which is limited 
to effective prescriptions for thermal feedback from stellar 
winds and supernovae and does not address radiative feedback 
or AGN feedback. The galactic mass outflow rate in our 
simulations is therefore only a fraction of the SFR, with mass 
loading factors ranging from zero to unity with an average of 
$\eta \sim 0.3$ at $0.5\Rv$, not reproducing observed strong 
outflows with mass loading factors of unity and above 
\citep{Steidel10,Genzel11,DK13}. The code also assumes a 
somewhat high SFR efficiency per free-fall time and does not 
follow in detail the formation of molecules and the effect of 
metallicity on the SFR \citep{KD12}. Therefore, the early SFR 
is overestimated, while the suppression of SFR in small galaxies 
is underestimated, resulting in excessive early star formation 
prior to $z \sim 3$, by a factor of order 2. This causes the 
typical gas fractions and SFRs at $z \sim 2$ to be underestimated 
by a factor of $\sim 2$ compared to observations of SFGs 
\citep{CDB,Tacconi10,Daddi10}. Furthermore, the fairly weak 
outflows lead to a stellar fraction of $\sim 0.1$ within the 
virial radius, a factor of 2-3 higher than the observationally 
indicated value \citep{Perez08,Guo10,Moster10,Moster13,Behroozi13}. 

These inaccuracies in the SFR, feedback and outflows induce a 
limitation on the generality of our analysis, while the low gas 
fractions suggest that our simulations may conservatively 
underestimate the effects of gravitational instability in real 
galaxies at these redshifts. However, while precise numerical 
values for the gas fractions and SFR in our simulated galaxies 
should not be trusted to within a factor of 2, the relative 
trends we detect between the galaxies and between the clumps 
contained within them are likely to be at least qualitatively 
representative. These simulations offer the additional advantage 
of allowing us to examine VDI in the weak (but non-zero) feedback 
limit, as the first step in a broader study which will include 
simulations with stronger radiative feedback (N. Mandelker 
et al., in preparation) in order to study its effect on VDI.

A further consequence of the limited feedback is an underestimate 
of outflows from clumps compared to observations \citep{Genzel11,Newman12}, 
which may artificially extend clump lifetimes. However, as it is far 
from clear whether more realistic feedback can cause disruption of the 
massive clumps \citep[][see discussion in \se{survival}]{DK13}, 
our simulations have the advantage of allowing us to test the extreme scenario 
of long lived clumps and to compare to an extreme scenario where clumps are 
rapidly disrupted. Prelimenary analysis of new simulations with non-thermal 
radiative feedback \citep{Ceverino13,Renaud13} reveals that winds with realistic 
mass loading factors of $\eta \gsim 1$ do not alter the frequency of massive 
clumps with ${\rm log(}\Mc {\rm )} \gsim 8.5$ (\citealp{Bournaud13,Moody14}; 
N. Mandelker et al., in preparation; 
A. Dekel, F. Bournaud and N. Mandelker, in preparation).

\subsection{The Clump Finder} 

\label{sec:algorithm} 
\begin{figure}
\begin {center}
\includegraphics[width =0.48 \textwidth]{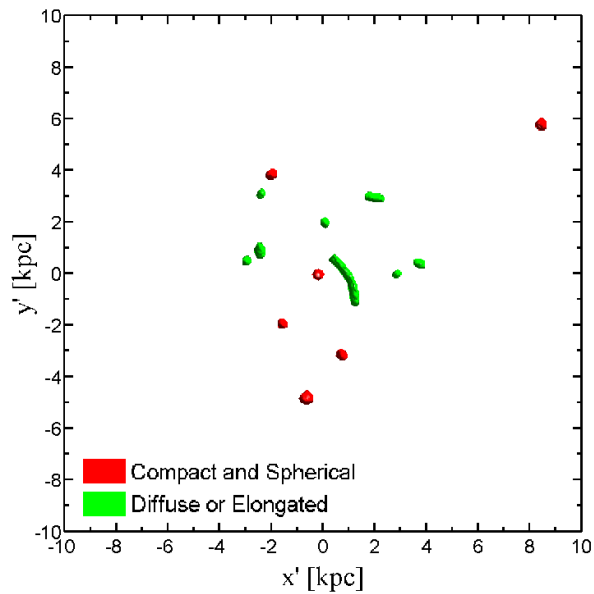}
\end {center}
\caption{
Clump identification. Face on projection of all cells in the 3D grid having 
density residulas above our adopted threshold. The 14 identified clumps, 
corresponding to the circles in \fig{disc_smoothing}, are visible. Compact, 
round clumps are colored red while diffuse or elongated clumps are colored 
green (see discussion in \se{clump_classification}). The clumps identified 
by the automatic algorithm coincide with the visual identification.
}
\label{fig:clump_find} 
\end{figure}

We identify clumps in the 3D gas distribution in and around the disc. 
Warps near the edge of the disc, such as those visible in \fig{vl02_edge} at late 
times, can make it difficult to properly define the disc boundaries. A lot of 
material in these warps likely does belong to the disc, despite being outside 
the defined cylinder. For this reason, we also probe a larger volume for clumps, 
and envision an {\it extended disc}, with radius $2\Rd$ and height $2\Hd$. We 
assign gas and stars to it using analogous criteria to those described above for 
the original, slim disc. This extended volume also allows us to identify additional 
\exsitu clumps above and below the disc plane. When comparing clump properties 
such as mass or SFR to those of their host disc, we refer exclusively to the 
extended disc, though the difference is small since 85\% of the cold gas is 
already contained in the slim disc. 

We briefly present here the main features of our clump finder. A more complete 
discussion can be found in appendix \se{clump_finding}. We smooth the 3D gas 
density field on two scales, first with a narrow Gaussian filter whose FWHM is 
taken to be $140\pc$, comparable to the spatial resolution of the simulation, 
then with a wide Gaussian filter whose FWHM is taken to be $2\kpc$, that will 
wash out features significantly smaller than the disc scale. We 
compute the residual density, and group together neighbouring cells above a threshold 
value. The local density peak defines the group center and each group is assigned 
a radius of a sphere that contains $90\%$ of its total mass. We eliminate all 
groups with volumes smaller than $(140\pc)^3$. Stellar and dark matter particles 
are placed on the same grid and we then eliminate all groups whose baryonic masses 
are less than $10^{-4} \Md$. All the remaining groups make up our sample of clumps. 
In practice, as can be seen in \fig{hist}, all compact clumps used in our analysis 
have masses larger than $\sim 10^{-3} \Md$. 

This process is illustrated in \fig{disc_smoothing} and \fig{clump_find} 
for one of the galaxies in our sample, MW3 at $a=0.30$ or $z\simeq 2.33$, 
shown as well in \fig{disc_frame}. In \fig{disc_smoothing} we plot the 
face-on surface density of the smoothed 3D gas density fields. The left 
panel relates to the narrow Gaussian which only washes out noise at the 
resolution level. Fourteen identified clumps have been marked. The 
apparently unidentified clump at $(x,y)\simeq (-3.5,5)\kpc$ is partly 
outside our volume, as can be seen in edge-on projections (\fig{disc_frame}) 
and was therefore ignored. The right panel relates to the wide Gaussian, 
which washes out features below the disc scale and leaves a good approximation 
for the local background in the disc. \Fig{clump_find} shows a face-on 
projection of the 3D residuals, after removing cells below the threshold 
and eliminating groups below the mass or volume limit. The identified clumps 
correspond to the circles in \fig{disc_smoothing}.

We note that the simulation resolution imposes an effective minimum mass 
for clumps, a crude estimate of which is as follows. With the smallest cell 
size being $35-70\pc$, and requiring as an absolute minimum at least $2^3$ 
cells per clump, the minimum clump volume is $(140 \pc)^3$.  Assuming a mean 
gas density of $30 \cmc$ within the clump (\se{art}, \se{clump_finding} and 
\fig{classification}), the minimum clump gas mass is 
$M_{\rm gas,\:c}^{\rm min}\gsim 10^{6.5} \msun$. With a typical gas fraction 
for \insitu clumps of 10\% (\fig{hist}), the minimum resolved baryonic 
mass in clumps is $\Mc^{\rm min}\sim 10^{7.5}\msun$. This minimum mass may vary 
with position in the disc and among different clumps as the disc density, the 
clump density contrast and the gas fraction may vary, and as the minimum cell 
size may be larger at the low-density regions in the outskirts of the disc.

\section{Classification of Clumps}
\label{sec:clump_classification}

In this section we detail how the clumps are classified based 
on their density (compact or diffuse), shape (round or elongated), 
position in the disc (central or off-center) and origin (\insitu 
or \textit{ex-situ}). The classification is summarized schematically at the 
end of the section, in \fig{diagram}.

\subsection{Density and Shape} 

We identify over 4000 clumps in our sample of 641 discs. Taken at face 
value, this gives an average of over 6 clumps per disc in the entire 
redshift range $1<z<4$. However, not all the identified clumps represent 
observed clumps. For example, the elongated ``worm-like" clump visible 
in \fig{clump_find} just to the right of the center (colored green) is 
clearly a different phenomenon than the smaller, rounder clumps visible 
in the same figure (colored red). Comparing to \fig{disc_smoothing}, we 
can see that this extended clump is part of a large-scale perturbation, 
perhaps indicative of the more linear phase of Toomre instability, before 
it is either disrupted by disc shear or stellar feedback, or collapses 
into smaller, rounder, denser clumps. Furthermore, a large fraction of 
the clumps in our sample are too small to be detected with the limited 
observational resolution.

\begin{figure}
\includegraphics[width =0.495 \textwidth]{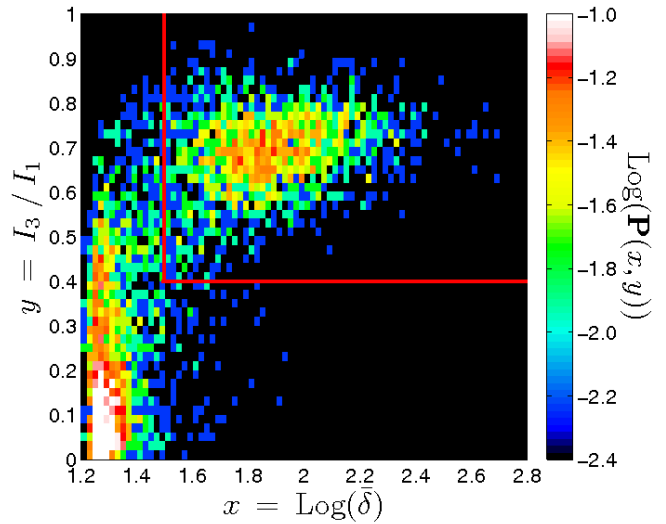}
\caption{Clump density and shape. Shown is the 2D probability 
density in the ${\rm log(}\bar{\delrho}{\rm )}-I_3/I_1$ plane for 
our population of clumps. There is a bi-modality between compact, 
oblate clumps (top right) and diffuse, elongated ones (bottom left). 
\textit{Compact} clumps are defined by ${\rm log(}\bar{\delrho}{\rm )}\ge 1.5$ 
and $I_3/I_1 \ge 0.4$.
}
\label{fig:classification} 
\end{figure} 

One expects the VDI perturbations in the disc to eventualy collapse 
into non-linear, self-gravitating, bound, spheroidal objects supported 
by rotation and pressure \citep[e.g. DSC09;][]{Bournaud07,KrumholzDekel,
Genzel11,Ceverino12}. We therefore wish to separate the compact clumps in 
our sample from more diffuse or elongated perturbations in the disc. 
For each clump, we measure its mean density residual $\bar{\delrho}$ 
(see \se{clump_finding} for the precise definition of the density residual 
${\delta}$) and the 3 eigenvalues of its inertia tensor $I_1\ge I_2\ge I_3$, 
computed from the unsmoothed gas density field in the region corresponding 
to the clump's location. We use as a shape parameter the ratio of minimum 
to maximum eigenvalue $I_3/I_1$, such that a value of 1 corresponds to a 
perfect sphere while a value of 0 corresponds to a 1D filament. An oblate 
rotating clump, as predicted by \citet{Ceverino12}, will have $I_3/I_1\simeq0.5$. 

\Fig{classification} shows the probability density in the plane defined by 
$x \equiv {\rm log(}\bar{\delrho}{\rm )}$ and $y \equiv I_3/I_1$, $P(x,y)$.
The probability for a clump to have values in an interval $\Delta x$ about $x$ 
and $\Delta y$ about $y$ is $P(x,y)\Delta x\Delta y$. 
There is a striking bimodality between a population of \textit{compact}, 
oblate and rather round clumps centered at $(x,y)\sim(1.85,0.70)$ 
and another population of diffuse, elongated clumps centered at 
$\sim(1.30,0.05)$. We hereafter define a clump as \textit{compact} if it 
has $x>1.5$ and $y>0.4$.

There are $\sim 1850$ \textit{compact} clumps making up $\sim 46 \%$ 
of the sample. However, they contain $\sim 91\% $ of the mass and 
$\sim 83 \%$ of the star formation of the entire clump population, 
and so are of much greater interest. Throughout the remainder 
of our current analysis, we focus solely on these \textit{compact} 
spheroidal clumps. 

\subsection{Insitu, Exsitu and Bulge Clumps} 
\label{sec:Is_Es}

\subsubsection{Bulge clumps}

As can be seen in \fig{vl02_face} to \fig{clump_find}, there is 
almost always a gas clump located at the disc center. We 
refer to these as \textit{bulge} clumps, defined as clumps whose 
centers are within 1 resolution element of the disc center. 
We identify a compact \textit{bulge} clump in $\gsim 91\%$ of the 
discs (and a diffuse \textit{bulge} clump in an additional $\gsim 8\%$). 
These are always the most massive clumps in their host disc, 
often by more than an order of magnitude, and have the highest 
SFR, though relatively low sSFR (\se{clump_prop}). The existence 
of such central clumps is not surprising, given that the disc center 
coincides with the peak of gas density, and that there is continuous 
gas inflow within the disc toward its center, driven either by VDI or 
by mergers \citep[][A. Zolotov et al., in preparation]{DekelBurkert13}.

Note that a bulge clump does not comprise the entire bulge --- it is 
limited to the bulge central region. Recall that we first identify 
clumps in gas, not stars, so the sizes and masses of the \textit{bulge} 
clumps are likely much smaller than those of the corresponding galactic 
bulges. On the other hand, since they do correspond to the peak of gas 
density within the bulge, it is not unreasonable to assume that the SFR 
in these clumps is indicative of the total SFR in our simulated bulges.

\begin{figure}
\includegraphics[width =0.495 \textwidth]{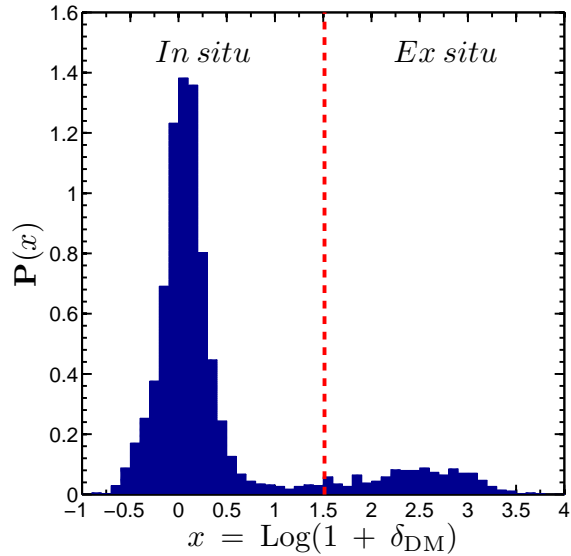}
\caption{
\Insitu vs \Exsitu clumps. Probability density of 
$x\equiv {\rm log(}1\:+\:{\delta}_{\rm DM}{\rm )}$ 
for all {\it off-center} clumps in our sample. 
A bi-modality is evident, and we 
define \exsitu clumps as having $x\ge 1.5$.
}
\label{fig:dm} 
\end{figure} 

\subsubsection{\Insitu and \Exsitu off-center clumps}

In addition to the \bulg clumps, $\sim 70\%$ of the discs in our sample 
contain off-center clumps. There are two possible origins for an off center 
clump: \insitu or \exsitu. \Insitu clumps form during VDI from gas and 
cold stars within the disc, in regions where $Q\lsim 1$. They are expected 
to bind only small amounts of dark matter, because the dissipationless 
dark-matter particles are "hot", with large and rather isotropic velocity 
dispersions, preventing them from participating in the disc instability which is 
driven by the cold gas and stars in the disc. \Exsitu clumps, on the other 
hand, merged in as external galaxies with their own gas, stars and dark matter 
components.

The distinction between \insitu and \exsitu clumps is important 
because it will allow us to ascertain the relative contributions 
of internal instabilities and mergers to various phenomena, 
such as disc morphology, bulge growth and SFR. In addition,  
correlations between properties of \insitu clumps and between them 
and their host discs, especially those that are not expected for 
\exsitu clumps, will offer insight into the nature of VDI and clump 
evolution.

Our major criterion for distinguishing between \insitu and \exsitu 
clumps is based on their dark matter overdensity with respect to 
the host halo. We calculate the mean dark matter density within 
the clump radius of each off-center clump and the mean background 
dark matter density in the host halo at the clump position, 
averaged over a spherical shell about the disc center with a width 
equal to the clump diameter. When calculating the background density, 
all clumps were removed from the shell. The ratio of clump density 
to background density is denoted $1+{\delta}_{\rm DM}$, analogous to 
$\delrho$ defined for the gas, \se{clump_finding}. \Fig{dm} presents 
the probability density, $P(x)$, where 
$x\equiv {\rm log(} 1+{\delta}_{\rm DM}{\rm )}$, for all off-center clumps. 
The distribution is clearly bi-modal, with a well defined peak at 
$x\gsim 0$ and a broader peak at $x\sim 2.5$. About $13 \%$ of the 
clumps contain no dark matter particles at all, placing them at 
$x=-\infty$, off the scale of \fig{dm}. Based on this bimodality, 
we define \exsitu clumps as having $x>1.5$. This happens to be 
similar to the selected threshold in ${\rm log(}\bar{\delrho}{\rm )}$ 
between compact and diffuse gas clumps. 

\begin{figure}
\includegraphics[width =0.495 \textwidth]{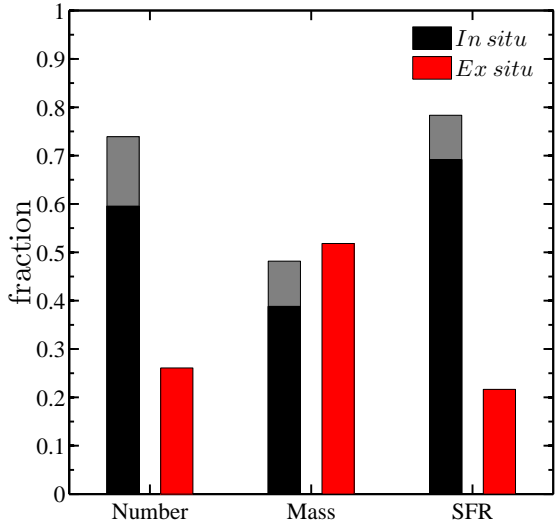}
\caption{Census of \insitu vs \exsitu clump properties. 
Relative contribution of \insitu and \exsitu clumps to the 
off-center clump population in terms of number, baryonic mass 
and SFR. In the 3 \insitu bars, the black section signifies our 
conservative estimate, removing those clumps that are not co-rotating 
with their background or that have significant contributions to their 
masses from external stars. The grey sections at the top of the bars 
signify the contribution of these suspicious clumps. 
The \exsitu clumps are roughly $20-25\%$ of the population in terms of number 
and SFR, though they contain $\gsim 50\%$ of the baryonic mass.
}
\label{fig:census} 
\end{figure} 

As a sanity check, we examined the stellar population of each \exsitu 
clump identified by its dark matter content. We find that in 94\% of 
them, more than half the baryonic mass consists of stars formed outside 
the disc, as defined at the snapshot closest to the star particle's 
birth time. This is in agreement with what is expected of an external 
merger. 

There is a small population of clumps in which more than half the 
baryonic mass consists of stars formed outside the disc while their 
dark-matter overdensity is low, which would have classified them as 
\insitu clumps. These constitute 8.5\% of the clumps that have only 
little dark matter. We defer the analysis of the origin of these 
clumps to future work, but can raise here three possibilities. First, 
they could be \insitu clumps, which accreted an unusually large 
amount of disc stars that joined the disc earlier during mergers. 
Second, they could be \exsitu clumps that were stripped of their 
dark matter when they merged with the disc. Finally, they could be 
clumps formed by thermal or hydrodynamical instabilities in the cold 
streams feeding the disc.

\subsubsection{Sharing the disc kinematics}

We also study the degree of co-rotation of the off-center clumps 
with the rotation of the cold gas in their host disc. For each 
clump, we measured the center-of-mass velocity of its gas component 
in cylindrical coordinates in the disc frame, $v_r,\:v_{\phi},\:v_z$. 
These were compared to the mass weighted 
mean velocity and velocity dispersion of the cold gas in a cylindrical 
shell about the disc center at the clump position. The shell extends 
the thickness of the disc in the vertical direction and the diameter 
of the clump in the radial direction and all clumps were removed from 
it. A clump is considered \textit{co-rotating} with the disc if all 
three components of its velocity are within $2{\sigma}$ of the local 
mean disc rotation velocity. 

\begin{figure}
\includegraphics[width =0.495 \textwidth]{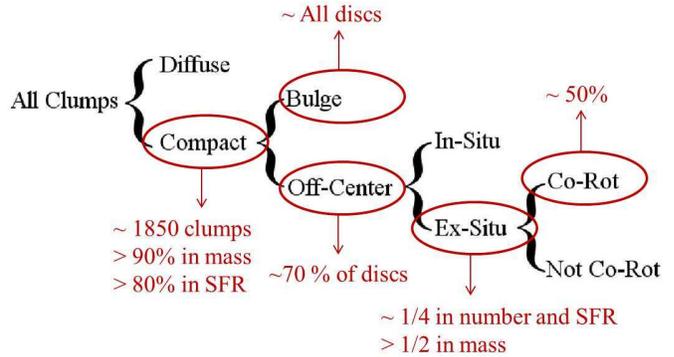}
\caption{
Schematic diagram of our clump classification scheme
}
\label{fig:diagram} 
\end{figure} 

We find that roughly half the \exsitu clumps are co-rotating 
with their host disc. Some of the properties of these co-rotating 
clumps are systematically different from the non-co-rotating ones, 
as will be discussed in \se{clump_prop}. Among the clumps with low 
dark matter contrast that are classified \textit{in-situ}, only 14\% are not 
co-rotating with their host disc, exhibiting strong radial or vertical 
velocity components. Most of these are possibly \insitu 
clumps whose rotation pattern was severely perturbed due to dynamical 
interactions with other clumps or the surrounding disc. About 22\% of 
those have large external stellar populations as well. 

We have thus divided the off-center clumps into two major populations 
plus a minor intermediate population. First are the pure \insitu clumps, 
defined as clumps of low dark-matter contrast that are co-rotating with 
their background and have only small contributions to their mass from 
stars that formed outside the disc. Second are the pure \exsitu clumps, 
which have high dark matter content. Lastly, a small population of clumps 
with low dark-matter contrast but with either significant contributions 
from external stars or significant kinematic deviations from the local 
disc rotation. This intermediate population between \insitu and \exsitu 
clumps is sometimes referred to as \textit{Is/Es}. However, since most 
of their properties are similar to those of the \insitu clumps (see 
\se{clump_prop}), we commonly treat them as part of the \insitu class. 

\Fig{census} shows the fractions in number, baryonic mass and SFR of each 
population among all the off-center clumps. We can conservatively estimate 
that the \insitu clumps make up $\sim 60-75\%$ of the off-center population, 
contain $\sim 40-50\%$ of their mass and $\sim 70-80\%$ of their 
SFR, with the \exsitu clumps comprising the remainder. It is interesting to 
note that there is a significant contribution from \exsitu clumps, which make 
up at least a quarter of the off-center clump population and contribute more 
than half of the mass found in clumps. The intermediate population is minor, 
making up $\sim 15\%$ of all the off-center clumps and containing $\lsim 10\%$ 
of their total mass and SFR. 

\subsubsection{Summary of clump classification}
\begin{figure*}
\begin {center}
\includegraphics[width =0.99 \textwidth]{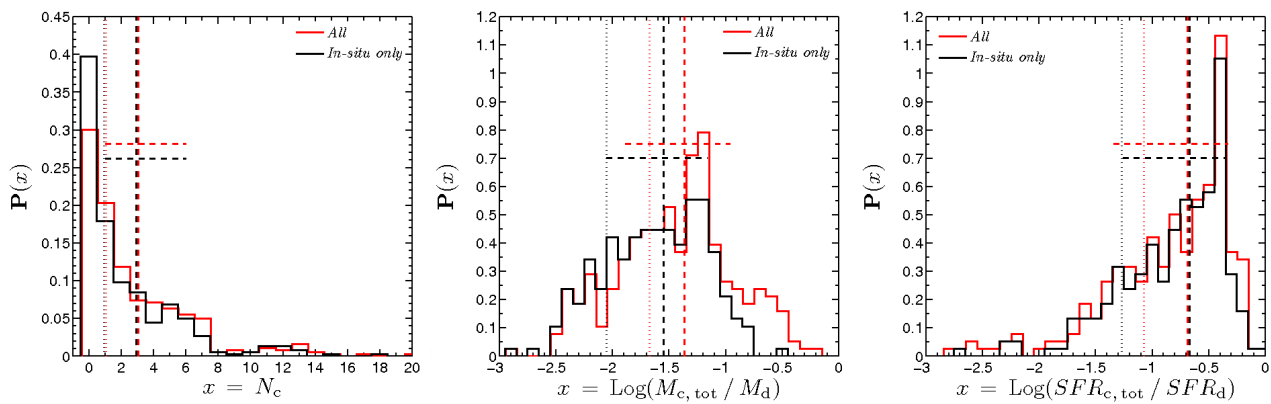}
\end {center}
\caption{
Clumpiness of our simulated discs. For the 380 discs in the 
redshift range $1.0\le z\le 3.0$ having 
${\rm log(}M_{\rm d}{\rm )}\ge 10.25$, we show the probability 
distributions of $x$: the number of clumps per disc (left), the 
log of the ratio of total baryonic mass in clumps to the 
disc baryonic mass (center) or the log of the ratio of 
total SFR in clumps to the disc SFR (right). The red 
histograms refer to all compact off-center clumps, while the 
black histograms refer only to \insitu clumps. Each 
histogram is normalized so that the integral over the entire 
disc population is equal to unity. However, in the center and right 
panels, those discs where $N_{\rm c} = 0$, so that $x = - \infty $, 
are beyond the scale of the figure. Thus, the integral of the red 
histogram is roughly $0.7$ and the integral of the black 
histogram is roughly $0.6$, as can be read from the left 
panel. For ``all" (\textit{insitu}) clumps, the vertical 
red (black) dotted lines mark the medians of the distributions 
including discs without any resolved clumps. The vertical red 
(black) dashed lines mark the medians considering only those 
discs where $N_{\rm c} \ge 1$, while the corresponding horizontal 
lines encompass 67\% of these discs about the median. The \exsitu 
clumps do not alter the distributions of the number of clumps per 
disc or the total SFR in clumps, but they can give rise to higher 
mass fractions.
}
\label{fig:clumpiness} 
\end{figure*}

\Fig{diagram} is a schematic diagram that summarizes our clump classification 
scheme. We begin by dividing all our clumps into {\it compact and round} versus 
{\it diffuse and elongated} based on their mean density residual and shape. The 
compact clumps contain $>90 \%$ of the mass and $>80\%$ of the star formation of 
the clump population. Focusing on the {\it compact} clumps, we identify central 
\bulg clumps versus {\it off-center} clumps. Nearly every galaxy hosts a compact 
\bulg clump, while $\sim 70 \%$ of the discs host {\it off-center} clumps. The 
{\it off-center} clumps are divided into \insitu and \exsitu clumps primarily 
based on their dark-matter density contrast with the host halo. The \exsitu 
clumps constitute roughly one quarter of the {\it off-center} clumps, but 
contain half the total mass. Finally, we find that the \exsitu clumps are 
divided roughly half and half into {\it co-rotating} and {\it non co-rotating} 
clumps. 

All the numbers quoted above refer to clumps that are found in the extended 
disc.  Restricting ourselves to the slim disc, we find that the number of 
\exsitu clumps is reduced by a factor of $\sim 3$. This is to be expected, 
as the \exsitu clumps may spend time as satellites of the central disc, and 
may have large velocity components vertical to the disc. The number of 
\insitu clumps in the slim disc is smaller than in the extended disc by 
$\sim 30\%$.  This is not surprising as many of the discs are warped away 
from the main disc plane (\fig{vl02_edge}) such that clumps located in these 
extended arms may have formed there \insitu. We find that the distributions 
of properties of the clumps are similar in the extended disc and in the slim 
disc, so we restrict the remainder of our discussion to results pertaining 
to the extended disc. Besides giving us better clump statistics, these results 
are closer to observations where the 2D images are integrated over more than the 
thickness of the gaseous disc.

\section{Disc Clumpiness}
\label{sec:clumpy_discs}
\begin{figure*}
\centering
\subfloat{\includegraphics[width =0.39 \textwidth]{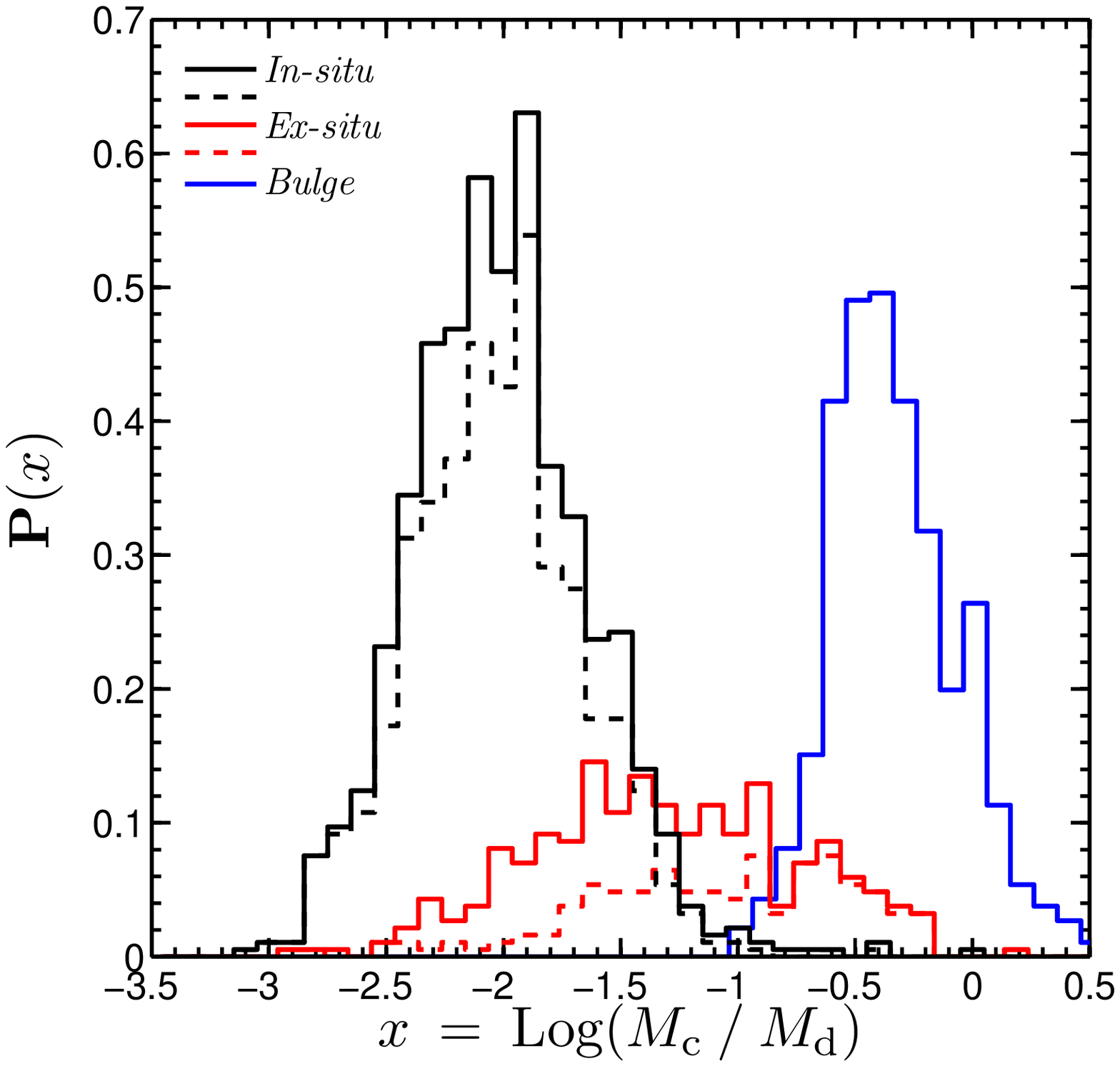}}
\subfloat{\includegraphics[width =0.39 \textwidth]{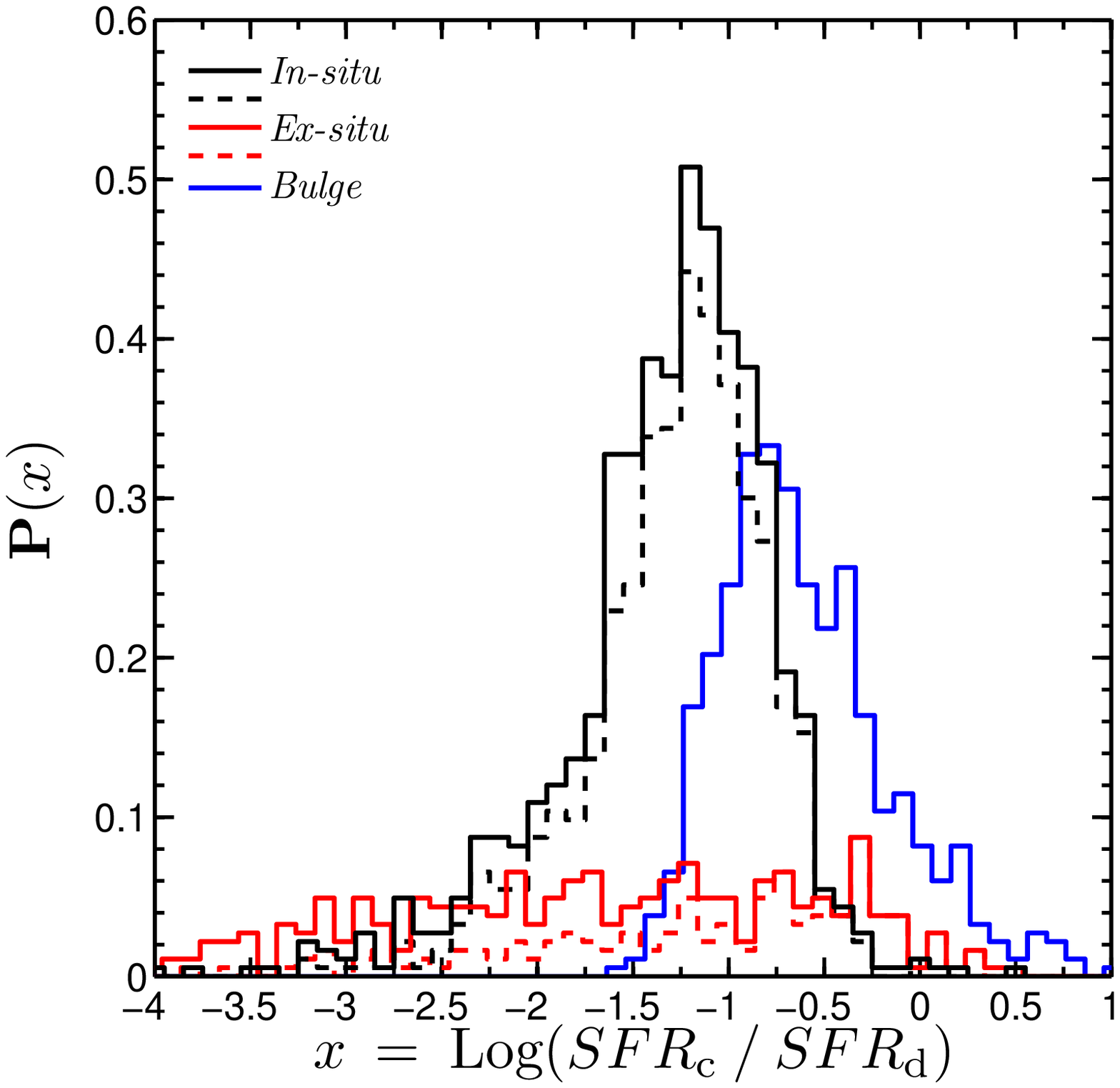}}\\
\subfloat{\includegraphics[width =0.39 \textwidth]{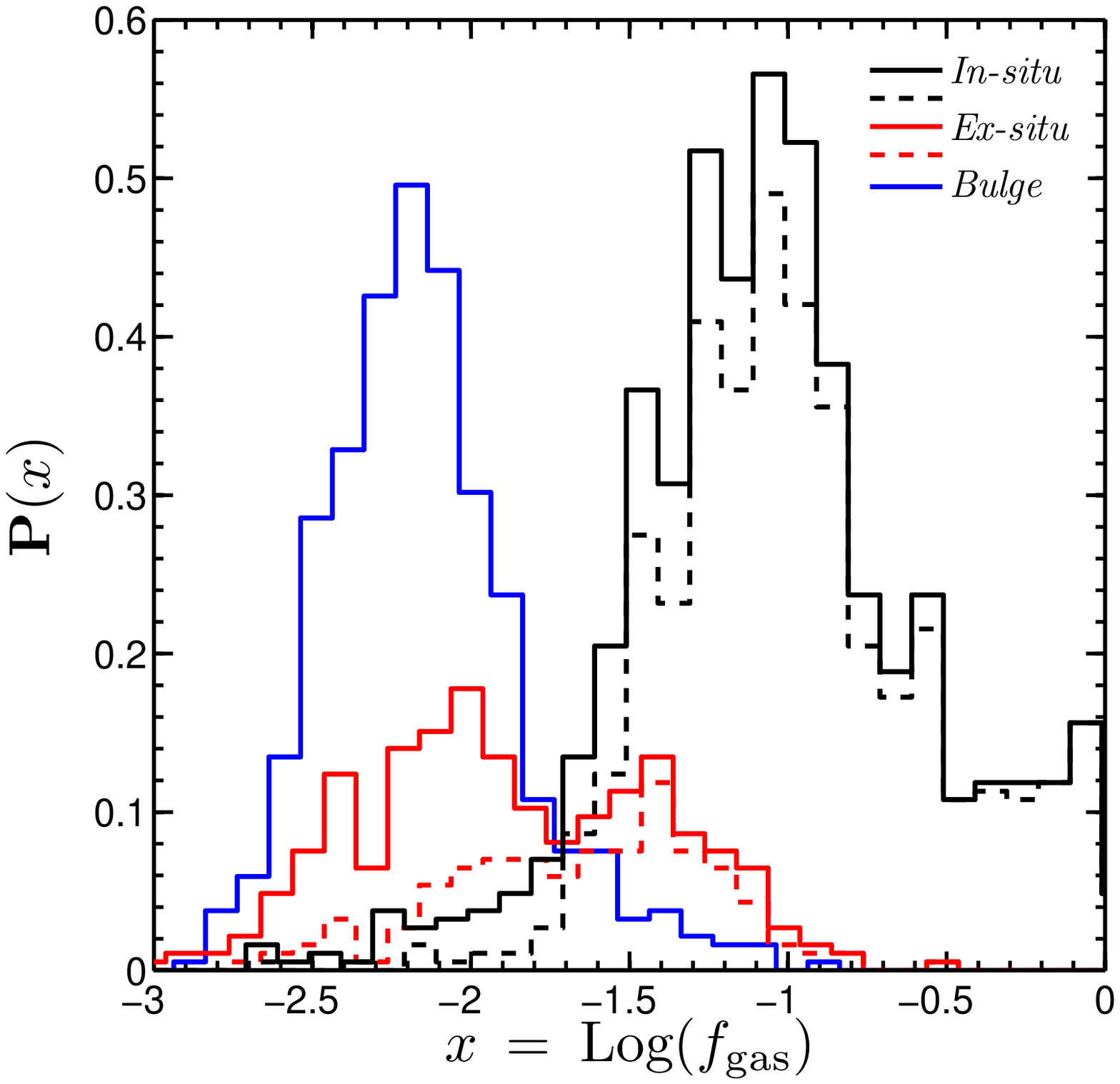}}
\subfloat{\includegraphics[width =0.39 \textwidth]{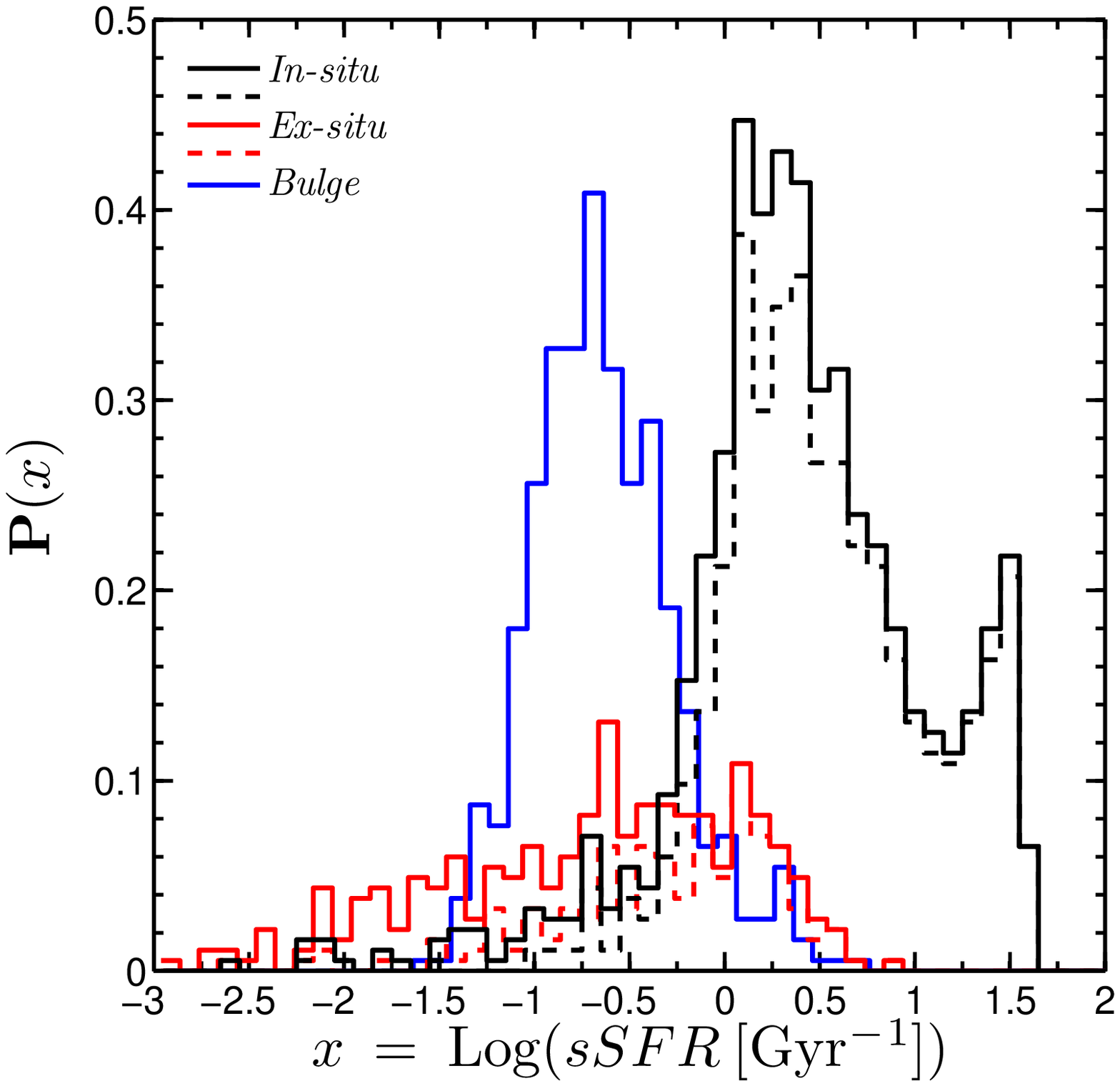}}\\
\subfloat{\includegraphics[width =0.39 \textwidth]{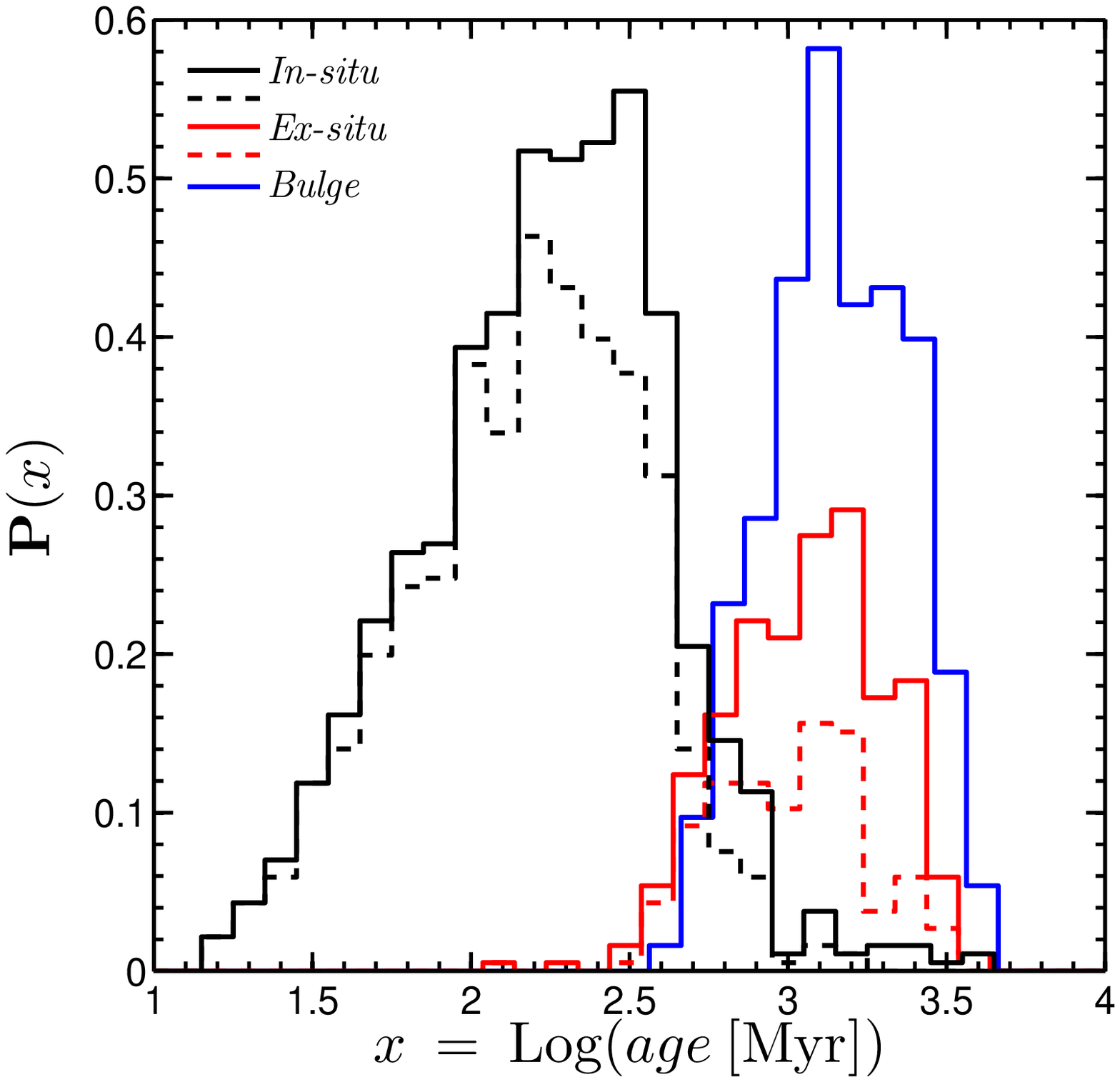}}
\subfloat{\includegraphics[width =0.39 \textwidth]{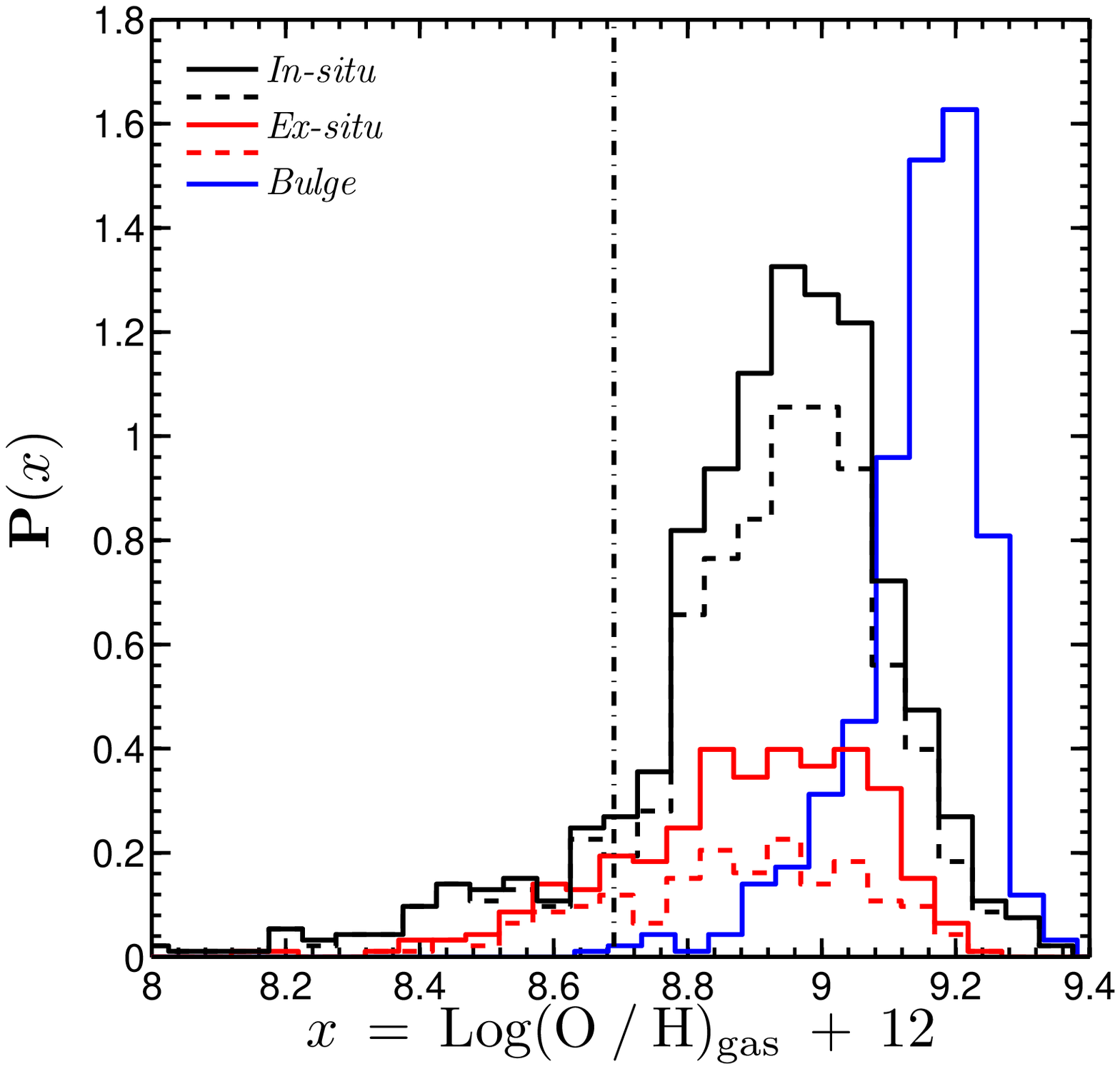}}
\caption{
Clump properties. Each histogram is the probability density for the quantity 
defined as $x$. The three solid histograms refer to \bulg clumps (blue), 
\insitu clumps (black) and \exsitu clumps (red). Also shown are the \insitu 
clumps without the uncertain cases (dashed black), and those \exsitu clumps 
that are co-rotating with the disc (dashed red). The distributions are 
normalized to the total number of compact clumps, such that the combined 
integrals under all three solid curves is equal to unity. Given that the \insitu 
clumps make up 51\% of the population, the \bulg clumps 31\% and the \exsitu 
clumps 18\%, each curve can be re-normalized to a probability density 
normalized to unity for each clump type. \textbf{Top row:} Clump 
mass relative to disc mass (left) and clump SFR relative to disc SFR (right). 
Disc properties were separated from bulge properties using the kinematic 
criterion defined in \se{sim}. 
\textbf{Middle row:} gas fraction (left) and sSFR (right). 
\textbf{Bottom row:} Mean stellar age (left) and gas metallicity 
(right, with solar metallicity marked by a dashed-dotted vertical line). 
\Insitu clumps tend to be less massive, with younger stars and higher gas 
fraction and sSFR.
}
\label{fig:hist} 
\end{figure*}

In this section we study the distribution of discs in terms of their 
off-center clumpiness properties. Recall that the resolution of our 
simulations imposes an effective minimum clump mass of 
$\sim 10^{7.5}\msun$ in baryons. Indeed, we do not detect any compact 
clumps below this mass (\fig{grad}). Since it appears clumps can have 
masses as low as $\sim 0.1\%$ of the mass of their host disc (\fig{hist}), 
we conclude that in low mass discs with $\Md\lsim 10^{10.5}\msun$, 
our sample of clumps must be incomplete. This effect is more severe 
at higher redshifts, since our galaxies are monotonically growing 
in time while the resolution scale remains fixed. In an attempt to 
minimize this effect, we limit our analysis in this section to the 
380 discs in the redshift bin $1.0\le z \le 3.0$ with masses 
${\rm log}(\Md) \ge 10.25$. In these discs, our sample of gas 
clumps should be nearly complete down to $10^{-2.5}\Md$. 

\Fig{clumpiness} shows the probablity density of (a) number of clumps 
per disc, (b) total baryonic mass in clumps relative to the 
disc baryonic mass (including the clumps), and (c) total SFR 
in clumps relative to the disc SFR (including the clumps). 
Separate histograms address the \insitu clumps alone and ``all" 
the off-center clumps, \insitu plus \textit{ex-situ}.

We read from the left panel of \fig{clumpiness} that $70\%$ of the discs 
host off-center clumps, and that $60\%$ specifically host \insitu clumps. 
If we associate the appearance of at least one off-center \insitu clump 
with a VDI phase, we can conclude that the fraction of discs undergoing 
VDI at a given time is about $60\%$. If only clumps more massive than 
$1\%$ of the disc mass are considered, the fraction of clumpy discs 
becomes $\sim 55\%$ while the fraction of discs undergoing VDI becomes 
$\sim 42\%$. This represents a lower limit to the clumpy fraction of 
galaxies, which should be robust to the resolution, and is consistent 
with observed clumpy fractions (\se{observ}).

The average number of \insitu clumps 
per disc is $\sim 2$, and the average number of ``all" off-center clumps 
is $\sim 3$, though the two distributions have a median value close to 
unity. Considering only the clumpy discs, the average (median) value of 
both distributions becomes $\sim 4$ (3).  Recall, however, that the number 
of clumps may be subject to the threshold minimum clump mass imposed by 
the resolution. 
The shapes of the distributions are also rather similar, both gradually 
declining till $N_{\rm c}\sim 8$ and then dropping sharply. In about 8\% 
of the clumpy discs there are more than 8 clumps per disc, and in about 
4\% the number is larger than 12, with the record in our sample being 20 
clumps per disc, 18 of them \insitu. This particular snapshot is visible 
in \fig{vl02_face} and \fig{vl02_edge} at expansion factor $a=0.34$, or 
$z\sim2$. The \insitu clumps span a mass range of $10^8-10^9\msun$ in 
baryons, while the disc mass is $\sim 3 \times 10^{10}\msun$.

The middle panel of \fig{clumpiness} refers to the total baryonic 
mass in off-center clumps compared to the disc baryonic mass. 
This is more robust than the number of clumps because it is dominated 
by the more massive clumps and is therefore less sensitive to the minimum 
mass imposed by resolution. This figure shows that for discs undergoing 
VDI the median fraction of disc mass in \insitu clumps is $\sim 3\%$ 
($\sim 1\%$ if all the discs are considered), though the probability 
density has a broad peak between 1\% and 10\%. Less than 3\% of the 
discs have mass fractions larger than 15\%. The value of the mass 
fraction addressed here is not straightforwardly constrained by Toomre 
instability theory. The fraction of mass in clumps compared to the cold 
mass in the disc, mostly gas and young stars, has been assumed in earlier 
works to be of order 20\% based on a crude argument (DSC09) and on estimates 
from simulations \citep{CDB,Elmegreen07}. Note that the current quote of 
a few to ten percent refers to the fraction of mass in clumps relative to 
the whole disc, which has a significant component of ``hot" old stars. 
Crudely defining ``hot" stars as being older than the time since the 
previous snapshot, roughly half an orbital time at the disc edge or 
$100 \Myr$, we find that they constitute on average $\lsim 70\%$ of the 
disc baryonic mass. Taking this into account, the mass fraction of clumps 
with respect to the cold disk becomes 3-30\%, consistent with the previous 
estimates.

Including \exsitu clumps as well, the median fraction 
of disc mass in ``all" off-center clumps is 4\% for the clumpy discs (and 
2\% for all the discs), with a peak near 6\%. Indeed, \exsitu clumps tend 
to be more massive than the \insitu clumps. This is reflected in the tail 
of non-negligible probability to have a fraction $0.1-0.4$ of the disc mass 
in clumps; about half \insitu and half \exsitu in the range $0.1-0.2$ and 
all \exsitu above a fraction of $0.2$.

The right panel of \fig{clumpiness} addresses the fraction of the SFR 
in off-center clumps compared to the whole disc. The distributions for 
\insitu clumps and ``all" off-center clumps are similar, since most 
\exsitu clumps tend to have low SFR. For \insitu clumps in VDI discs 
the median SFR fraction is 22\% (6\% if all the discs are considered), 
and for ``all" the off-center clumps in clumpy discs, the median is 20\% 
(8\% if all the discs are considered). Both distributions have simillar 
modes, at $\sim 40\%$. In about 13\% of discs undergoing VDI more than 
half the SFR is in clumps, 
and in about 3\% the SFR fraction in clumps is more than 75\%. 

We attempted to address dependencies of the above statistics on disc mass 
and redshift, but detected mostly marginal or null trends. This is partly 
because of our limited sampling and partly because of the bias due to the 
fixed resolution scale in low mass disks and at high redshifts. The analysis 
is presented in an appendix \se{marginal}. The most significant result is that 
the fraction of clumpy discs peaks for intermediate mass discs of 
$(1-3)\times 10^{10}\msun$ at all redshifts (\fig{mass_dep}). The fraction also 
seems to be higher at $z=1-2$ than at $z=2-4$. No systematic trend with either 
disc mass or redshift is apparent for the contribution of \insitu clumps to the 
total disc mass, suggesting that discs undergoing VDI turn a roughly constant 
fraction of their mass into clumps (see DSC09). \Exsitu merging clumps contribute 
more to the disc mass at higher redshift. This is consistent with the theoretical 
estimate that the timescale for mergers of a given mass ratio, in terms of the 
galaxy dynamical time, is shorter at higher redshift \citep{NeisteinDekel08}.

\section{Distributions of Clump Properties}
\label{sec:clump_prop}

We now turn to studying the properties of individual clumps 
and their distributions among the populations of clump types 
and within each population. The 1850 compact clumps in our 
sample are divided to $31\%$ central \bulg clumps, $51\%$ 
off-center \insitu clumps (including $10\%$ that are uncertain 
\textit{Is/Es} clumps), and $18\%$ off-center \exsitu clumps 
(half of which are co-rotating with the disc). \Fig{hist} 
shows the distributions of six clump properties 
for the different clump types. The probability densities for 
the \bulg clumps, \insitu clumps and \exsitu clumps are 
together normalized to the total number of clumps. The 
quoted fractions of clump types 
allow a re-normalization of each histogram to a probability 
density normalized to unity for each clump type.
\vspace{-10pt}

\subsection{Mass}
The top left panel of \fig{hist} refers to baryonic masses of 
individual clumps relative to their host disc, $\Mc/\Md$. 

\textbf{In-situ clumps:} 
For the \insitu clumps, the distribution of relative clump mass is 
close to lognormal, rather symmetric about a mean at $\Mc/\Md=0.01$, 
with a standard deviation of $\sim 0.38$ dex (FWHM$\sim 0.7$ dex). 
About 8\% of the \insitu clumps are more massive than $0.03\Md$, and 
less than 1\% are more massive than $0.1\Md$. The average of $0.01$ 
is consistent with Toomre instability theory which predicts that the 
\insitu clumps should each be a few percent of the cold mass in the 
disc (see DSC09 and references therein). We note that the least 
massive clumps are about $10^{-3}\Md$, higher than the minimum 
mass imposed by the resolution. 
The distribution of masses for the \textit{Is/Es} clumps is similar to that 
of the whole \insitu population and they appear less massive than the \exsitu 
clumps, consistent with most of them being \insitu clumps. 

\textbf{Ex-situ clumps:} 
The \exsitu clumps tend to be more massive than the \insitu clumps, 
with an average mass of $0.04\Md$. However, the scatter is large, 
roughly 1.5 dex FWHM, indicating that there is only a weak correlation 
between the mass of an \exsitu clump and that of its host disc, as 
expected from mergers. The co-rotating \exsitu clumps have systematically 
higher masses than the non-co-rotating ones. The average relative masses 
are 0.08 for the co-rotating clumps and 0.03 for the non-co-rotating 
ones. The possible origin of this phenomenon will be discussed in 
\se{discussion}.

\textbf{Bulge clumps:} 
The central \bulg clumps are rather massive. Since they may 
contain stars that do not kinematically belong to the disc, namely 
with $j_z\:/\:j_{max}<0.7$ (\se{sim}), $\Mc/\Md$ can be 
larger than unity. The distribution of $\Mc/\Md$ for the \bulg clumps 
is close to lognormal, with the mean at $\sim 0.4$ and a standard deviation 
of $\sim 0.27$ dex (FWHM = 0.7 dex). Recall that the mass of a \bulg 
clump as defined here is a severe underestimate of the actual bulge 
mass (\se{clump_classification}). 

\subsection{SFR}

The top right panel of \fig{hist} refers to 
the SFR of individual clumps relative to 
their host disc, SFR$_{\rm c}$/SFR$_{\rm d}$.

\textbf{In-situ clumps:} 
For the \insitu clumps, the distribution of relative SFR is somewhat 
skewed; it has a narrower peak at high values and a tail extending to 
low values. In general, individual clumps contribute more to the disc 
SFR than to its mass. The mean SFR fraction is $\sim 0.05$ and the FWHM 
spread is $\sim 0.9$ dex. About 24\% of the clumps contribute more that 
10\% of the disc SFR, and 2\% contribute more than 30\%. At the low end, 
about 13\% of the \insitu clumps have SFRs smaller than 1\% of the disc 
SFR. The low SFR tail is largely due to \insitu clumps older than $100\Myr$ 
located near or beyond $\Rd$, mostly with below average 
gas fractions. Many of them are \textit{Is/Es} clumps with external 
stellar populations, which tend to be among the oldest of the \insitu 
clumps, with ages of $300-800\Myr$. Several of these have been 
tracked through time using their stellar particles and were found 
to have unusually large vertical motions and long migration times, 
thus exhausting their gas supply while still near the disc edge.

\textbf{Ex-situ clumps:} 
The distribution of relative SFR in \exsitu clumps appears to be even 
broader than the distribution of relative mass, and is nearly uniform 
in the range $0.001-1$. The co-rotating \exsitu clumps have systematically 
higher SFRs than the non-co-rotating ones, in addition to higher masses. 
The average relative SFRs are 0.12 and 0.004 for the co-rotating and 
non-co-rotating clumps respectively. Possible imlications of this will 
be discussed in \se{discussion}.

\textbf{Bulge clumps:} 
The SFR in the \textit{bulge} clumps tends to be higher than in the off-center 
clumps, with the average of SFR$_{\rm c}$/SFR$_{\rm d}$ at 0.23 
and a FWHM of 0.8 dex. In about 10\% of the cases the SFR in the 
central $\bulg$ clump is higher than the disc total. Such high SFRs 
characterize the \bulg clumps as compact "blue nuggets", resulting from 
intense gas inflow into the disc center \citep{Cheung12,Barro13,DekelBurkert13}. 

\subsection{Star-Formation in Clumps I: Gas Fraction}

The second and third rows of \fig{hist} present distributions of 
properties relating to star-formation within the clumps: gas fraction, 
sSFR, stellar age and metallicity. We do not attempt to produce 
reliable colors prior to implementing dust and radiative tranfer.
However, ignoring dust reddening, blue colors should be associated 
with high gas fractions, high sSFRs, young ages and low metallicities. 

The left-center panel in \fig{hist} refers to the gas fractions 
with respect to the baryonic mass within the clumps. While the 
overall gas fractions at $z \sim 2$ are underestimated in the 
current simulations (\se{sim}), the relative values and 
distributions for the different clump types are likely to be 
more reliable. 

\textbf{In-situ clumps} 
The \insitu clumps are fairly gas rich, far more so than the other clump 
types, as expected from their formation by instability in gas-rich discs. 
The median of the distribution of gas fraction is at roughly 0.08, with the 
peak at $\sim 0.09-0.1$ and a FWHM of $\sim 0.8$ dex. There is a tail 
extending to low gas fractions, as well as a significant population of very 
gas rich clumps. Roughly 14\% of the clumps have gas fractions below 0.03 and 
3\% have fractions below 0.01. This gas-poor tail is mostly due to {\it Is/Es} 
clumps with old stellar ages, corresponding to the low SFR tail discussed above. 
At the gas-rich end, 12.5\% of the clumps have gas fractions above 0.3, 8\% have 
values above 0.50 and 3.5\% have gas fractions above 0.75. These very gas rich 
clumps have very low stellar masses (though not unusually large gas masses) and 
young stellar ages, with high sSFR and low metallicity. They are located near 
the edge of the disc and appear to be newly formed clumps undergoing an initial 
burst of star-formation.

\textbf{Ex-situ clumps} 
The \exsitu clumps have on average much lower gas fractions than the \insitu 
clumps, with a median value of about 0.01. However, the co-rotating clumps 
have systematically higher gas fractions than the non-co-rotating ones, giving 
rise to an almost bi-modal distribution. The co-rotating clumps have an average 
gas fraction of roughly 0.02, with the peak of their distribution at 0.04 and 
10\% of them having gas fractions above 0.06. On the other hand, 
the average gas fraction of non-co-rotating \exsitu clumps is $\sim 0.007$, 
with the peak roughly at 0.01. Only 4\% of them have gas fractions greater 
than 0.06, while over 15\% have values below 0.003, compared to only 3\% of 
the co-rotating clumps. This may indicate late gas accretion onto co-rotating 
\exsitu clumps, as discussed further in \se{discussion}. 

\textbf{Bulge clumps} 
The $\bulg$ clumps have a simillar distribution to the non-co-rotating 
\exsitu clumps. The average gas fraction is 0.006 and the FWHM = 0.6 dex. 
There is a tail towards higher values, with roughly 3\% 
of the clumps having gas fractions above 0.03. These correspond to 
young bulges at redshifts $z>3$. 

\subsection{Star-Formation in Clumps II: sSFR}

The right-center panel in \fig{hist} refers to the sSFR in the clumps, 
defined as SFR divided by stellar mass. Given our crude estimate 
for SFR, this amounts to $(M_{\rm *,\:c}(<\Delta t)/\Delta t)/M_{\rm *,\:c}$, 
where $\Delta t = 30\Myr$ and $M_{\rm *,\:c}(<\Delta t)$ is the mass in stars 
younger than $\Delta t$ within the clump. This has a maximum possible value 
of $1/\Delta t=33.33 \Gyr^{-1}$, attained when 
$M_{\rm *,\:c}(<\Delta t)=M_{\rm *,\:c}$, 
or in other words when all of the stars in the clump are younger than $30\Myr$. 

\textbf{In-situ clumps} 
The \insitu clumps have very high sSFRs, with a median value of 
$\sim 2.1 \Gyr^{-1}$ ($\sim 2.5 \Gyr^{-1}$ without the \textit{Is/Es} 
clumps), a natural result of their high gas fractions. The distribution 
peaks at around $1.7 \Gyr^{-1}$ and has a FWHM spread of $\sim 0.9$ dex. 
With such high sSFRs and gas fractions, the \insitu clumps should appear 
quite blue in observations, provided that there is no significant dust 
reddening. 

There is, however, a tail towards low sSFR, with 4\% of the clumps 
having values below $0.1 \Gyr^{-1}$. These clumps also occupy the 
tail of low gas fractions discussed above, and are dominated by 
{\it Is/Es} clumps whose stellar ages are over $300 \Myr$ near the 
outskirts of the disc. Note that there are also 
many {\it Is/Es} clumps with high sSFRs of $2 \Gyr^{-1}$ or more, 
similar to the bulk of the \insitu population and higher than the 
typical values for \exsitu clumps. 

Also interesting is the prominence of a second peak in the distribution, 
at around $\sim 30 \Gyr^{-1}$. The peak itself and its sharp break at higher 
values is a numerical artifact, due to our crude estimate of SFR, which does 
not resolve star-formation on timescales shorter than $30\Myr$. This causes 
all the clumps which might have had sSFRs above $30 \Gyr^{-1}$ to pile up 
around this value, even though the intrinsic distribution is more likely an 
extended tail to very high sSFR values. A robust statistic is that roughly 
16\% of the \insitu clumps have sSFRs higher than $10 \Gyr^{-1}$. These 
extremely efficient star-forming clumps coincide with the very gas rich 
clumps discussed above, with an average gas fraction of 0.46. They have 
very young ages of $65\Myr$ on average. Their stellar masses are low, 
with an average of $8\times 10^7\msun$. Their metallicities are sub solar, 
averaging to (${\rm log(}(O/H)+12\simeq 8.63$ and $8.40$ for the gas and 
stars, respectively. Finally, they tend to be located near the edge of 
the disc. In other words, these are newly formed clumps undergoing their 
first burst of star formation. 

\textbf{Ex-situ clumps} 
The \exsitu clumps have much lower sSFR than the \insitu clumps, 
similar to their relative gas fractions. We thus expect \exsitu 
clumps to appear redder in observations. The median sSFR is 
$0.2 \Gyr^{-1}$, but the distribution is very broad and skewed 
towards low values. About 12\% of the \exsitu clumps have sSFR 
below $0.01 \Gyr^{-1}$ while 6\% have values above $2\Gyr^{-1}$. 
As expected from their bi-modality in gas-fraction, the co-rotating 
\exsitu clumps have significantly higher sSFR values than the 
non-co-rotating ones, with median values of 0.5 and 0.06$\Gyr^{-1}$ 
respectively. 

\textbf{Bulge clumps} 
The \bulg clumps have a median sSFR of $0.2\Gyr^{-1}$ and a fairly symmetric 
distribution about this value, with a standard deviation of 0.37 dex (FWHM = 
0.7 dex). There is a tail towards high sSFR, with 4.5\% of the clumps having 
values above $1\Gyr^{-1}$, corresponding to those with high gas fractions.

\subsection{Star-Formation in Clumps III: Stellar Age}

The bottom left panel of \fig{hist} refers to the mass-weighted 
mean stellar age of the clumps, calculated with all the stars 
present in the clump.

\textbf{In-situ clumps} 
The \insitu clumps are fairly young, with a median age of 
$\sim 160\Myr$ and a broad peak from $\sim 150-300\Myr$, 
on the order of 1-2 orbital times at the edges of the discs, 
which is the expected migration time of clumps to the disc 
center \citep[DSC09;][]{CDB,Ceverino12}. The FWHM of the distribution 
is 0.65 dex, but it is skewed towards younger ages. 30\% of 
the clumps are younger than $100\Myr$ and 13\% are younger 
than $50\Myr$. On the other hand, only 20\% of the clumps are 
older than $300\Myr$ and 7\% are older than $500\Myr$, most 
of them {\it Is/Es} clumps, where more than 50\% of the mass 
is in stars formed outside the disc. \Insitu clumps are not 
expected to survive for much longer than a couple of orbital 
times, which explains the sharp drop of the distribution towards 
older ages. We note here that a model where clumps distrupt 
shortly after they form due to intense feedback would predict 
much younger ages for clumps, of order $\lsim 50\Myr$, with a 
much smaller age spread (see discussion in \se{survival}). 

We find the \insitu clumps to be, on average, older at lower 
redshift. At $4.0<z<2.5$, their average age is $\sim 100\Myr$, 
as opposed to $\sim 200\Myr$ at $1.0<z<1.5$. The standard error 
of the mean in these two redshift bins is only 0.04 dex and 0.02 
dex, respectively, so the difference appears significant. This 
is interpreted as an increase in the disc orbital time at lower 
redshifts, due to the cosmological density decrease, and a 
corresponding increase in the clump migration time.

\textbf{Ex-situ clumps} 
The \exsitu clumps are much older than the \insitu clumps. 
Their median age is $\sim 1.1\Gyr$, while their distribution 
peaks at $1.6 \Gyr$ with a FWHM spread of 0.7 dex. The 
distribution is very similar to that of the \bulg 
clumps, unsurprising, as the \exsitu clumps are  
largely bulges of little merging galaxies. We also note an 
increase in the average age of \exsitu clumps from high to 
low redshifts. At $4.0<z<2.5$ the average age is $\sim 620\Myr$ 
while at $1.0<z<1.5$ it is $\sim 2200\Myr$, with a standard 
error of the mean of less than 0.02 dex. This variation was 
expected, since the \exsitu clumps are simply external, 
merging galaxies which age along with the Universe. At all 
redshifts, the \exsitu clumps have systematically lower gas 
fractions and sSFRs and much older ages than the \insitu clumps, 
and will likely appear systematically redder. 

\textbf{Bulge clumps} 
Unsurprisingly, the \bulg clumps contain old stellar populations 
with median ages of $\sim 1.3 \Gyr$ (FWHM = 0.5 dex). At all 
redshifts, \exsitu and \bulg clumps have very simillar age 
distributions. 

\subsection{Star-Formation in Clumps IV: Metallicity}
The bottom right panel of \fig{hist} refers to gas-phase 
metallicity in units of ${\rm log}(O/H)+12$. 
Solar metallicity in these units is 8.69 \citep{Asplund09}. 
Note that the same effect which causes an underestimate of 
gas fractions in the current simulations may also cuase 
elevated metallicity values in our galaxies, but we can 
still trust the relative values of the different clump types. 

\textbf{In-situ clumps}
The metallicity in the \insitu clumps is high, 
and nearly all clumps have super solar metallicity in both gas 
and stars. They have a median 
gas-phase metallicity of ${\rm log}(O/H)+12 \sim 8.9$, with a 
FWHM spread of 0.35 dex. There is also a tail extending to 
sub-solar values, containing 13\% of the clumps, and dominated 
by the young, gas-rich, star-forming clumps discussed above.

\textbf{Ex-situ clumps}
The distribution of gas-phase metallicity for the \exsitu 
clumps closely resembles that of the \insitu clumps in the 
median, peak and width of the distribution. The \exsitu clumps 
also exhibit a simillar tail to sub-solar values, dominated by 
the co-rotating clumps. This result suggests that most of the gas 
found in those \exsitu clumps which have gas, i.e. the 
co-rotating ones, was accreted from the disc and is the same gas 
that formed the \insitu clumps.

\textbf{Bulge clumps} 
The \bulg clumps have the highest metallicity values, with a median value 
of 9.1 and a narrow FWHM of 0.3 dex. Less than 1\% of the clumps have sub-solar 
metallicity.

\section{Gradients of Clump Properties}
\label{sec:clump_grad}

\begin{figure*}
\centering
\subfloat{\includegraphics[width =0.39 \textwidth]{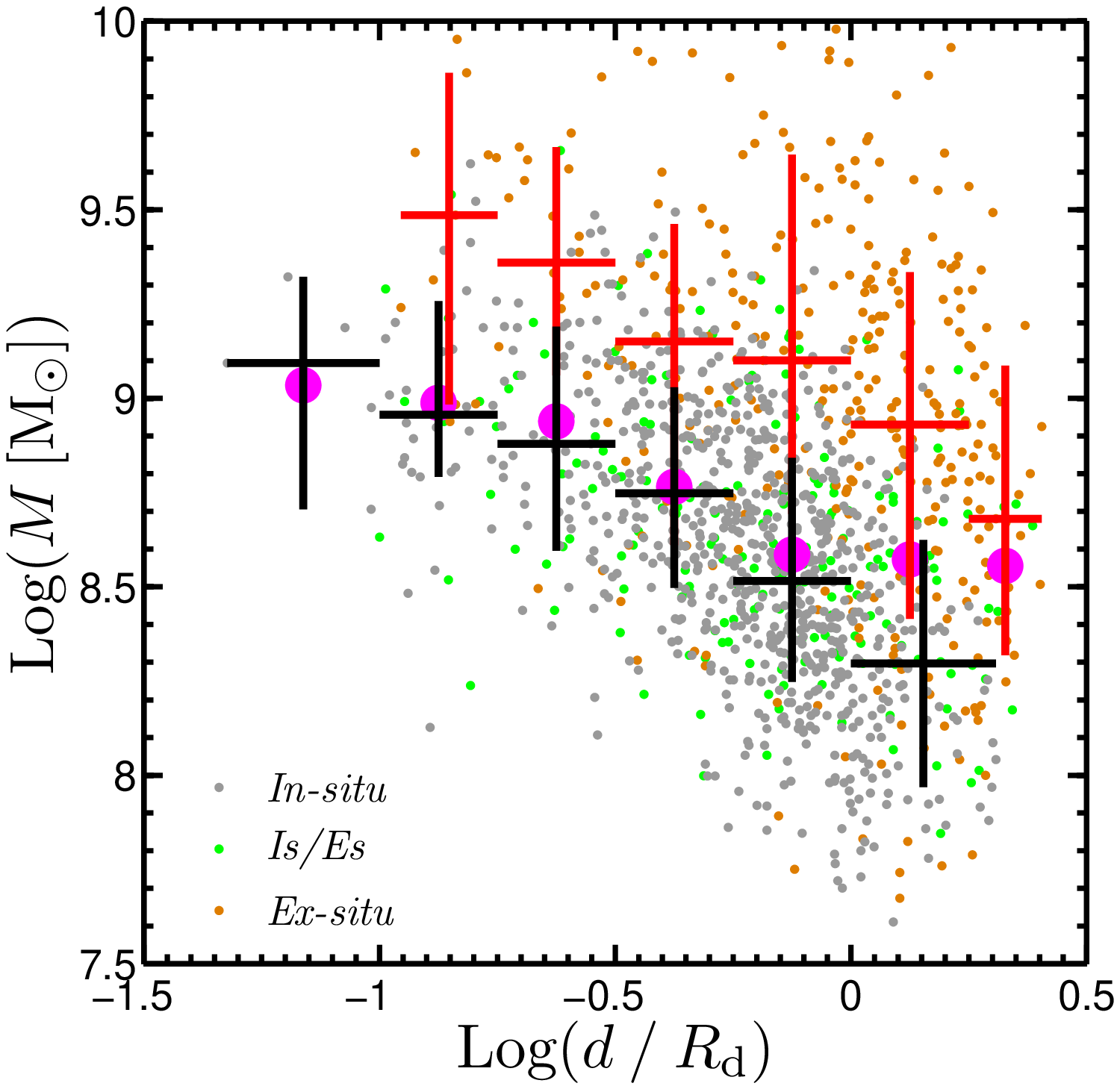}}
\subfloat{\includegraphics[width =0.39 \textwidth]{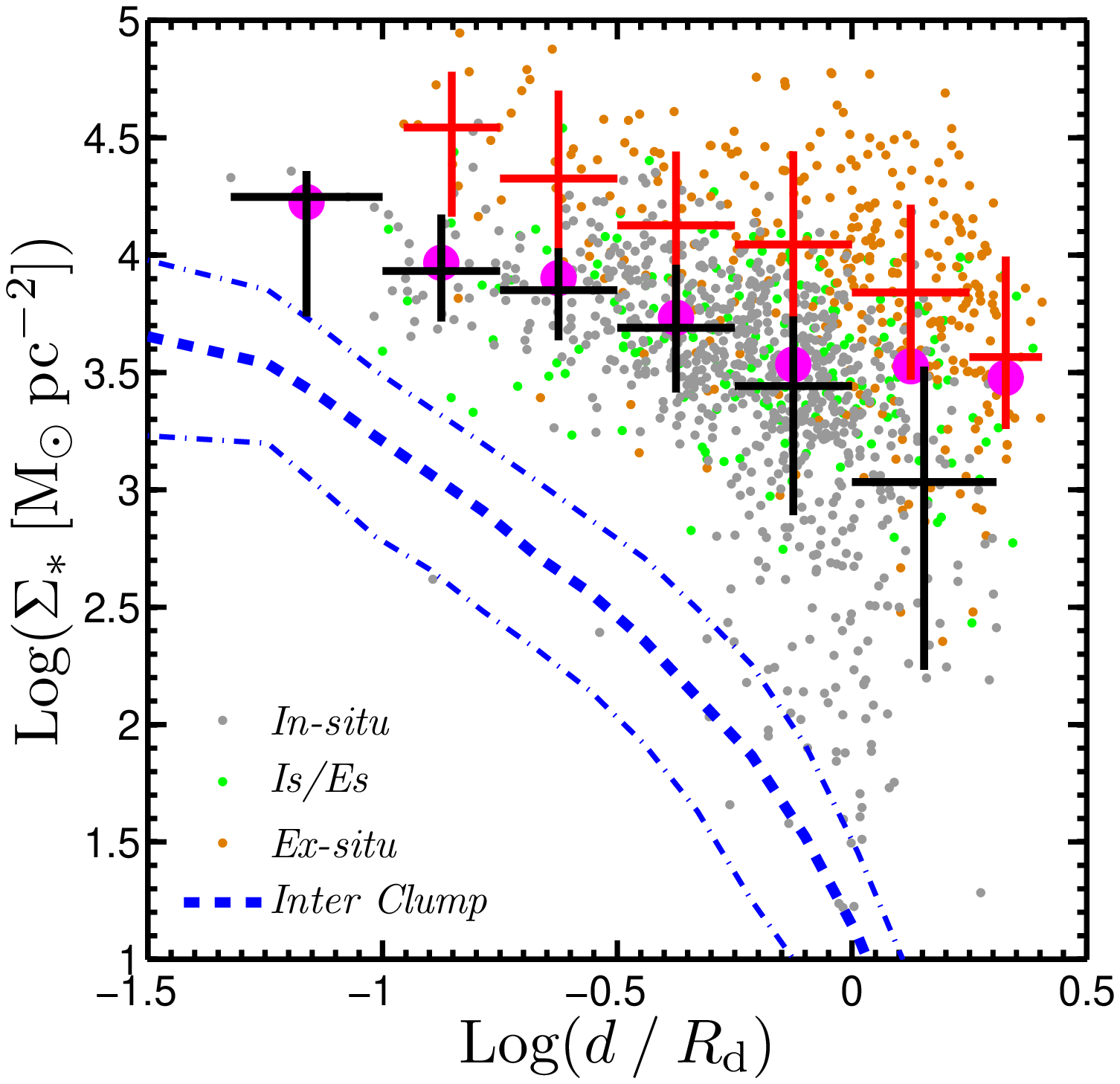}}\\
\subfloat{\includegraphics[width =0.39 \textwidth]{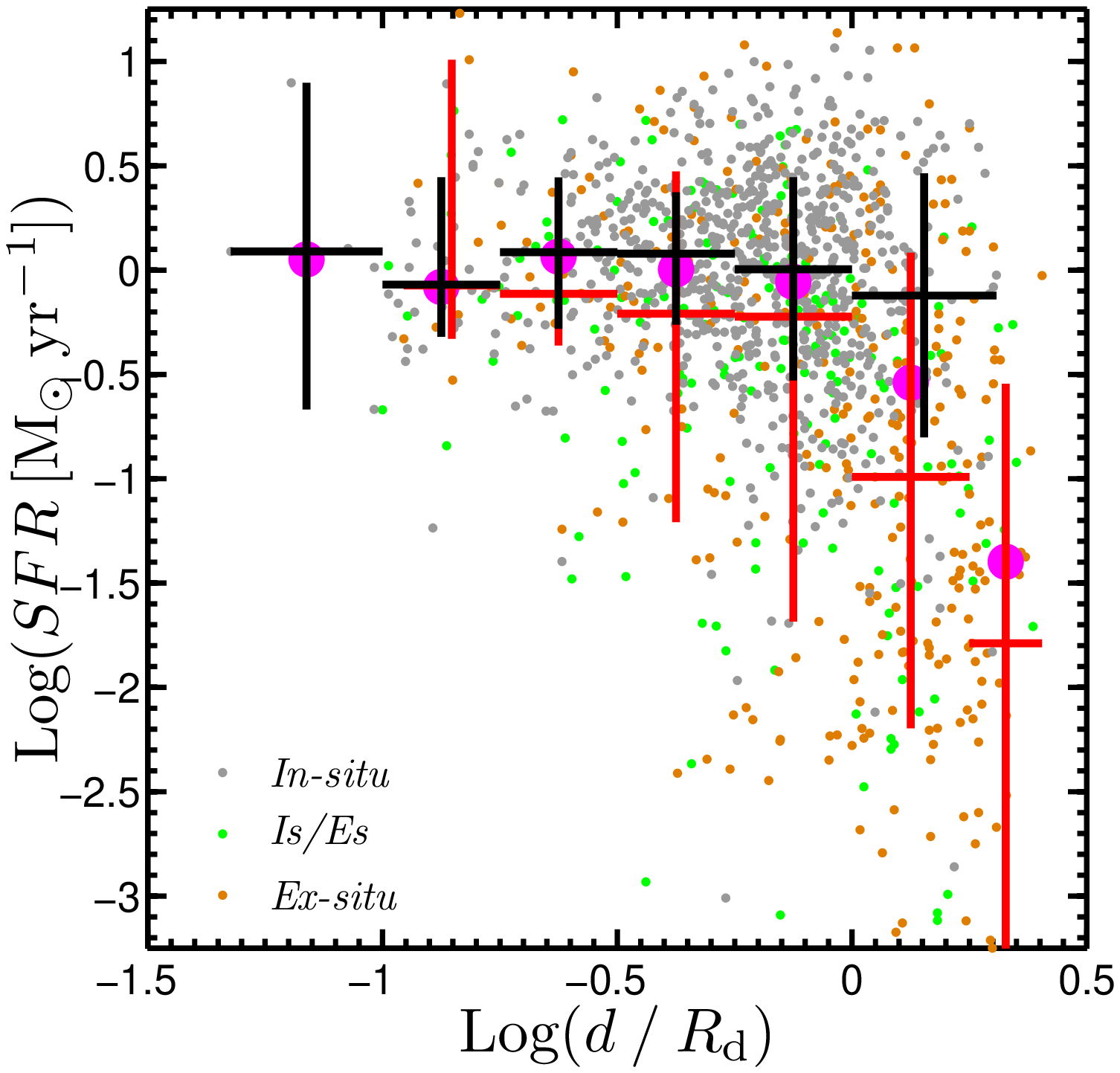}}
\subfloat{\includegraphics[width =0.39 \textwidth]{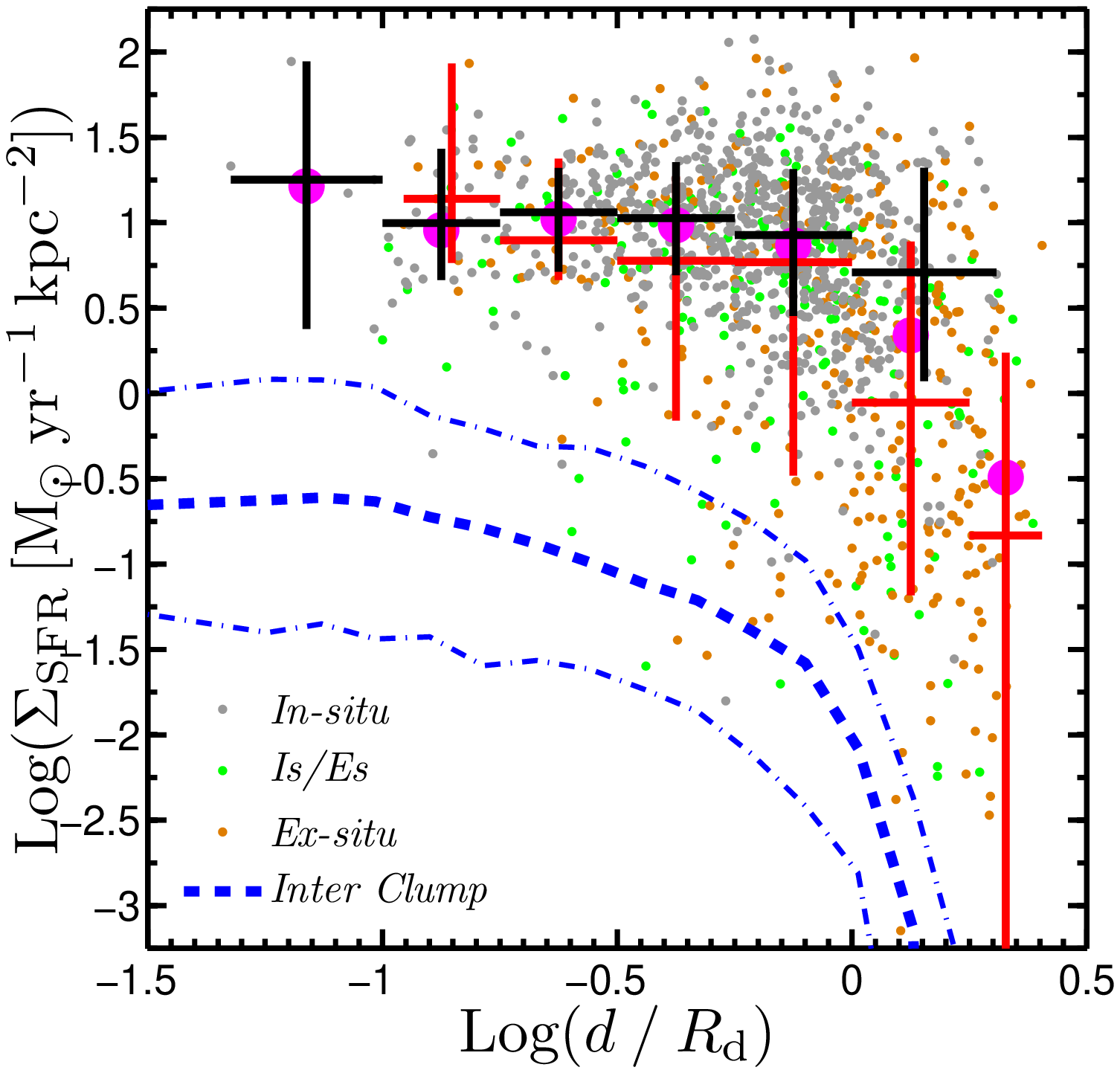}}\\
\caption{Gradients of various properties across the disc. 
The x axes are all ${\rm log(}d\:/\: \Rd{\rm )}$, where 
$d$ is the galactocentric distance and $\Rd$ is the radius 
of the host disc. The grey points are the values for the 
\insitu clumps without the suspicious {\it Is/Es} cases, 
which are marked in green. The orange points are for the 
\exsitu clumps. The black crosses show the median values 
for the \insitu clumps in 6 equally spaced bins of 
${\rm log(}d\:/\: \Rd{\rm )}$. The horizontal bar spans 
the width of the bin while the vertical bar marks the 
$67 \%$ scatter about the median. The red crosses show 
the same, but for the \exsitu clumps. Large magenta 
circles show the median values for all off-center clumps. 
The thick blue dashed lines in the two right-hand panels 
show the median profiles for the inter-clump medium in all 
our clumpy discs (the discs hosting the plotted clumps). 
These values were calculated in rings extending the thickness 
of the disc in the vertical direction with all clumps removed 
from them. The thin dot-dashed blue lines mark the $67 \%$ 
scatter about the median. 
\textbf{Top row:} Baryonic mass (left) and stellar 
surface density (right). 
\textbf{Bottom row:} SFR (left) and SFR surface density 
(right). 
Note that the masses and SFRs are absolute values, not normalized to 
the disc total. The clump surface densities were calculated in a 
face on projection of the disc, using their projected surface areas. 
The \insitu clumps are more massive nearer the disc center though 
their SFR values have no systematic radial variation. 
}
\label{fig:grad} 
\end{figure*} 

\begin{figure*}
\centering
\subfloat{\includegraphics[width =0.39 \textwidth]{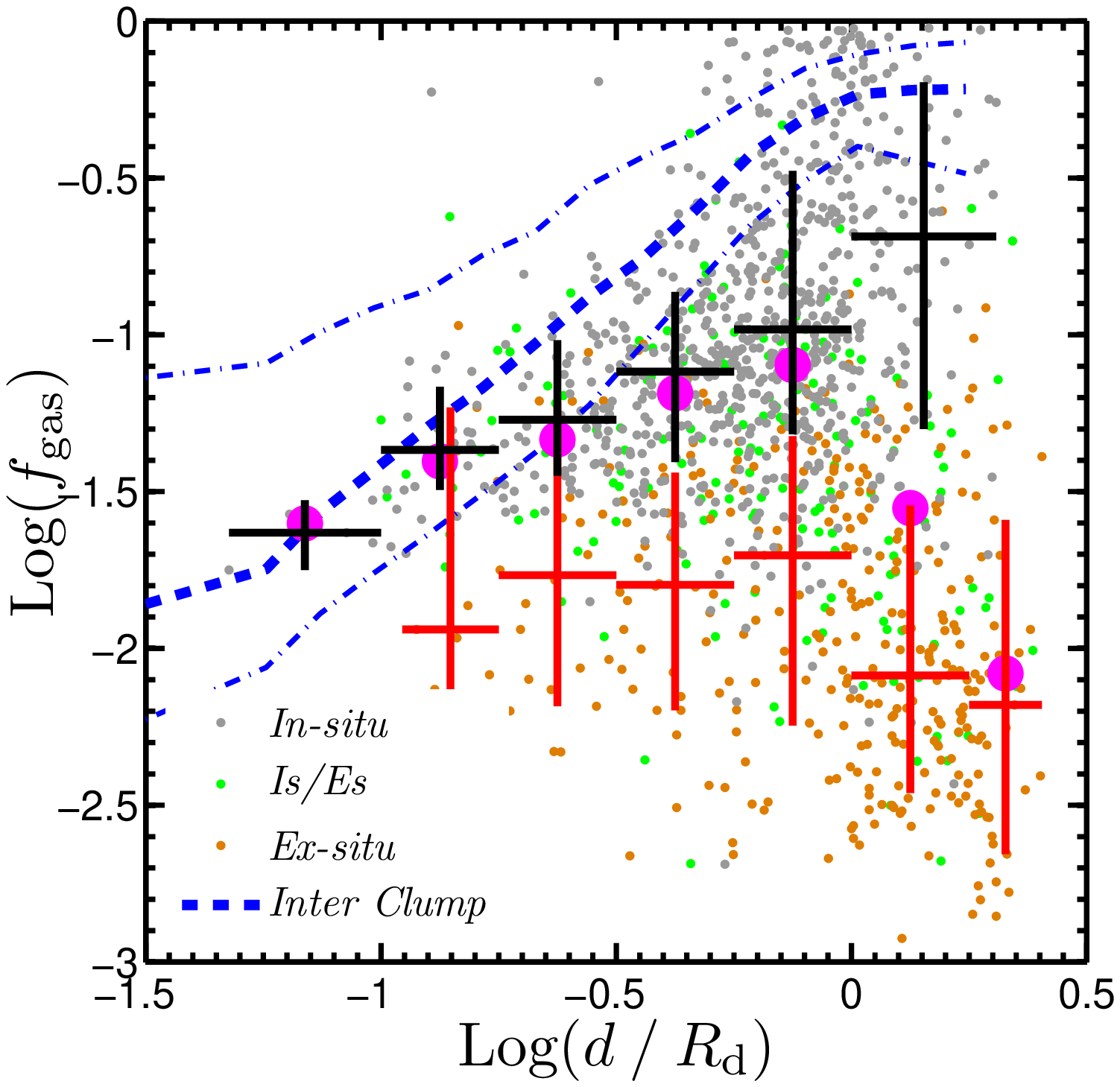}}
\subfloat{\includegraphics[width =0.39 \textwidth]{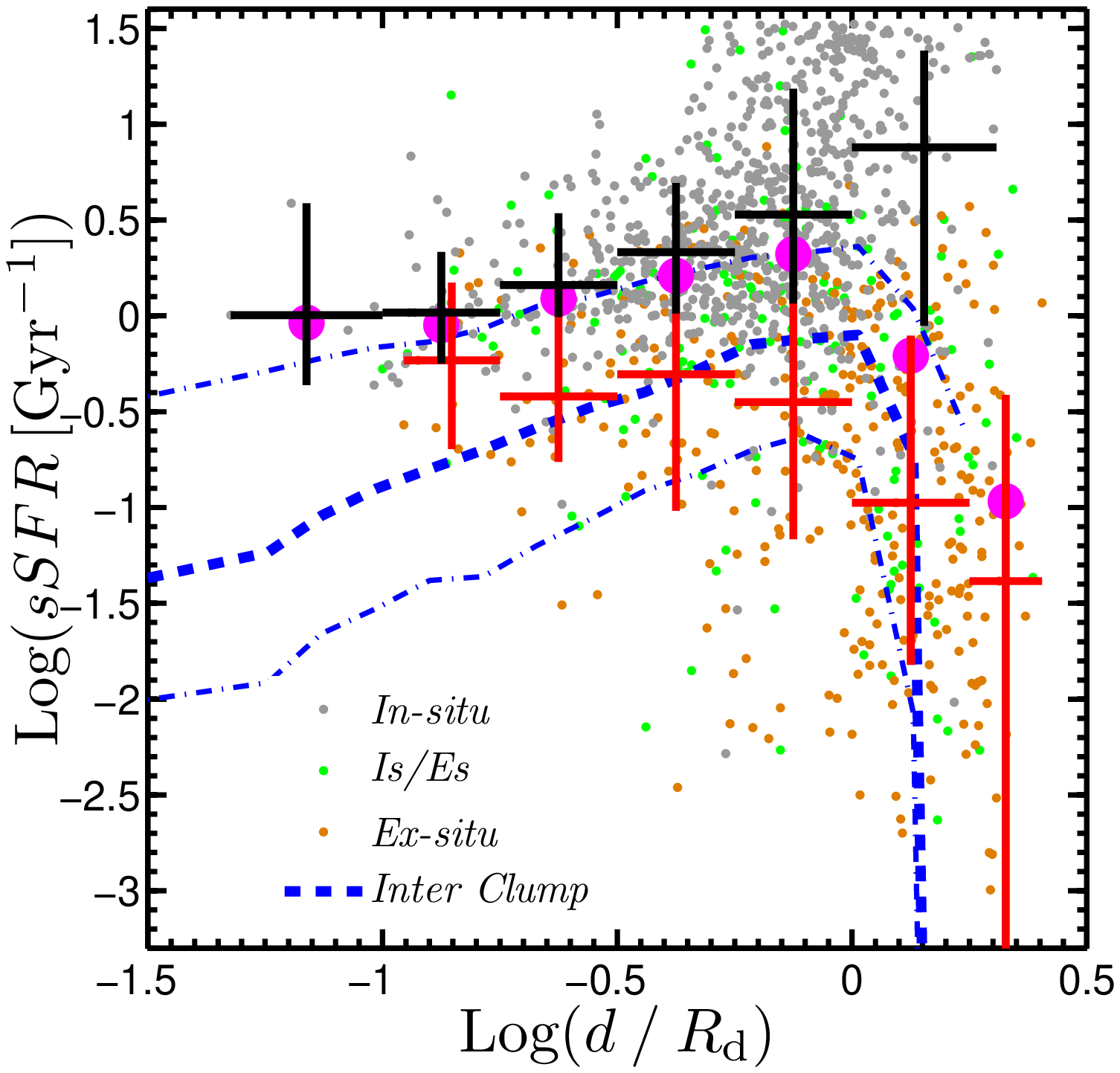}}\\
\subfloat{\includegraphics[width =0.39 \textwidth]{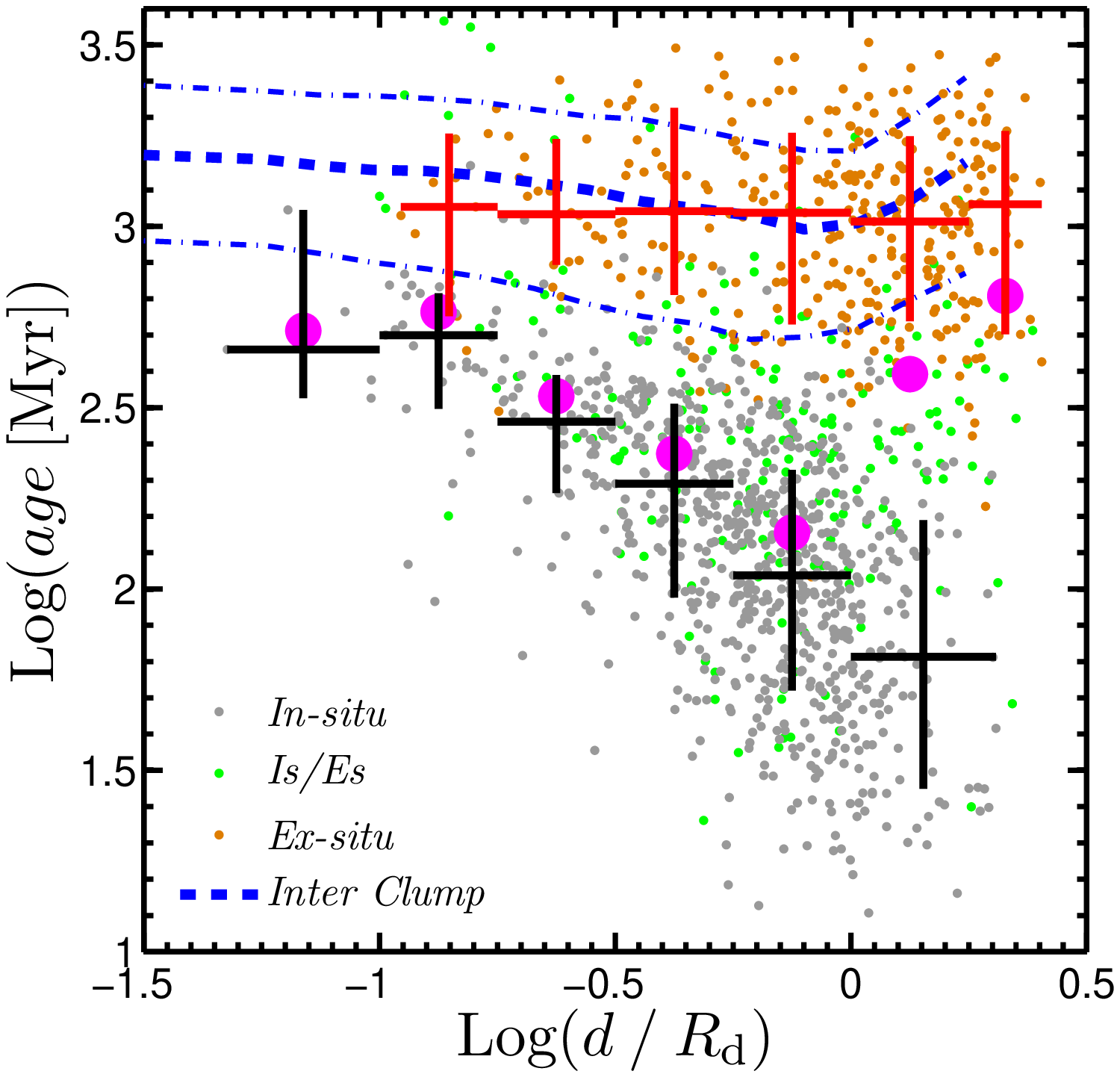}}
\subfloat{\includegraphics[width =0.39 \textwidth]{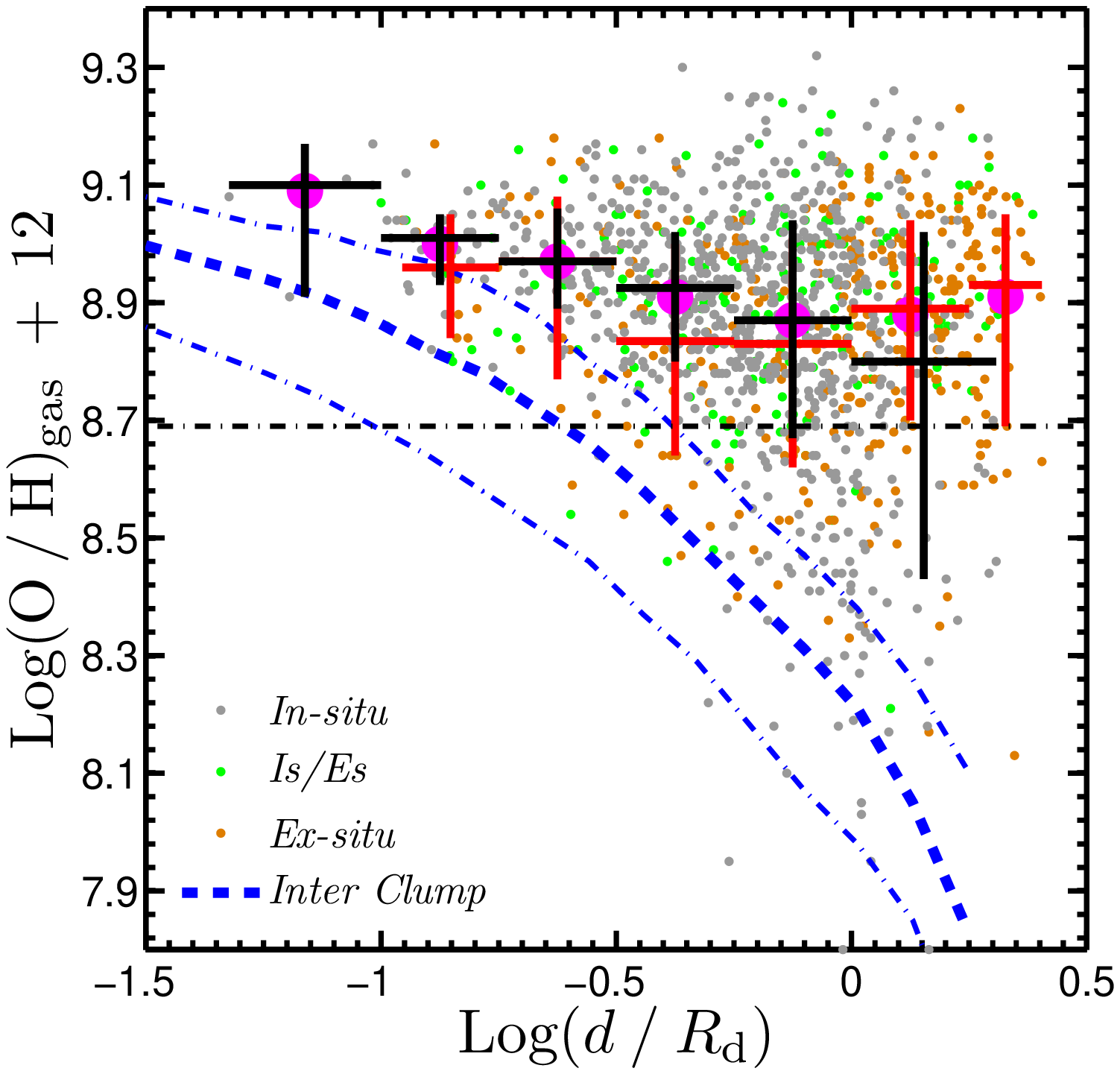}}
\caption{Same as \fig{grad}, gradients of clump properties across the disc. 
\textbf{Top row:} Gas fraction (left) and sSFR (right). 
\textbf{Bottom row:} Stellar age  (left) and gas metallicity 
(right, with solar metallicity marked).
The \insitu clumps are older, with lower specific SFRs, lower gas 
fractions and higher metallicities nearer the disc center and the 
gradient is steeper than the underlying smooth component. Near the disc edge, 
the \exsitu clumps have older stellar ages, lower gas fractions and lower sSFR 
values.
}
\label{fig:grad2} 
\end{figure*} 

Here we investigate the properties of the off-center clumps as a 
function of distance from the disc center, and especially search 
for systematic radial variations that characterize the \insitu 
and the \exsitu clumps in comparison with the corresponding 
gradients in the background disc. The theoretical implications 
of such variations for distinguishing between clump migration 
versus rapid disruption are discussed in \se{survival}. In the 
profiles shown in \fig{grad} and \fig{grad2}, the radial distance 
$d$ is measured in terms of the disc radius $\Rd$. Points represent 
the \insitu clumps (grey circles and green diamonds) and \exsitu 
clumps (red squares). The corresponding medians and 67\% scatter 
within bins of $d/\Rd$ are marked. The medians for all off-center 
clumps (large magenta circles) follow the values for \insitu clumps 
within the disc radius. Significant contamination from \exsitu clumps 
is visible only at $d \ge \Rd$. The radial profiles in the inter-clump 
disc are shown for comparison (dashed blue).

\subsection{Mass}

The top left panel of \fig{grad} refers to the baryonic 
mass of clumps and the top right panel refers to the 
profie of stellar surface density within the clumps.  

\textbf{In-situ clumps}
The \insitu clumps show a strong gradient in mass, with the 
median values exhibiting a logarithmic slope of $-0.60 \pm 0.06$. 
The profile is better fit by a broken power law, with a slope of 
$-0.85 \pm 0.07$ exterior to $0.3\Rd$ and a slope of $-0.40 \pm 0.08$ 
interior to this radius. The median mass increases from 
$2-3\times 10^8 \msun$ near $\Rd$ to $\sim 8\times 10^8 \msun$ near 
$0.1 \Rd$. This is consistent with the clumps preferentially forming 
at large radii and accreting mass from the disc as they migrate inward 
(\se{survival}). Alternatively, clumps may also form at small radii and 
be more massive there to begin with, possibly due to the larger fraction 
of cold mass at small radii. However, the gradient in clump ages (\fig{grad2}) 
argues against this, as there are almost no clumps younger than $100\Myr$ 
interior to $0.4 \Rd$.

Nearly all the \insitu clumps have stellar surface densities higher than 
the local background disc by an order of magnitude or more. This is 
reassuring, since our clumps were identified in gas and it was not clear 
a priori that they would contrast strongly with the stellar background 
in the disc and could be associated with observed stellar clumps. 
The exceptions, with stellar surface densities only a few times greater 
than the background, are all found at large radii and correspond to young, 
gas rich clumps with high sSFR. These clumps may not be observable in 
stellar maps of the galaxy, but may be detectable in \Halpha. 
The best-fit power law to the profile of medain values in the range 
$0.1<d/\Rd<1$ yields a slope of $-0.65 \pm 0.13$, shallower 
than the corresponding slope of $-1.88 \pm 0.1$ in the background disc. 
The median clump surface density increases from 
$\Sigs \sim 10^3 \msun \pc^{-2}$ at $\Rd$ to 
$\Sigs \sim 10^4 \msun \pc^{-2}$ near $0.1\Rd$.

\textbf{Ex-situ clumps}
The \exsitu clumps tend to be more massive than the \insitu clumps at 
every radius, by a factor of 3-4 in the median. The median profiles 
appear to have similar slopes. This may indicate that \exsitu clumps, 
especialy the co-rotating ones, also migrate inward and accrete mass 
from the disc. However, the scatter of masses for the \exsitu clumps 
is larger, as massive \exsitu clumps are naturally present at all radii.
Simillarly, the median surface densites of the \exsitu clumps are a 
factor 3-4 higher than those of the \insitu clumps at all radii. At 
large radii one can find \exsitu clumps with surface densities three 
orders of magnitude above the background. This would be a clear signal 
for an \exsitu clump, as \insitu clumps are not expected to exhibit 
such high stellar surface densities, especially near the disc edge. 
Closer to the center, the difference between the \insitu and \exsitu 
clumps becomes less pronounced, as the \insitu clumps have formed many 
more stars during their migration. 

\subsection{SFR}
The bottom left panel of \fig{grad} refers to the SFR in clumps 
and the bottom right panel refers to the profie of star-formation 
rate surface density.  
The sharp drop in the disc ${\Sigma}_{\rm SFR}$ near $\Rd$ corresponds 
to the edge of the gas disc as defined using the density profile 
of cold gas, beyond which stars simply cannot form. 

\textbf{In-situ clumps}
The \insitu clumps do not show a systematic radial variation of the SFR, 
with the median values quite constant at $1-2\sy$. 
This seems to require a replenishment of gas supply during migration inward, 
again hinting at accretion of fresh gas from the disc into the clumps. 
The \insitu clumps also exhibit a fairly constant SFR surface density, 
${\Sigma}_{\rm SFR}\sim 10-20 \sy \kpc^{-2}$. This is a factor of $\sim 50-100$
above the local background, making the clumps appear as distinct peaks of 
enhanced SFR. 
The outliers with low $\Sigs$ discussed above 
have particularly high ${\Sigma}_{\rm SFR}$ values. 
In both mass and SFR, the {\it Is/Es} clumps are indistinguishable from the 
\insitu clumps. 

\textbf{Ex-situ clumps}
The \exsitu clumps tend to have lower SFRs, and especially so
near or outside $\Rd$, where the median SFR drops to below $0.1 \sy$. 
Interior to $\Rd$, the median SFR in the \exsitu clumps is on the order of 
$0.5 \sy$, albeit the scatter is much larger than for the \insitu clumps.
Their ${\Sigma}_{\rm SFR}$ exhibits a simillar behaviour.
At all radii interior to $\Rd$, the \exsitu clumps have values a factor of 
$\sim 2$ below the \insitu clumps, but still more than an order of magnitude 
above the background. 

\subsection{Gas Fraction, sSFR and Age}

The top two panels of \fig{grad2} refer to the gas fraction (left) and 
sSFR (right), while the bottom left panel refers to the stellar age. 
These three properties trace the star-formation within the clumps and 
best distinguish the \insitu from the \exsitu clumps.

\textbf{In-situ clumps}
Between $0.1 \Rd$ and $\Rd$, the \insitu clumps' 
median gas fraction has a logarithmic slope of $+0.52 \pm 0.04$. 
Inclduing clumps at larger and smaller radii gives a slope of 
$0.67 \pm 0.05$. This is shallower than the underlying gradient in 
the inter-clump background, which has a slope of $1.24 \pm 0.02$. 
Near $0.1\Rd$, the median gas fraction in \insitu clumps is roughly 
3\%, comparable to the background (but recall that the absolute 
values in the current simulations at $z \sim 2$ are underestimates, 
\se{sim}). Near $\Rd$, however, the median gas fraction in \insitu 
clumps is about 10-20\%, roughly a factor of 3 below the background.
Many clumps near $\Rd$ are almost totally gaseous, and thus undetectable 
in stellar maps; these are the clumps with very low $\Sigs$, which make 
up the high end of the gas fraction distribution in \fig{hist}. 
On the other hand, we note that the gas mass of the \insitu clumps 
(not shown) does not exhibit a clear trend with distance. The above is 
consistent with a preferred formation of \insitu clumps at large radii, 
followed by accretion of gas from the disc onto the clumps such that 
the decreasing gas fractions at smaller distances result from increasing 
stellar masses due to star formation. 

The \insitu clumps constitute peaks in the sSFR distribution, with median 
values a factor of $\sim 5$ above the disc at all radii. In the range 
$0.1\Rd<d<\Rd$, the logaritmic slope of the median sSFR in \insitu clumps 
is $+0.68 \pm 0.04$, though this steepens to $+0.82 \pm 0.09$ including 
clumps at $d \ge \Rd$. For the inter-clump material the slope is $+0.84 \pm 0.03$. 
Near $\Rd$ the \insitu clumps exhibit a median sSFR of $\sim 5 \Gyr^{-1}$, 
though many clumps have values above $10 \Gyr^{-1}$, as we saw in \fig{hist}. 
These are newly formed clumps with very high gas fractions and young ages. 
Near $0.1\Rd$, the median value drops to $1\Gyr^{-1}$. 

The age gradient of the \insitu clumps is rather steep. The median age exterior 
to $0.1\Rd$ has a logaritmic slope of $-0.86 \pm 0.03$, rising from $\sim 80\Myr$ 
near $\Rd$ to $\sim 500\Myr$ near $0.1 \Rd$. This is consistent with the clumps 
starting to form stars in the outer disc and then gradually migrating inwards. 
The age gradient in the smooth inter-clump component is much shallower, with a 
slope of $-0.20 \pm 0.01$ in the same range. This suggests a separate evolution 
of the clumps and the smooth component. 

\textbf{Ex-situ clumps}
The \exsitu clumps show no clear radial gradient in any of the properties 
related to star formation, except for a tendency of clumps interior to 
$\Rd$ to have higher gas fractions and sSFRs than those outside $\Rd$. 
Interior to $\Rd$, the \exsitu clumps exhibit median gas fractions of 1-2\%, 
sSFRs of $\sim 0.4 \Gyr^{-1}$ and ages of $\gsim 1\Gyr$, roughly independent 
of distance. Beyond $\Rd$, the gas fractions and sSFR drop sharply while 
the age remains constant. The profiles of age and sSFR in the \exsitu clumps 
appear very similar to those of the background disc at all radii. 

Since the \insitu clumps dominate the off-center clump population 
within $\Rd$, the gradients in this region do not change dramatically 
even if the \insitu and \exsitu populations are mixed together. The gradients of 
the median values for clump gas fraction, sSFR and stellar age including 
all the off-center clumps in the range $0.1\Rd < d < \Rd$ are, respectively, 
$0.43 \pm 0.04$, $0.49 \pm 0.02$ and $-0.79 \pm 0.05$, only slightly shallower 
than the values quoted above for the \insitu clumps only. The signal should 
therefore still be detectable observationally, provided the clumps are properly 
resolved from the inter-clump disc component. At large distances $d \ge \Rd$, 
however, significant contamination from \exsitu clumps is evident, which causes 
a distinct break in the profile.

The difference between the properties associated with gas and star 
formation of the \insitu and \exsitu clumps can provide an observable 
distinction between them. We see that massive clumps near the disc edge 
that have old ages, low sSFR and low gas fractions, namely are red in 
color, are almost exclusively \exsitu clumps.  \Insitu clumps at 
these large radii are dominated by young stars and have high sSFR and gas 
fractions, namely they should be blue unless reddened by dust. These 
differences are less pronounced at smaller radii, as the stars in the 
\insitu clumps have partly aged and their sSFR and gas fractions have 
decreased, while the \exsitu clumps have not evolved noticeably. 

The population of {\it Is/Es} clumps have values in between the values for 
the two main populations. However, they exhibit a large scatter and there is 
no significant gradient. This population could thus weaken the observable 
gradients of the \insitu clumps, but this is hopefully a small effect.

\subsection{Metallicity}
The bottom right panel of \fig{grad2} refers to the gas metallicity. 
As we saw in \fig{hist}, the \insitu and the \exsitu clumps have fairly 
similar distributions. However, while the \exsitu clumps show no 
clear gradient with distance, the \insitu clumps systematically become 
more metal rich as they near the center, as expected from clump migration. 
Their median value increases from ${\rm log}(O/H)+12 \sim 8.8$ near $\Rd$ 
to a value of $\gsim 9$ near $0.1\Rd$, with a logarithmic slope of 
$-0.20 \pm 0.01$. This gradient is much shallower than in the inter-clump 
gas. While the metallicity is simillar for the clumps and the disc near 
the center, at large radii the clumps appear much more metal rich than 
the background, suggesting self enrichment of the clumps independently 
from the disc. Note that the lower envelope of clump 
metallicity, both for \insitu and \exsitu clumps, closely follows the 
background disc median, consistent with accretion of fresh gas from 
the disc onto the clumps which they then use to form new stars. The 
large tail of sub-solar metallicity clumps visible in \fig{hist} 
exists only at large radii, and corresponds to the very young, very gas 
rich and highly star forming clumps. 

\section{Discussion} 
\label{sec:discussion} 

\subsection{Clump Survival versus Disruption} 
\label{sec:survival} 

One could compare two extreme scenarios concerning the life and fate 
of high-redshift giant clumps formed by VDI. 
In one scenario (I), as in our current simulations, the clumps, 
despite udergoing outflows, remain intact and even grow by 
accretion as they migrate into the disc center on an orbital 
timescale, $\sim 250\Myr$ at $z\sim 2$. In the competing 
scenario (II), the clumps disrupt on a dynamical timescale, 
$\sim 50\Myr$ at $z \sim 2$, well before they complete their 
migration. 

On the theory side, DSC09, based on \citet{DekelSilk86}, estimated 
that supernova feedback may not have enough power to disrupt the 
clumps on a dynamical timescale. \citet{murray10} argued that 
momentum-driven radiative stellar feedback could disrupt the clumps 
on a dynamical timescale, as it does in the local giant molecular 
clouds. Then \citet{KrumholzDekel} showed that such an explosive 
disruption is not expected to occur in the high-redshift giant clumps 
unless the SFR efficiency in a free-fall time is 
$\epsilon_{\rm ff}\sim 0.1$, much larger than what is implied by 
the Kennicutt relation at $z=0$. Such a deviation from the local 
Kennicutt law is inconsistent with observations 
\citep{Tacconi10,Tacconi12,Freundlich13}.
 
\citet{DK13} proposed instead that the observed outflows from
high-redshift clumps \citep{Genzel11,Newman12}, with mass loading 
factors of order unity, are driven by steady momentum-driven outflows 
from stars over many tens of free-fall times. Their analysis was 
based on the finding from simulations that radiation trapping is 
negligible because it destabilizes the wind \citep{KT12,KT13}. 
Each photon can therefore contribute to the wind momentum only once, 
so the radiative force is limited to $\sim L/c$. When combining 
radiation, protostellar and main-sequence winds, and supernovae, 
\citet{DK13} estimated the total direct injection rate of momentum 
into the outflow to be $\sim 2.5 L/c$. The adiabatic phase of 
supernovae and main-sequence winds can double this rate. They 
conclude that most clumps are expected to complete their migration 
prior to gas depletion. Furthermore, as described below, the clumps 
are expected to double their mass in a disc orbital time by accretion 
from the disc and clump-clump mergers, so their mass actually grows 
in time and with decreasing radius within the disc.

Given the uncertainties in the theoretical analysis, it is worthwhile 
to consider, simulate and compare different scenarios where the feedback 
strength is pushed to extreme limits, such as scenarios I and II. 
\citet{Genel12a} simulated scenario II by implementing in their SPH 
simulations a phenomenological, ``sub-grid", model for producing extreme 
galactic super-winds, and indeed got the clumps disrupted. \citet{Hopkins12} 
obtained similar results in isolated disc SPH simulations which implemented 
both short and long range radiative feedback. Trapping factors were calculated 
using the optical depth from the center of the clump to the surface, which is 
likely to overestimate the realistic effect of radiative feedback by a an 
order of magnitude \citep{KT12,KT13,DK13}. 
Detailed statistics of clump properties in scenario II are yet 
to be performed. The clump properties in our current simulations of scenario I, 
especially their gradients and relation to the off-clump gradients in the 
disc as discussed in \se{clump_grad}, can help us come up with observables 
that will distinguish between the two extreme scenarios.

The accretion of material from the disc onto the clumps during their 
migration is another distinct prediction of scenario I. Ignoring 
outflows from clumps, we can crudely estimate this accretion rate as 
follows. A collapsed clump accretes material from its tidal radius, 
which in a $Q\sim 1$ disc is roughly its Toomre radius, $R_{\rm T}$. 
This is the initial size of the proto-clump patch prior to collapse, 
where self gravity overcomes pressure and rotation support. 
Assuming that the surface density in the proto-clump was roughly equal 
to the mean surface density in the disc, the initial clump mass is 
$\Mc = \Md (R_{\rm T}/\Rd)^2$, where $\Md$ and $\Rd$ are the disc mass 
and radius. We parameterize the background density at the clump position 
as a factor ${\alpha}$ times the mean density in the disc: 
${\rho}=\alpha \Md/(2\pi \Rd ^2 \Hd)$. If the disc is exponential with 
a constant scale height, and $\Rd$ is roughly twice the effective radius 
(which contains 85\% of the mass) then ${\alpha}\sim 0.23$ at the disc 
edge and 1.23 at $0.5\Rd$. During its migration, the relative velocity 
between the clump and the surrounding medium is roughly $\sigma$, the 
velocity dispersion in the disc, which is assumed to be isotropic and 
the same for all the cold disc components. 
Vertical stability in the disc thus tells us that 
${\sigma}/{\Vd} \sim {\Hd}/{\Rd}$, where $\Vd$ is the rotation 
velocity in the disc.  The clump accretion rate thus becomes 
${\dot{\Mc}} = {\pi}R_{\rm T}^2{\rho}{\sigma} = \Mc (\alpha/2) (\Vd/\Rd)$, 
so the accretion time is $t_{\rm acc} \equiv \Mc/{\dot {\Mc}} = 2\alpha^{-1}\td$,
where $t_{\rm d} = \Rd/\Vd$ is the crossing time. Following section 
4 of DSC09 we approximate the clump migration time as $\tm \sim 8\td$, 
so the relative increase in clump mass during migration is 
${\Delta\Mc}/{\Mc} \sim 4{\alpha}$, which is roughly 2 if the 
effective ${\alpha}$ for the clump during its migration is roughly 0.5. 
Thus, the clump can triple its mass during migration from accretion alone. 
This is consistent with the mass gradient for clumps in \fig{grad}. A 
steady accretion of fresh gas onto the clumps during their migration may 
also explain the lack of radial trend in their average SFR, as the gas 
supply is constantly replenished. The fact that the \exsitu clumps show 
a similar mass gradient, may suggest that at least the co-rotating ones 
accrete mass in a similar fashion.

The other gradients can be understood in the following way.
We assume that the \insitu clumps form from patches within the disc having 
initial gas fractions and stellar distributions similar 
to the local background where they formed. However, as the SFR in clumps is 
more intense than in the inter-clump medium  
\citep{Bergh96,Elmegreen04a,Elmegreen05b,Forster06,Genzel08,Guo12,Wuyts12}, 
they quickly become dominated by a young stellar population, which  
outweigh and outshine the older stars that the clump inherited from the 
disc. During its migration, the clump continuously accretes fresh gas and 
forms stars, increasing its stellar mass and decreasing its gas fraction and 
sSFR, even if the total gas mass and SFR do not change much. New generations 
of stars enrich the clump with metals, causing metallicity to increase 
during the migration. When the clump reaches the center at the end of its 
migration, its mass-weighted mean stellar age is of the order of the migration 
time, namely several hundred $\Myr$. Thus, the clump ages are predicted to be 
between several tens and several hundreds of $\Myr$ and their masses should 
have a spread of about a factor of 3, with clumps nearer the galactic center 
being more massive and older, with lower gas fractions and sSFRs but higher 
metallicities. 

This is in stark contrast to the expectations from Scenario II, 
where the stellar populations in the clumps do not have a chance to
evolve much or develop such an age spread, as they are all recycled 
back into the disc within less than $100\Myr$. \citet{Genel12a} report 
that their simulated clump masses are all within a factor of 2 of each 
other, at the scale of the turbulent Jeans mass of the disc, with no 
clear gradient in the clump mass to distance relation. Their clumps do 
exhibit an age gradient, but this closely follows that of the inter-clump 
stars and is fairly shallow, with a logarithmic slope of $-0.57 \pm 0.14$, 
compared to $-0.86 \pm 0.03$ in our simulations. Only $\sim20\%$ of the stellar mass 
in their clumps was formed in the clump itself, the remaining $80\%$ being 
background disc stars, while in the majority of our \insitu clumps, 
more than 50\% of the clump mass consists of stars formed internally in 
the clump. Since observations of high-$z$ SFGs indeed reveal rather weak 
gradients in both color and mass-to-light ratios across the disc 
\citep{Forster11a}, the gradients in stellar age of disc stars are 
not expected to be very strong, and neither are the gradients of clump 
properties in scenario II.

In principle, we could imagine a stronger radial variation of clump 
age even in scenario II, if we allow the efficiency of clump formation 
to vary with radius. However, in this case we should still expect the 
age gradient for the clumps to be close to that of the background disc 
stars, because clumps only experience one major episode of star-formation 
before disrupting and depositing all their newly formed stars in the 
surrounding disc. This will cause the disc to appear younger in regions 
where clump formation/disruption is efficient, and somewhat older 
elsewhere, where few young clumps are found. The exact gradient of 
the disc age will then be determined by the ratio of the clump formation 
timescale to the migration timescale of the stars, but we can likely expect 
it to be steeper than what we see in our simulations, where the stellar 
disc age is nearly constant with radius, and to follow more closely the 
\insitu clump age gradient. 

In order to distinguish between the two scenarios regarding the fate of the 
giant clumps, detailed observations of clump properties within the disc 
are needed. The dominant distinguishing feature between the two scenarios 
is the age spread among the clumps. Observations revealing spreads of more than 
$100\Myr$ would provide strong evidence against scenario II. 
A further test calls for detailed observations of the radial trends of 
clump properties within the disc. These should then be compared to the underlying 
gradients in the smooth disc component, to determine if the clumps evolved 
seperately, indicating long lived stages of equilibrium. Since contamination by 
\exsitu clumps will tend to smooth the observed gradients if they are not separated 
from the \insitu clumps, any observed gradient in clump properties beyond 
that of the background disc may be considered a lower limit on the intrinsic 
gradient, and evidence against scenario II. 

It is important to keep in mind that reality likely lies in between the two 
extreme scenarios outlined above. Future work examining simulations with more 
realistic feedback, as described in \citet{Ceverino13}, 
will determine how robust the various gradients are to steady outflows from 
clumps with mass loading factors of order unity, rather than $\eta \sim 0.3$ 
as in our current simulations. Recent results from isolated disc simulations 
incorporating radiative feedback \citep{Bournaud13} as well as analytic estimates 
\citep{DK13} suggest that with realistic outflows, the mass gradient may be suppressed.


\subsection{Comparison with Observations} 
\label{sec:observ} 

We summarize here some of the main observational results concerning
the giant clumps in high-redshift SFGs, and comment on their theoretical 
implications in view of our findings in the current paper. A list 
of relavent references includes \citet{Elmegreen04a,Elmegreen05b,
Forster06,Genzel08,Elmegreen09,Forster11b,Jones10,Genzel11,Guo12,Wuyts12}. 

\subsubsection{General Evidence for VDI}

When "clump-cluster" and "chain" galaxies were first observed at 
$z = 1-2$ \citep{Cowie95,Bergh96}, many believed them to be merging 
systems. However, studies of the distributions of shapes of such 
galaxies 
at $z=1-3$ \citep{Elmegreen04b,Elmegreen05a}, followed by resolved 
kinematic studies of massive SFGs at $z\sim 2$ using integral field 
spectroscopy \citep{Forster06,Forster09,Genzel06,Shapiro08}, revealed 
many of these systems to be extended rotating disc galaxies showing no 
signatures of major mergers. Toomre gravitational disc instability in 
its violent phase is an obvious alternative mechanism for the formation 
of the clumpy phase (e.g. DSC09 and references therein). 
Indeed, the observations indicated indirectly that these discs may be 
gravitationally unstable, with $Q \leq 1$ \citep[e.g.][]{Genzel06,Genzel11}, 
though it is hard to learn about the linear instability from the nonlinear 
conditions in the observed discs.

Estimates by \citet{Wuyts12} of the overall fraction of clumpy discs 
at redshifts $z=1.5-2.5$ range from $\sim 40-75\%$, depending on the 
band used to identify clumps. However, we note that in this study, clumps 
are not detected individually as in our work, but are rather defined as 
off-center pixels with heightened surface density in stacked, pixelated 
images of the galaxy population. 
Observations of a large sample of galaxies from the CANDELS survey, where 
clumps are detected both visually and using an automated algorithm simillar 
in spirit to the one implemented in this work (Y. Guo et al., in preparation), 
reveal clumpy fractions of $\sim 40-60\%$ for galaxies in a simillar mass and 
redshift range as studied in this work. Both these results are consistent with 
the clumpy fractions predicted by our simulations.

Subsequent estimates of the masses of clumps in the redshift 
range $z=1-3$ reveal them to be one to a few percent of the disc 
mass each, consistent with Toomre instability theory, and altogether 
contributing 10-20\% \citep{Elmegreen05b,Elmegreen09,Forster11b,Genzel11,
Guo12,Wuyts12}. These same studies also find that the contribution 
of clumps to the total SFR in the disc is higher, roughly 10-20\% 
individually and up to 40\% altogether, though there is significant 
scatter. \citet{Wuyts12} finds that not only do clumps contribute less 
to their host galaxy's mass than to their light, they also contribute 
more light in bluer wavelengths, indicative of younger stellar 
populations and active star formation. The results from our simulations 
share these general trends, though the simulated clump masses tend 
to be slightly below the observed values. This can be largely explained 
by observational beam smearing causing the clumps to appear larger 
than they actually are (see section 6 in \citet{Ceverino12}, especially 
their figure 19). Combined with improper background and foreground 
subtraction, this raises the clumps' contribution to the disc mass and SFR. 
Taking this into account, the fractions we find are in agreement with 
the clump masses deduced from the observations. 

\subsubsection{\Insitu Clumps vs. Mergers}

We have learned that the populations of VDI \insitu clumps and 
merging \exsitu clumps are distinguishable. If the two populations 
are not properly separated, the testing of the theoretical predictions 
for the \insitu clumps in the competing scenarios may prove difficult, 
as the \exsitu clumps can have qualitatively different properties. 
Efforts must therefore be made to distinguish between the two populations 
observationally, rather than by dark-matter cotrast as in this work. 
Our results suggest that any off-center clump containing more than 
$\gsim 10\%$ of the total disc mass, such as observed by (e.g.) 
\citet{Genzel11}, is likely to be an \exsitu clump, since such high 
mass \insitu clumps are rare. Another possible distinction relies on 
observing clumps close to the disc edge. If such a clump has a stellar 
population of age $\sim 1\Gyr$ and a low sSFR $\sim 0.1 \Gyr^{-1}$, it 
is almost definately an \exsitu clump.  All the \insitu clumps at these 
large radii should be dominated by a young stellar population of 
$\lsim 80\Myr$ and have high sSFR $\sim 5 \Gyr^{-1}$. These differences 
become less pronounced at smaller radii. 

In many cases, young stellar ages and high sSFRs and gas fractions 
should translate to observed blue colors and vice versa. However, 
realistic luminosities and colors will depend on the effects of dust. 
In an ongoing work, dust and radiative transfer is incorporated into 
the simulated galaxies using the SUNRISE code, thus creating realistic 
mock observations comparable to CANDELS data \citep{Moody14}. 
Prelimenary results show that \insitu clumps are not affected 
much by dust near the disc edge in face-on images and they indeed tend 
to appear very blue, so the distinctions between \insitu and \exsitu 
discussed above should be observable. However, in edge on views and 
toward the central parts of the disc, the \insitu clumps are reddened, 
and the differences between the clump types become less pronounced.

Observations find most off-center clumps to be blue  
\citep{Elmegreen05b,Forster11b,Guo12,Wuyts12}. 
Based on our analysis of the simulations and the 
preliminary SUNRISE images, this suggests that the 
majority of the observed off-center clumps are, indeed, 
\insitu rather than \exsitu in origin. The observations 
also suggest that the off-center clumps are peaks in the 
distribution of sSFR, similar to what we find for the 
\insitu clumps, which have sSFR values a factor of 
$\sim 5$ higher than the median of the smooth disc.

A few of the observed off-center clumps are redder and more massive 
\citep{Forster11b}, and 
these may well be \exsitu clumps. It is also possible that more 
\exsitu clumps have been observed, but classified as mergers rather 
than clumps. 
In addition, central massive red clumps have been observed 
\citep{Elmegreen09,Guo12,Wuyts12} 
which seem to resemble our \bulg clumps. 

Given the mounting theoretical and observational evidence that most of the 
high-$z$ SFGs are extended discs undergoing VDI and that external mergers are 
responsible for only a part of the clump population, it would not make sense 
to classify the high-z SFGs using the familiar classification schemes used at 
low redshifts. In particular, the high-$z$ VDI phase with giant clumps is 
unlikely and therefore unaccounted for at low redshift, where disc instability 
takes the form of a bar and spiral arms associated with secular evolution. 
This calls for a new classification scheme for high redshift galaxies, which 
recognizes the dominance of VDI systems and explicitly differentiates between 
VDI galaxies and merging systems. Such a scheme will be devised using the SUNRISE 
images of the simulated galaxies together with the complete merger history of all 
the clumps. 

\subsubsection{Clump Survival: Scenario I vs. Scenario II}

The stellar ages of clumps and their predictred lifetimes are 
being estimated observationally, though with very large uncertainties.
\citet{Elmegreen05b} observed ten clump-cluster galaxies in 
the HUDF at $1.6<z<3.0$ with 5-10 clumps each, and estimated 
clump ages of $100-800 \Myr$, with an average of $340 \Myr$, 
hosted in older discs of $1.4-3 \Gyr$. Then \citet{Elmegreen09} 
found a very large range of ages for star-forming clumps, centered 
around $100\Myr$ but reaching values as high as $1\Gyr$. 
\citet{Genzel11}, who examined five $z\sim 2.2$ clumpy SFGs with 
SINFONI, estimated clumps to be between 10 and a few hundred 
$\Myr$ old, with typical clumps having stellar ages of $100-200\Myr$ 
and an upper envelope of $300\Myr$. Additional considerations 
led them to estimate the average lifetime of clumps to be $\sim 500 \Myr$. 
\citet{Forster11b} obtained ages for 7 clumps 
in one galaxy from the SINS survey, which ranged from a few 
tens to about $\sim 250\Myr$, centered on just below $100\Myr$. 
\citet{Wuyts12} used a mass complete sample of 326 SFGs at $1.5<z<2.5$ and 
another 323 SFGs at $0.5<z<1.5$, pixelated and stacked the images and defined 
clumps as off-center pixels with elevated surface brightness above the 
background. At $z\sim 2$, the clump pixels have ages of $\sim 200 \Myr$, 
far younger than the off-clump pixels. At $z\sim 1$, both the clumps and 
the discs are older by about a factor of 2. \citet{Guo12} collected data 
on $\sim 40$ clumps from ten galaxies in the HUDF and found the distribution 
of clump ages to be roughly lognormal, centered on $\sim 300\Myr$, but 
covering a wide range from $10 \Myr$ to a few $\Gyr$, while the disc ages 
were concentrated in the range $0.3-1 \Gyr$. 

These studies all use different methods for estimating the clump stellar ages, 
but all agree that the uncertainties of the models are very large. 
Nevertheless, the fact that the clumps exhibit such a wide range of ages in 
any particular study and that even low estimates on clump ages are rarely far 
below $100 \Myr$ seem to favour scenario I over scenario II. In other words, 
it appears that clumps are indeed able to survive for durations of several 
hundred $\Myr$ and do not disrupt on dynamical timescales of $\sim 50-100\Myr$. 

Some of these studies \citep{Forster11b,Guo12} have even attempted to measure 
radial variations of 
clump properties within the discs. They find evidence for older, redder 
and more massive clumps to be located closer to the disc center. 
\citet{Forster11b} find a logarithmic slope for the ages of seven clumps in 
a single galaxy of $-2.06 \pm 0.63$, much steeper than the radial variation 
of the background disc. They note that even if the absolute values for clump 
ages are wrong, the relative trend should hold. In another galaxy from 
their sample, also containing seven clumps, they found clumps to become 
redder closer to the center, with increased mass-to-light ratios. Clumps 
near the disc edge have masses of $\sim 10^8\msun$ while close to the 
center the masses are $\gsim 6\times 10^8\msun$. \citet{Guo12} find that 
clumps closer to the disc center have lower sSFRs, older ages, higher dust 
extinctions and higher stellar surface densities. Moreover, they find the 
radial variation in clump properties steeper than the global gradients in 
the background disc, and deduce that the clumps must have evolved seperately 
from the disc in a state of quasi-equilibrium. They find clumps within 0.1 
times the disc radius to have sSFR values 5 times lower and stellar surface 
densities 25 times higher than clumps in the outer half of the disc, very 
simillar to the results from our simulations of scenario I. The clumps near 
the disc center have ages of roughly $700 \Myr$ as opposed to $\sim 100 \Myr$ 
in the outer disc. These studies, while crude and prelimenary, are in good 
agreement with the predictions of clump migration in scenario I as simulated 
in the current paper, which seems to indicate that clumps survive for extended 
periods longer than an orbital time and evolve as they migrate inwards. 

\subsection{Co-Rotation of \Exsitu Clumps} 

We found that roughly half the \exsitu clumps are co-rotating with their 
background in the disc, and some of their properties are systematically 
different from those \exsitu clumps with significant vertical or radial 
velocity components. The median masses and SFRs of the former  
are larger by factors of $\sim 2$ and $\sim 25$ respectively, their 
median gas fraction is $\sim 3$ times higher and their medain sSFR 
is $\sim 8$ times higher.

The co-rotation and associated clump properties may partly be a result of 
the initial properties of the incoming satellite. The degree of co-rotation may 
correlate with the initial orbit of the incoming satellite, and with the time 
since it first hit the disc. It may also depend on the gas fraction in the 
incoming clump, as ram pressure could be the mechanism that forces the satellite 
to join the disc kinematics. Finally, it may correlate with the clump total mass, 
if the mechanism that brings the satellite into the disc is tidal torques or 
dynamical friction. Indeed, the co-rotating clumps are distinguished by they higher 
total mass and gas mass. 
Perhaps even more important in producing the distinct properties of the 
co-rotating clumps are the later evolutionary effects within the disc. 
Once co-rotating for whatever reason, the clump would tend to migrate 
inwards like the \insitu clumps and to grow in mass by accretion of 
gas from the disc (\se{survival}). The fresh gas will naturally increase 
the SFR. The origin of the kinematics bimodality of the \exsitu clumps 
will be addressed using the simulations along these lines in a future work.

The kinematic bimodality of \exsitu clumps should be detected observationally. 
One can identify a subsample of \exsitu clumps by their distinct properties 
described in the previous subsection, and measure their kinematics with 
respect to the host disc. One can then search for systematic difference in 
mass, gas fraction, sSFR and position within the disc between the two kinematic 
subsamples.

\subsection{Caveats and Future Prospects} 

We can mention 3 main caveats regarding our simulations. 
First, despite the high resolution that allows the clump 
analysis in the first place, it is limited to only marginally 
resolving the clumps and not accounting for their sub-structure. 
\citet{Ceverino12} compared a small sample of clumps from the 
simulations described here to a few clumps from disc galaxies 
simulated in isolation with a varying resolution up to $\sim 1\pc$. 
The high-resolution clumps, showing a rich substructure, appear 
to have simillar masses and sizes to those from our simulations, 
and they also remain in Jeans equilibrium while migrating towards 
the disc center. We therefore do not expect the resolution to have 
a major effect on the global properties of individual clumps. 
The effect of resolution on VDI and on the clump properties is 
being tested using a new suite of simulations with twice higher 
resolution (\citealp{Moody14}; N. Mandelker et al., in preparation).

Second, as mentioned in \se{sim}, the SFR efficiency in a free-fall 
time as assumed in the current simulations led to an overestimate of 
the SFR at $z>4$ and thus to an underestimate of the gas fraction and 
SFR at $z\sim 2$, by a factor of order two. It seems reasonable that 
underestimating the gas fraction may cause us to underestimate the 
importance of VDI and the degree of clumpiness in our discs, since 
these are mostly driven by the gas and young stars, thus making our 
studies of VDI at $z \sim 2$ rather conservative. 
This will be tested using a new suite of simulations with higher 
resolution and reduced star-formation efficiency.

Third is the fairly weak feedback incorporated in the current simulations, 
which helps the overestimate of SFR at very high z, and underestimates the 
effect of outflows on clump survival. This will be examained (N. Mandelker et 
al., in preparation) using a new suite of cosmological simulations 
\citep{Ceverino13} that employ stronger momentum-driven feedback, at a 
physicaly realistic level of a few times $L/c$ \citep{DK13}. 
While a systematic study of outflows from clumps in our existing 
simulations has not yet been performed, we note that a few individual 
cases of clumps exhibiting strong outflows have been identified. These 
clumps were found near the edge or outside the slim disc, where the ambient 
gas density is low. The outflows caused the clumps to lose most of their 
gas in less than $100 \Myr$. However, the stellar clump remained intact, 
and after reentering the slim disc, was observed to accrete fresh gas 
and continue its migration to the disc center.

There are several additional interesting issues concerning the 
clump properties that we will explore in future work. These 
include mutual correlations between different clump properties, 
and between the clump properties and the characteristics of the host 
galaxy. Both these topics are relevant for comparison with observations.
A further avenue for future work involves tracking individual clumps 
through time in the simulations. Recall that in our current simulations, 
clumps may be present in the disc for 2-3 consecutive snapshots before 
completing their migration. During this time, they can be tracked based 
on their stellar particles and insight can be gained on their individual 
formation and evolution. Such a detailed analysis is currently being 
performed on clumps produced in our new simulations (A. Dekel, F. 
Bournaud and N. Mandelker, in preparation; N. Mandelker et al., in 
preparation). Preliminary results suggest that the frequency and longevity 
of clumps with baryonic masses $\gsim 10^8 \msun$ are unaffected by the 
inclusion of radiative feedback. The age gradient is still present and is 
only mildly suppressed compared to our current simulations. However, the mass 
gradient may be strongly suppressed, as outflows due to feedback can balance 
the inflow of gas from the disc onto the clumps, so the clump mass does 
not evolve much during migration. 
In addition, several simulations are beeing run with much denser time 
spacing between outputs to allow individual clumps to be followed more 
accurately.

Also interesting will be to address the validity and properties of VDI at 
very high $z>3$ and low $z<1$ redshifts. Pioneering observations of massive 
galaxies out to $z\sim 10$ may start exploring the extent to which VDI 
operates also earlier than $z \sim 3$. This question will be addresed 
using simulations in future work, where we will have to struggle 
with the limitation that only the most massive galaxies are properly 
resolved in our current simulations before $z \sim 4$. Prelimenary visual 
inspection of the existing simulations do reveal at least a few cases of 
massive galaxies with a clumpy appearance already at $z\sim 5-6$. In order 
to address the fate of VDI at $z <1$, and the low-redshift decendants of 
the VDI galaxies seen at $z \sim 2$, we will push the simulations to after 
$z \sim 1$. 

The analysis presented here has been performed in 3D, taking full 
advantage of the information available in the simulations. While 
this method has helped us develop a theoretical understanding 
of VDI, one should now worry about how to directly compare our 
results with realistic 2D observations that also suffer from dust 
effects, background and foreground contamination, and beam smearing. 
As described in \se{observ}, efforts are being made to ``observe" these 
simulated galaxies in a realistic way, e.g. for comparison with CANDELS 
data. Using these ``observed" simulated galaxies and their merger 
histories will allow the development of more accurate observable criteria for 
distinguishing between \insitu VDI clumps and merging clumps, and thus devise 
a new theory-motivated classification scheme for high-$z$ galaxies, which 
explicitly accounts for VDI.

\section{Conclusion}
\label{sec:conc} 

We have studied the properties of giant clumps in $1<z<4$ disc 
galaxies using a suite of high resolution hydro-cosmological 
simulations. The simulations used here, being characterized by 
moderate stellar feedback, may underestimate the outflow mass 
loss from clumps. The result may be an overestimate of the clump 
mass growth during its migration, but we expect most other clump 
properties to be recovered in a qualitatively robust way. Our main 
results can be summarised as follows: 
\begin{enumerate}
\item 
On average, $\sim 70\%$ of the discs host off-center clumps, 
while $\sim 60\%$ specifically host \insitu clumps, formed by VDI. 
The fraction of clumpy discs peaks at intermediate disc masses 
of $\Md \sim 10^{10.5} \msun$ at $z\lsim 2$. 
Considering only the clumpy discs, there are on average 3.2 \insitu 
clumps per disc, together contributing 1-7\% of the disc mass (with 
a median of 3-4\%) and 5-45\% of the disc SFR (with a maximal probability 
of 40\% and a median of 22\%). Considering only the discs undergoing VDI 
the average number of \insitu clumps per disc is 3.7.

\item 
Among the off-center clumps, \insitu clumps make up roughly 75\% 
in number, $\lsim 80 \%$ in SFR and $\lsim 50\%$ in baryonic 
mass. Individual \insitu clumps contain $\sim 1-2\%$ of the disc 
baryonic mass, in agreement with standard Toomre analysis. 
Their contribution to the disc SFR is greater, with individual 
clumps having $\sim 5-6 \%$ of the disc total. They have high gas 
fractions, typically $\sim 10\%$, though the youngest clumps have 
more than $50\%$. As a result, they have high sSFRs, typically 
$\gsim 2\Gyr^{-1}$, with the youngest clumps having sSFRs above 
$10\Gyr^{-1}$. Their median stellar age is $\sim 160\Myr$ with a 
broad peak at $150-300\Myr$, and a sharp decline at older ages. 
This is comparable to 1-2 orbital times at the disc edge which is 
the expected migration time for clumps (DSC09). Typical metallicity 
values are ${\rm log}(O/H)+12 \sim 8.9$, though the youngest, most 
actively star forming clumps still have sub-solar values. The \insitu 
clumps exhibit radial trends, where clumps closer to the disc center 
are more massive and have lower gas fractions, lower sSFR, older stellar 
ages and higher metallicity.

\item  
\Exsitu clumps, which joined the disc as minor mergers, were primarily 
identified by their dark matter contrast with respect to the host halo. 
The average number of \exsitu clumps per disc is 0.5 among all the discs, 
0.7 among the clumpy discs, and 0.6 among the discs unergoing VDI. The 
\exsitu clumps turn out to be about 25\% of all the off-center clumps in 
number, over 20\% in SFR and more than 50\% in mass. A typical \exsitu 
clump contains about 4\% of the disc mass, though it is not uncommon to 
find massive clumps having 10-30\% of the disc mass. Their SFR is 
uncorrelated with the host disc, and the SFR distribution among the 
clumps is uniform in the range 0.001-1 times the total disc SFR. Their 
gas fractions are typically $\lsim 1-3 \%$, lower than the \insitu clumps, 
resulting in lower sSFR values of $\sim 0.25 \Gyr^{-1}$ and old stellar 
ages of order $1 \Gyr$. On the other hand, the metallicities are similar 
to those of the \insitu clumps, especially in the gas phase. The \exsitu 
clumps are more massive closer to the disc center, but show no significant 
radial trends in their other properties. The exception is a tendency for 
clumps interior to $\Rd$ to have higher gas fractions, SFRs and sSFRs than 
those outside $\Rd$. Roughly half of the \exsitu clumps are co-rotating with 
their host disc, the remainder exhibiting large vertical or radial 
velocities. These co-rotating clumps typically have higher masses, SFRs, 
gas fractions and sSFR than the non-co-rotating ones, and they can have 
sub-solar metallicities simillar to the \insitu clumps. This is consistent 
with accretion of fresh gas from the disc during migration inwards. 

\item 
In addition to the off-center clumps, there is a compact \textit{bulge} clump 
at the center of $\sim 91\%$ of our discs. These \textit{bulge} clumps are 
typically smaller in size than the stellar bulge, representing gaseous 
overdensities at the bulge center. On average, \bulg clumps have masses 
of $\sim 0.4\Md$, an order of magnitude more massive than a typical 
off-center clump. The SFR in the $\bulg$ clumps is typically $\sim 20\%$ 
of the disc value, a factor of $\lsim 4$ higher than typical \insitu clumps. 
They typically have low gas fractions of less than 1\%, low sSFR of 
$\sim 0.2 \Gyr^{-1}$, and old stellar populations with ages of order 
$1 \Gyr$. They are also metal rich with ${\rm log}(O/H)+12 \sim 9.1-9.2$. 
At high redshifts they are often gas rich and star forming, 
with SFR values greater than the disc itself, resembling the observed "blue 
nuggets" \citep{Barro13,DekelBurkert13}.

\item 
The \exsitu merged galaxies and the \insitu clumps can be partly 
distinguished by their masses. For example, a clump that is as 
massive as 10\% or more of the disc is almost certainly \exsitu 
in origin. Assuming the clumps are not severely reddened by dust, 
one can partly distinguish \exsitu from \insitu clumps based on 
color, especially in the outer parts of the disc. \Insitu clumps 
are expected to be blue, due to their high  sSFRs and their young 
stellar ages, while the \exsitu clumps should exhibit redder colors 
characteristic of older stars and low sSFR. Prelimenary analysis 
using SUNRISE suggest that the clumps are not significantly 
affected by dust in face-on images of the disc near the edge. 

\item 
If the clumps survive intact for a migration time they are expected 
to accrete gas from the surrounding disc and increase their mass 
during the migration. The accretion promotes additional star formation 
during the migration. If the accretion rate and SFR remain roughly 
constant, the gas fraction and sSFR will decrease 
as the clump migrates inwards, while the metallicity increases. This would 
imply that the accretion of gas from the disc does not grow near the disc 
center, possibly due to the strong gradient of gas fraction in the disc 
or to the strong tidal field near the center. The gradient of \insitu clump 
masses within the disc is consistent with the accretion scenario, as is the 
age gradient which is too steep to be explained by the gradient in the 
background disc. Prelimenary observations of such radial gradients in clump 
properties tend to favour the clump survival and migration picture, as they 
are not expected in models where clumps are short lived. Observed gradients 
are only lower limits on the intrinsic evolution of \insitu clumps, as the 
presence of \exsitu clumps may contaminate the signal, though this is not 
expected to be a strong effect in the inner disc. 
Detailed observations of radial variations of clump properties accross a large 
sample of galactic discs, possibly taking care to remove \exsitu mergers on a 
statistical basis, are needed to properly address the question of the lifetime 
of observed giant clumps.

\end{enumerate}

\section*{Acknowledgments} 
 
We acknowledge stimulating discussions with Yicheng Guo. 
The simulations were performed in the astro cluster 
at HU, at the National Energy Research Scientific 
Computing Center (NERSC), Lawrence Berkeley National 
Laboratory, and at NASA Advanced Supercomputing (NAS) 
at NASA Ames Reserach Center. This work was partially 
supported by ISF grant 6/08, by GIF grant G-1052-104.7/2009, 
by a DIP grant, by NSF grant AST-1010033 and by MINECO 
grants AYA2012-31101, and AYA-2009-13875-C03-02. DC is 
a Juan de la Cierva fellow.

\bibliographystyle{mn2e} 

\appendix 

\section{Cosmological simulations with the ART code} 
\label{sec:art} 

The cosmological simulations utilize the ART code 
\citep{Kravtsov97,Kravtsov03}, which accurately follows the evolution of a 
gravitating N-body system and the Eulerian gas dynamics using an adaptive mesh 
refinement approach. Beyond gravity and hydrodynamics, the code incorporates 
many of the physical processes relevant for galaxy formation, as described in 
\citet{Ceverino09} and in \citet{CDB}. These processes, representing subgrid 
physics, include gas cooling by atomic hydrogen and helium, metal and molecular 
hydrogen cooling, and photoionization heating by a UV background, with partial 
self-shielding. 
Cooling and heating rates are tabulated for a given gas 
density, temperature, metallicity and UV background based on the CLOUDY code 
\citep{Ferland98}, assuming a slab of thickness 1 kpc. A uniform UV background 
based on the redshift-dependent \citet{HaardtMadau96} model is assumed, 
except at gas densities higher than $0.1\cmc$, where a substantially 
suppressed UV background is used
($5.9\times 10^{26}{\rm erg}{\rm s}^{-1}{\rm cm}^{-2}{\rm Hz}^{-1}$) 
in order to mimic the partial self-shielding of dense gas. 
This allows the dense gas to cool down to temperatures of $\sim 300$K. 
The assumed equation of state is that of an ideal mono-atomic gas. 
Artificial fragmentation on the cell size is prevented by introducing 
a pressure floor, which ensures that the Jeans scale is resolved by at least 
7 cells (see \citet{CDB}). 

\begin{table*} 
\caption{Properties of our 29 dark matter halos. 
The box sizes are in $\hmpc$. 
$Target\:\Mv$ refers to the virial mass at the target redshift 
of the selected halo from the low resolution N-body simulation and 
is given in units of $10^{11}\msun$. The halos were selected at $z=1$, 
except for MW6-MW9, which were selected at $z=0$. $z_{f}$ is the final 
redshift that the simulation reached. The virial properties are given 
at $z=2$, except for MW5, for which they are given at $z=2.23$ where 
the simulation was stopped. The virial mass is given in units of 
$10^{11}\msun$, the virial radius in units of $\kpc$ and the 
velocities in $\kms$. 
}
 \begin{center} 
 \begin{tabular}{ccccccc} \hline 
$Gal$ &  $Box\:Size$ & $Target\:\Mv$ & $z_{f}$ & $\Mv$ & $\Rv$ & $\Vv$ \\ 
   &   $\hmpc$  & $10^{11}\msun$  &  &   $10^{11}\msun$  &  $\kpc$  &  $\kms$ \\ \hline 
MW01 & 20 & 15.3  & 1.38 & 8.1  & 101.6 & 185.2 \\ 
MW02 & 20 & 12.1  & 1.94 & 8.9  & 104.9 & 190.5 \\ 
MW03 & 20 & 19.3  & 1.38 & 7.3  & 98.7  & 178.4 \\ 
MW04 & 40 & 40.1  & 1.38 & 14.2 & 122.9 & 222.9 \\ 
MW05 & 80 & 103.7 & 2.23 & 35.0 & 155.5 & 311.2 \\ 
MW06 & 40 & 40.9  & 0.00 & 9.2  & 106.0 & 192.9 \\ 
MW07 & 40 & 17.0  & 0.33 & 3.0  & 73.3  & 133.4 \\ 
MW08 & 40 & 14.1  & 0.35 & 2.8  & 71.2  & 129.4 \\ 
MW09 & 40 & 11.0  & 0.00 & 1.6  & 59.4  & 108.0 \\ 
MW10 & 20 & 15.3  & 1.00 & 8.2  & 101.8 & 185.6 \\ 
MW11 & 20 & 14.2  & 1.50 & 5.3  & 88.4  & 161.2 \\ 
MW12 & 20 & 16.9  & 1.08 & 17.0 & 130.0 & 237.2 \\ 
\hline
VL01 & 40 & 20.0  & 1.00 & 12.3 & 117.1 & 212.2 \\ 
VL02 & 40 & 20.0  & 0.96 & 8.1  & 101.4 & 184.9 \\ 
VL03 & 40 & 20.4  & 1.00 & 11.9 & 115.8 & 209.8 \\ 
VL04 & 40 & 20.6  & 0.96 & 10.1 & 109.2 & 199.5 \\ 
VL05 & 40 & 20.0  & 0.92 & 12.8 & 118.2 & 215.4 \\ 
VL06 & 40 & 20.1  & 1.00 & 7.5  & 99.2  & 180.5 \\ 
VL07 & 80 & 26.1  & 1.85 & 16.6 & 129.0 & 235.4 \\ 
VL08 & 80 & 26.6  & 1.22 & 10.9 & 111.8 & 204.9 \\ 
VL09 & 80 & 25.9  & 1.93 & 4.9  & 85.8  & 155.9 \\ 
VL10 & 80 & 25.9  & 1.08 & 8.1  & 101.5 & 185.6 \\ 
VL11 & 80 & 26.4  & 1.00 & 17.2 & 130.1 & 238.7 \\ 
VL12 & 80 & 26.1  & 1.00 & 9.0  & 104.7 & 191.9 \\ 
\hline
SFG1 & 40 & 33.0  & 1.17 & 16.6 & 128.7 & 235.3 \\ 
SFG4 & 40 & 32.9  & 1.38 & 10.9 & 112.3 & 204.3 \\ 
SFG5 & 40 & 33.3  & 1.00 & 13.8 & 123.0 & 219.6 \\ 
SFG8 & 80 & 65.9  & 1.86 & 13.8 & 121.1 & 221.5 \\ 
SFG9 & 80 & 51.7  & 1.03 & 18.9 & 134.8 & 245.3 \\ 

\hline 
 \end{tabular} 
 \end{center} 
\label{tab:halos} 
 \end{table*} 
 
Star formation is assumed to occur at densities above a threshold of $1\cmc$ 
and at temperatures below $10^4$K. More than 90\% of the stars form at 
temperatures well below $10^3$K, and more than half the stars form at 300~K 
in cells where the gas density is higher than $10\cmc$. 
The code implements a stochastic star-formation model that yields a 
star-formation efficiency per free-fall time of 5\%. At the given resolution, 
this efficiency roughly mimics the empirical Kennicutt law \citep{Kennicutt98}. 
The code incorporates a thermal stellar feedback model, in which the combined 
energy from stellar winds and supernova explosions is released as a constant 
heating rate over $40\Myr$ following star formation, the typical age of the 
lightest star that explodes as a type-II supernova. 
The heating rate due to feedback may or may not overcome the cooling 
rate, depending on the gas conditions in the star-forming regions 
\citep{DekelSilk86,Ceverino09}. No shutdown of cooling is implemented. 
We also include the effect of runaway stars by assigning a velocity kick of 
$\sim 10 \kms$ to 30\% of the newly formed stellar particles. 
The code also includes the later effects of type-Ia supernova and 
stellar mass loss, and it follows the metal enrichment of the ISM. 
 
The selected halos were drawn from an N-body simulation. 
The initial conditions corresponding to each of the selected haloes 
were filled with gas and refined to much higher resolution on an adaptive 
mesh within the Lagrangian volume that encompasses the mass within twice 
the virial radius at the redshift when the halo was selected ($z =0$ for 
MW6-MW9, $z=1$ for the remaining halos). This is roughly a sphere of 
comoving radius $1\Mpc$, and it was embedded in a comoving cosmological 
box with length ranging from $20-80\hmpc$. 

\tab{halos} summarizes the properties of our simulated halos, including 
the target halo mass, final redshift and virial properties at $z=2$. Note 
that the most massive galaxy in our sample, MW5, did not reach redshift 2, so 
its virial properties are given at $z=2.23$ when the simulation was stopped. 
The halo virial radius, $\Rv$, is defined as the radius of a sphere within 
which the average density is a factor ${\Delta}_c$ times the universal 
mean. The overdensity is given by \citet{Bryan98} as:
\be
{\Delta}_c = 18{\pi}^2 - 82x - 39x^2,\;\;\; x{\equiv}1-{\Omega}(z)
\ee

A standard $\Lambda$CDM cosmology has been assumed, with the 
WMAP5 cosmological parameters 
$\omm=0.27$, $\oml=0.73$, $\omb= 0.045$, $h=0.7$ and $\sigma_8=0.82$ 
\citep{WMAP5}. 
The zoom-in regions have been simulated with 
$\sim (4-24)\times 10^6$ dark-matter particles of mass 
$6.6\times 10^5\msun$ each, and the particles representing stars 
have a minimum mass of $10^4\msun$. 
Each galaxy has been evolved forward in time with the full hydro ART and 
subgrid physics on an adaptive comoving mesh refined in the dense regions 
to cells of minimum size 35-70 pc in physical units at all times. 

Each AMR cell is refined to 8 cells once it contains a mass in stars and 
dark matter higher than $2\times 10^6\msun$, equivalent to 3 dark-matter 
particles, or it contains a gas mass higher than $1.5\times 10^6\msun$. This 
quasi-Lagrangian strategy ends at the highest level of refinement 
that marks the minimum cell size at each redshift. 
In particular, the minimum cell size is set to 35 pc in physical units 
at expansion factor $a=0.16$ ($z=5.25$), but due to the expansion of the 
whole mesh while the refinement level remains fixed, 
the minimum cell size grows in physical units and becomes 70 pc by 
$a=0.32$ ($z=2.125$). 
At this time we add a new level to the comoving mesh, so the minimum cell 
size becomes 35 pc again, and so on. 
This maximum resolution is valid in 
particular throughout the cold discs and dense clumps, allowing cooling to 
$\sim 300$K and gas densities of $\sim 10^3\cmc$. 

\section{Defining the Disc Frame} 
\label{sec:disc-frame} 

As described in the text, we model the discs as cylinders with radius 
$\Rd$ and height $\Hd$ (total thickness $2\Hd$). We describe here 
how the disc plane and dimensions are determined.

We begin by iteratively defining the disc center. The initial 
estimate of the center is at the minimum of the potential well. 
This is then refined by computing the center 
of mass for the stars in spheres of decreasing radii from $r_{max}=600\pc$ 
to $r_{min}=130\pc$. After each iteration, we update the center 
and decrease the radius by a factor of $1.1$. The algorithm stops 
once the radius gets below $r_{min}$ or the number of stars in the 
sphere drops below $20$. A visual inspection of each snapshot was 
performed to ensure the correctness of the center.

\begin{figure}
\includegraphics[width=0.495\textwidth]{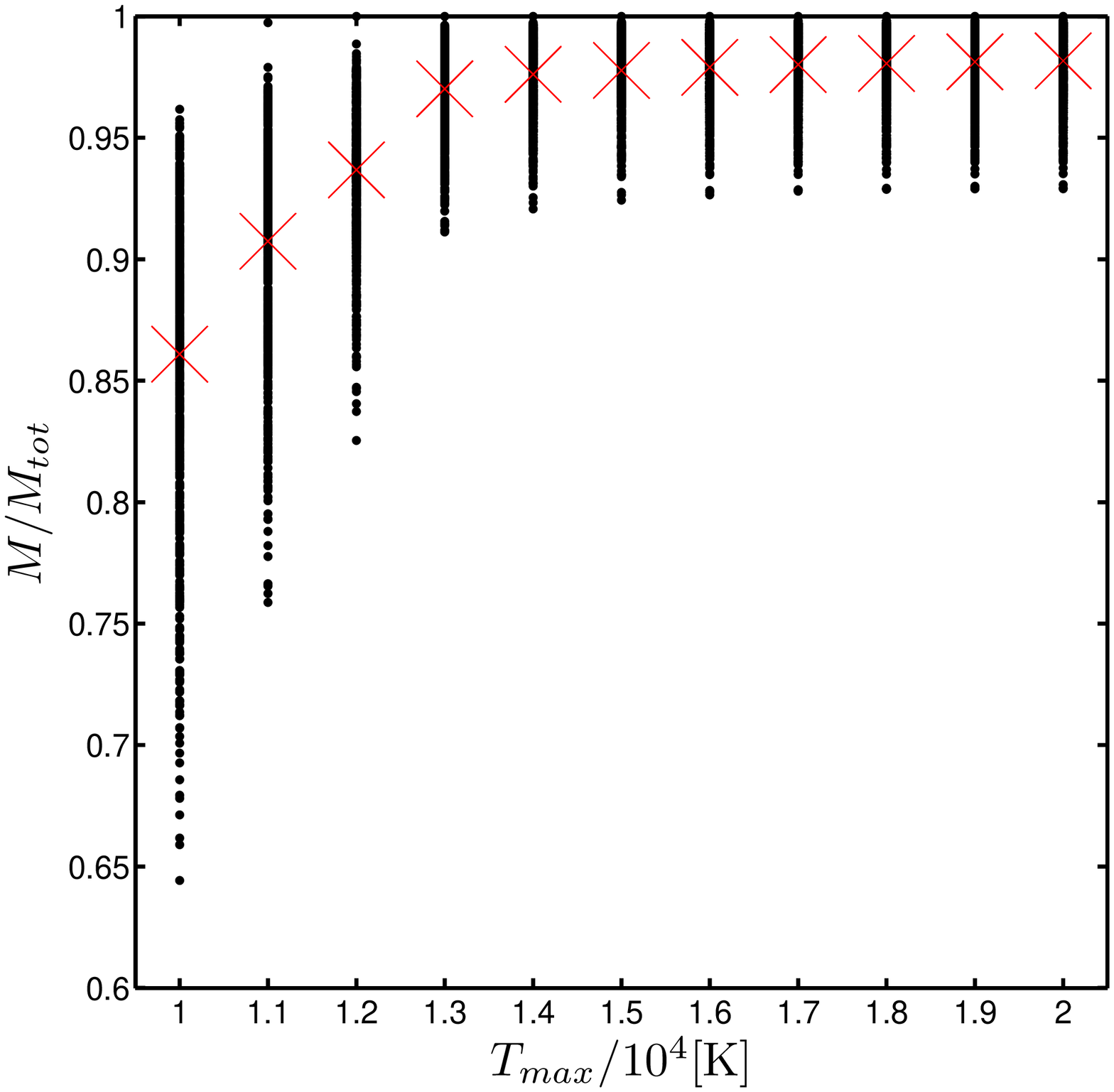}
\caption{Temperature threshold for cold gas. Mass in gas with $T<T_{max}$ 
relative to the total gas mass contained within the disc ($r<\Rd, |z|<\Hd$), as 
a function of $T_{max}$. Black dots represent individual galaxies, while red 
crosses represent the medians. On average, $\sim 97\%$ of the gas has 
$T<1.5{\times}10^4 {\rm K}$. 
}
\label{fig:Tmax} 
\end{figure}

We then procede to determining the disc rest frame, axes 
and dimensions. Here we limit ourselves to cold gas with 
temperature $T<1.5{\times}10^4 {\rm K}$. This cut in 
temperature is important especially for calculating the 
angular momentum because while the warm / hot gas may have 
relatively little mass, it can have extremely high velocities 
and need not be co-rotating with the disc (e.g. gas escaping 
the disc as a result of feedback). All the cold gas, on the 
other hand, is assumed to have settled into a rotating disc. 
Our adopted temperature threshold must remove the non-co-rotating 
gas, while not removing too much of the disc mass. $10^4 {\rm K}$ 
is near the peak of the cooling curve for neutral atomic Hydrogen. 
The numerical factor $1.5$ was determined by examining cumulative 
mass profiles of the gas in our simulated galaxies as a function 
of maximum temperature. For each galaxy, we calculated the fraction 
of mass in gas having $T<T_{max}$ compared to the total gas mass 
of the disc. The results are displayed in \fig{Tmax}. On average, 
we find that $\sim 97 \%$ of the gas in the discs has 
$T<1.5{\times}10^4 {\rm K}$. The median remains roughly constant 
up to at least $5{\times}10^4 {\rm K}$. This result is fairly 
insensitive to the exact dimensions of the cylinder within which 
we compute the mass ($r\le \Rd,\:2\Rd;\; |z|\le \Hd, 2\Hd, 1\kpc$).

Using the cold gas, we iteratively compute the disc rest frame 
velocity $\vcm$, angular momentum $\jd$, radius $\Rd$ and height 
$\Hd$. Our initial estimate for $\Rd$ is the half mass radius of 
the cold gas within a sphere of radius $0.15\Rv$. The disc rest 
frame velocity, $\vcm$, is then taken to be the center of mass velocity 
of cold gas within a sphere of radius $\Rd$ and the angular momentum of 
cold gas in this sphere, computed in the rest frame, defines the disc axis 
$\hat{z}'$. At this stage we initialize $\Hd = \Rd$.

The disc axis and dimensions are now refined, explicitly taking into 
account the cylindrical geometry of the disc. The procedure is as follows: 
\begin{enumerate} 
\item Rotate to the frame defined by $\hat{z}'$ and examine the 
cylinder with radius $r=0.15\Rv$ and height 
$h=$min($\Hd,1\kpc$) (total thickness $2h$). 
Define the disc radius $\Rd$ as the radius which contains 85\% of 
the cold gas mass in this cylinder.
\item Examine a cylinder where both the radius and the height are $r=h=\Rd$. 
Define the disc thickness $2\Hd$ as containing 85\% of the cold gas 
mass within this cylinder.
\item Update $\vcm$ and $\hat{z}'$ within the cylinder defined by $\Rd$ and $\Hd$.
\item Repeat steps (i)-(iii) until $\Rd$, $\Hd$ and $\hat{z}'$ all converge to 
within $5\%$.
\end{enumerate} 
Convergence is usually acchieved in 3-4 iterations, and in any case the 
process is stopped after 5 iterations (not including the initial spherical one).


\begin{figure*}
\begin {center}
\includegraphics[width =0.995 \textwidth]{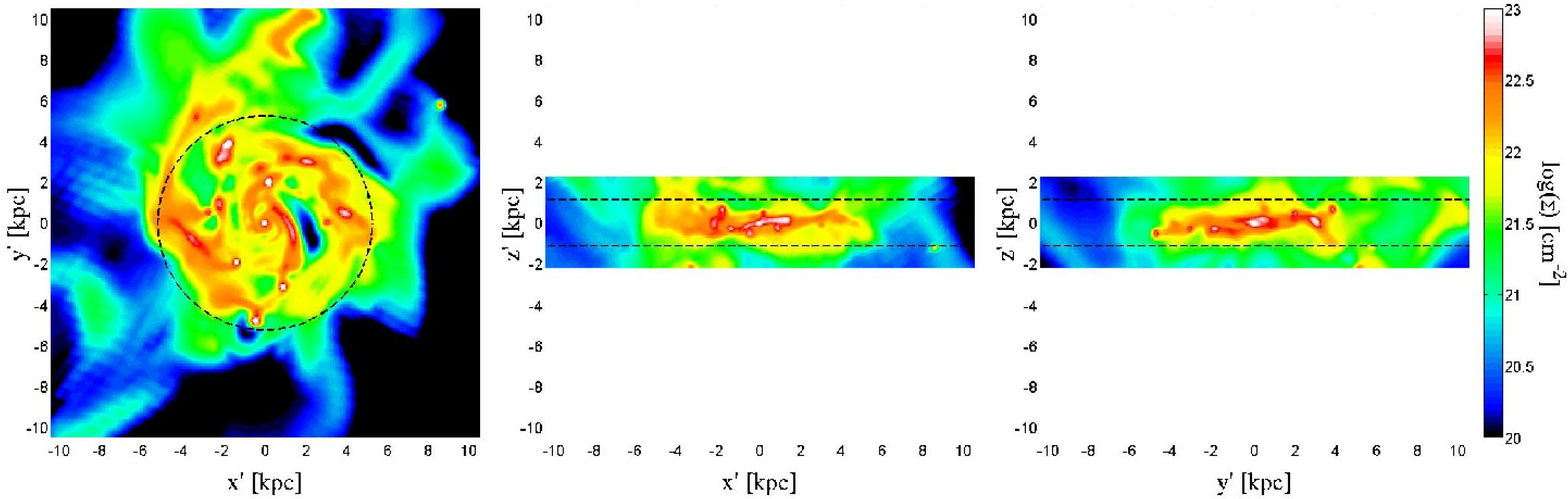}
\end {center}
\caption{Disc frame. Face on and edge on views of one of 
our studied galaxies, MW3 at expansion factor $a=0.30$ (redshift 
$z\simeq 2.33$).
The axes are marked ${\bf {x'}}$,${\bf {y'}}$ and ${\bf {z'}}$ in accordance 
with \equ{zprime} - \equ{yprime}. In the face on view, the dashed circle 
marks the disc radius at $\Rd=5.3\kpc$, and in the two edge on views, the 
horizontal dashed lines mark the disc thickness at $z'=\pm \Hd = \pm 1.1\kpc$. 
The face on image has been integrated over $\pm 2\Hd$ and the two edge on images 
have been integrated over $\pm 2\Rd$. The automated algorithm agrees well with 
the visual impression.
}
\label{fig:disc_frame} 
\end{figure*} 

To summarize, after the iterations have completed, the disc axis $\hat{z}'$ 
is alligned with the angular momentum of cold gas ($T<1.5{\times}10^4 {\rm K}$) 
within a cylinder of radius $\Rd$ and height $\Hd$. $\Rd$ contains $85 \%$ of 
the cold gas mass in a cylinder having radius $r=0.15\Rv$ and height 
$h=$min($\Hd,1\kpc$). The thickness of the disc contains $85\%$ of the 
cold gas mass in a cylinder with both radius and height $r=h=\Rd$. 

The 3 axes in the disc frame are defined as follows:
\be
\label {eq:zprime}
{\bf{\hat z'}} = {\rm sin(}{\theta}{\rm )cos(}{\phi}{\rm )}{\bf {\hat x}} + {\rm sin(}{\theta}{\rm )sin(}{\phi}){\bf {\hat y}} + {\rm cos(}{\theta}{\rm )}{\bf {\hat z}}
\ee
\be
\label {eq:xprime}
{\bf{\hat x'}} = {\rm cos(}{\theta}{\rm )cos(}{\phi}{\rm )}{\bf {\hat x}} + {\rm cos(}{\theta}{\rm )sin(}{\phi}){\bf {\hat y}} -{\rm sin(}{\theta}{\rm )}{\bf {\hat z}}
\ee
\be
\label {eq:yprime}
{\bf{\hat y'}} = -{\rm sin(}{\phi}{\rm )}{\bf {\hat x}} + {\rm cos(}{\phi}{\rm )}{\bf {\hat y}}
\ee
{\no}where ${\bf {\hat x}}$, ${\bf {\hat y}}$ and ${\bf {\hat z}}$ are the three 
unit vectors in the simulation box frame and $\theta$ and $\phi$ are the standard 
polar and azimuthal angles.
Note that ${\bf {\hat z'}}$ is alligned with the disc angular momentum, 
and the set $\bf {{\hat x'},{\hat y'},{\hat z'}} $ form a right hand basis.

\Fig{disc_frame} shows an example of the disc frame and dimensions 
for one of the galaxies in our 
sample: MW3 at expansion factor $a=0.30$ (redshift $z\simeq 2.33$), 
(\fig{disc_smoothing}). The radius and 
height for this galaxy are $\Rd=5.3\kpc$ and $\Hd=1.1\kpc$. In the figure 
we examine a box of dimensions $\pm 2\Rd \times \pm 2\Rd \times \pm 2\Hd$ 
around the disc center, which is the volume probed for clumps in our 
analysis (\se{algorithm}). 

\section{Identifying Gas Clumps in 3-d} 
\label{sec:clump_finding} 

In what follows we present in detail our method for detecting 
clumps in the simulations. The method is based on a two level 
smoothing of the 3D gas density field, first with a narrow 
gaussian of Full Width at Half Maximum (FWHM) $\Fn=140\pc$ and 
then with a wide gaussian of FWHM $\Fw=2\kpc$. Recalling that 
the maximal AMR resolution in our simulations is between $35-70\pc$ 
at all times, we see that $\Fn$ is always between 2 and 4 spatial 
resolution elements (or 1-2 force resolution elements) and serves 
only to wash out noise at the resolution level, having no serious 
effect on our results. The purpose of the wide gaussian is to 
approximate the large scale disc background and wash out 
any clumps. Observations which have attempted to directly estimate 
the sizes of high-z giant clumps suggest that the largest of them 
have diameters of $\sim 2\kpc$ \citep{Forster11b,Genzel11,wisnioski12}, 
while numerical studies of clump properties in cosmological simulations 
indicate that clump diameters are even smaller \citep[e.g.][]{Ceverino12,
Genel12a}. Therefore, while our value for $\Fw$ is somewhat arbitrary, 
it should be a good estimate for our purposes. We have experimented 
with varying the value of $\Fw$ in the range $1.5-3 \kpc$ and found 
the properties of \textit{compact} clumps to vary at the $\sim 30\%$ level, 
while the sizes and masses of \textit{diffuse} or elongated perturbations 
may vary by a factor of $\sim 2$.

Using the two smoothed density fields, ${\rho}_{\rm N}$ and ${\rho}_{\rm W}$ 
respectively, we define the density residual 
$\delrho=\frac{{\rho}_{\rm N}-{\rho}_{\rm W}}{{\rho}_{\rm W}}$ 
and zero out all cells having $\delrho<\delmin=15$. The selected value for 
$\delmin$ depends on the selected value for $\Fw$. Based on experimentation 
and visual inspection of $\sim 20$ snapshots from our sample, we found our 
adopted values to work well with each other and adopted them for all galaxies. 
Our objective was on the one hand to be able to detect all the clumps, even 
those with relatively low density contrast, but on the other hand to avoid 
situations where neighboring clumps were grouped together or where a clump 
was "buried" in an extended transient feature.

Neighbouring cells having $\delrho>\delmin$ are grouped together and the 
cell with maximal ${\rho}_{\rm N}$ within each group defines the center. 
We assign a radius to each group, defined as the radius of a sphere 
containing $90 \%$ of the group mass. For groups which are roughly spherical, 
this provides a good estimate of their size, while for more elongated or 
filamentary ones it is not terribly meaningful. It must be stressed that 
while we do {\it not} strictly impose spherical geometry on our clumps but 
rather adhere to their identified shape, we do remove from them any material 
lying outside this radius.

The final step involves applying a cut both in mass and in size. 
We impose a minimum volume for clumps of $\Fn^3 = (140\pc)^3$. 
Our mass cut is {\it baryonic}, so we first add the 
stellar data onto the same grid used for the gas and compute the total 
baryonic mass for each identified group. We remove all groups with mass 
less than $10^{-4} \Md$, where $\Md$ is the baryonic mass of the host disc. 
Note that this mass cut is very low, and does not affect any of our results, 
since all the compact, spherical clumps used in our analysis have masses 
$\gsim 10^{-3} \Md$ (\fig{hist}). 
Groups that pass both the volume and mass thresholds define our sample of clumps. 

\section{Marginal mass and redshift dependence of disc clumpiness} 
\label{sec:marginal} 

\begin{figure*}
\begin {center}
\includegraphics[width =0.6 \textwidth]{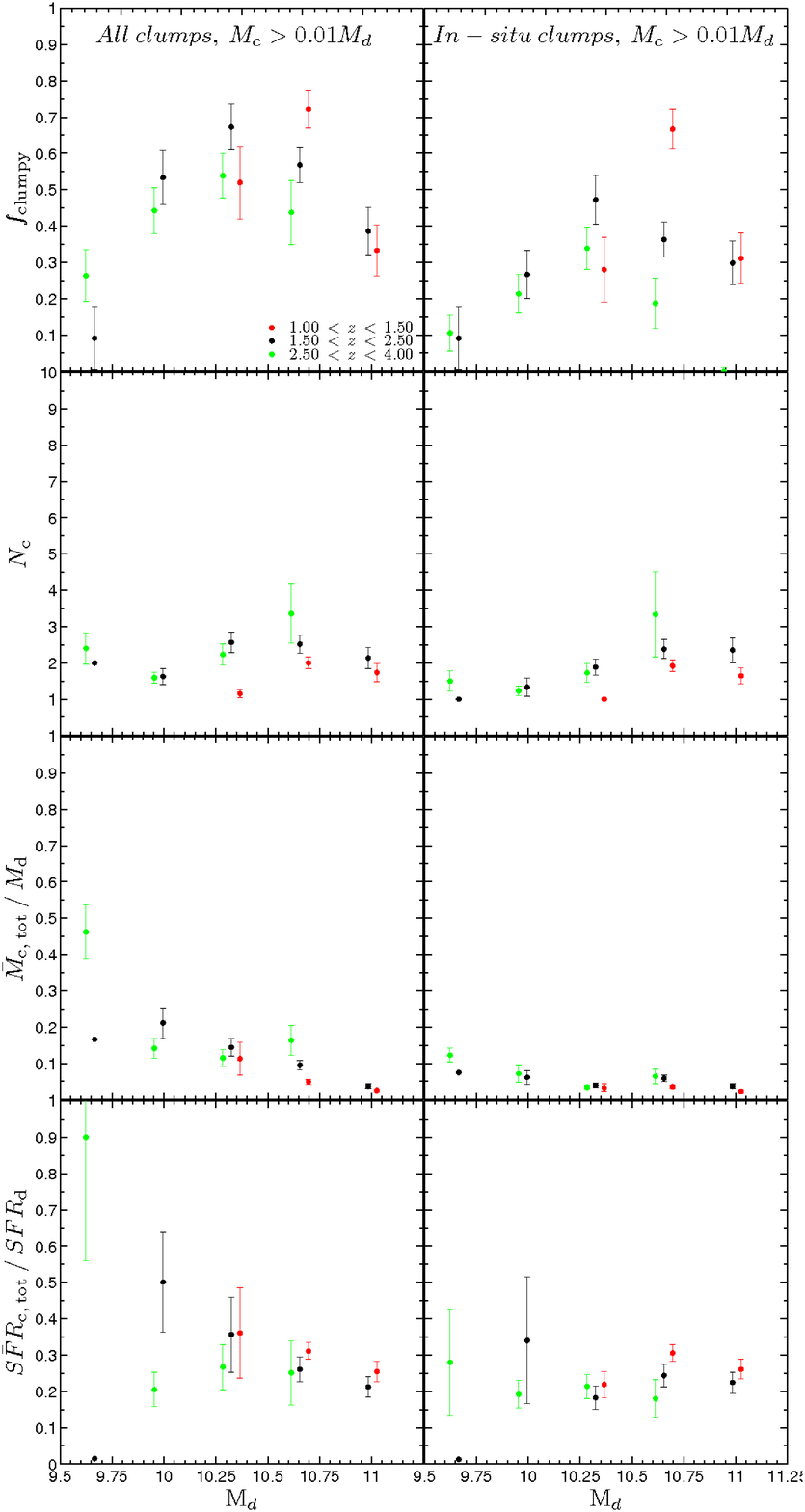}
\end {center}
\caption{Clumpiness properties of our sample of discs as a 
function of disc mass and redshift. Discs are divided into 
three redshift bins: $2.5<z<4.0$, green; $1.5<z<2.5$, 
black; $1.0<z<1.5$, red; and their properties 
are plotted as a function of mass in 5 equally spaced bins from 
$9.5<{\rm log(}\Md{\rm )}<11.25$ Only clumps with $\Mc>0.01\Md$ 
were considered. Left panles consider all off-center clumps while 
right panles consider only \insitu clumps. 
\textbf{Top row:} Fraction of clumpy discs. Error bars mark 68\% 
confidence levels of the standard error of percentage. 
\textbf{Second row:} Average number of clumps per disc, only for those 
discs with at least one clump. Error bars mark the standard error of 
the mean.
\textbf{Third row:} Average contribution of clumps to the disc baryonic 
mass, only for those discs with at least one clump. Error bars mark 
the standard error of the mean.
\textbf{Bottom row:} Average contribution of clumps to the total disc 
SFR, only for those discs with at least one clump. Error bars mark 
the standard error of the mean.
The clumpy fraction peaks for intermediate mass discs. \Insitu clumps 
contribute equally to disc mass and SFR at all disc masses.
}
\label{fig:mass_dep} 
\end{figure*} 

\begin{figure*}
\begin {center}
\includegraphics[width =0.60 \textwidth]{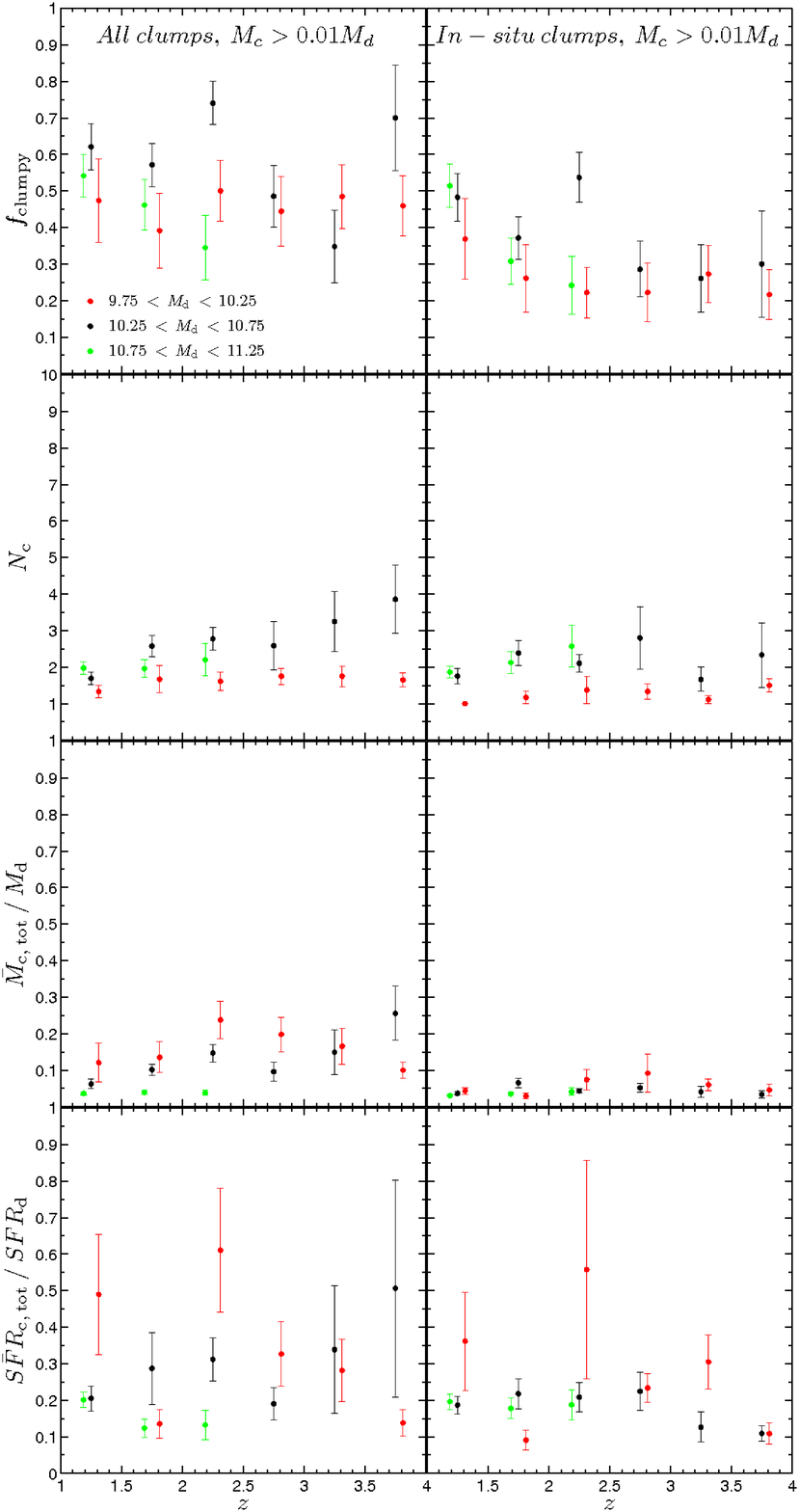}
\end {center}
\caption{Clumpiness properties of our sample of discs as a 
function of disc mass and reshift. Discs are divided into 
three mass bins: high mass ($10.75<{\rm log(}\Md{\rm )}<11.25$, 
green diamonds), intermediate mass ($10.25<{\rm log(}\Md{\rm )}<10.75$, 
black circles) and low mass ($9.75<{\rm log(}\Md{\rm )}<10.25$, red squares) 
and their properties are plotted as a function of redshift in 
6 bins from $1<z<4$ with ${\Delta}z=0.5$. 
Only clumps with $\Mc>0.01\Md$ were considered. 
Rows, columns and error bars are as in \fig{mass_dep}.
More discs appear clumpy at lower redshift, though the total mass and 
SFR in \insitu clumps does not seem to vary much.
}
\label{fig:z_dep} 
\end{figure*} 

In \se{clumpy_discs} we addressed the distribution of discs in terms of 
their off-center clumpiness properties. In this section, we attempt to 
examine a possible dependence of the disc clumpiness on disc mass and 
redshift. We restrict our analysis here to discs with masses 
${\rm log(}\Md{\rm )}>9.5$ and only account for clumps more massive 
than $0.01\Md$. Observations of high redshift clumpy galaxies are not 
expected to be sensitive to clumps below this mass ratio (Y. Guo, private 
communication). By limiting, as we have, the mass range of the discs 
examined, we should have a fairly complete sample of clumps, eliminating 
the bias caused by the simulation resolution. However, we caution that 
in the low mass discs the number of clumps may be subject to the resolution 
threshold. 

We have divided our sample into three redshift bins, as shown in 
\fig{mass_dep}. For each bin, we show as a function of disc mass the 
fraction of clumpy discs, mean number of clumps per disc, mean contribution 
of clumps to the disc baryonic mass and mean contribution of clumps to 
the total disc SFR. We do this seperately for all off-center clumps 
and for \insitu clumps only. We have seperately divided our sample 
into three bins of disc mass, and show the same four clumpiness properties 
as a function of redshift in \fig{mass_dep}. In both figures, we consider 
only bins that contain at least 10 galaxies.

At all redshifts, the fraction of clumpy discs is highest among 
intermediate mass discs with ${\rm log(}\Md{\rm )} \sim 10.5$. 
This remains true if considering all off-center clumps or only 
\insitu clumps, and is also seen in preliminary CANDELS observations 
at $z<3$ (Y. Guo et al, in preparation; M. Mozenna et al., in preparation). 
For high mass discs, the clumpy fraction appears to increase with time, 
also consistent with CANDELS observations. For intermediate mass discs, 
the clumpy fraction is higher at $1<z<2.5$ than at $2.5<z<4$ and reaches 
a maximum at $z\sim 2-2.5$, the origin of which is unclear at this time. 
The clumpy fraction is high as well at $3.5<z$, though this is due primarily 
to \exsitu mergers and is not evident when examining only \insitu clumps. 
The clumpy fraction of low mass discs shows no apparent trend with redshift 
when examining all off-center clumps, also consistent with CANDELS data, 
though there may be a very marginal trend for increased clumpy fraction at 
$1<z<2$ when examining \insitu clumps only. If real, this may suggest an 
increase in the prevalence of VDI at lower redshifts for all masses. We also 
note that at $z<2.5$, the intermediate mass discs appear to have the highest 
clumpy fraction.

The number of clumps per disc is perhaps the least certain statistic, due to 
the limitations of the simulation resolution. We note no significant trend with 
redshift for low mass discs or high mass discs. Intermediate mass discs contain 
more clumps on average at higher redshifts, due mainly to the higher frequency of 
\exsitu clumps. It also appears that lower mass discs contain fewer clumps on average, 
though this may well be an artifact of the resolution. 

The most robust statistic is the contribution of clumps to the total disc mass, 
as this will be dominated by the most massive, well resolved clumps. Examining 
only the \insitu clumps, there does not appear to be any trend with disc mass 
or redshift, suggesting that during VDI, the disc will turn a constant fraction 
of its mass into clumps, $\sim 5-10\%$. \Exsitu clumps contribute more to the disc 
mass at $2<z$, consistent with the theoretical estimate that the timescale for 
mergers of a given mass ratio, in terms of the galaxy dynamical time, is shorter 
at higher redshift \citep{NeisteinDekel08}.

Finally, the contribution of clumps to the disc SFR appears highest for low mass 
discs. We note, however, that including less massive clumps with $\Mc<0.01\Md$ 
significantly weakens this trend. Therefore, we cannot draw any significant 
conclusions at this time. No additional systematic trends with disc mass or 
redshift are evident. 
 
\section{Measuring Oxygen Fractions in the Simulations} 
\label{sec:metals} 

Our simulations track the mass fraction in metals released from type Ia SNae 
(${\rm z}_{\rm SNIa}$) and from type II SNae (${\rm z}_{\rm SNII}$). These values are given 
for the gas as well as for the steallr particles. 
The metallicity values quoted in the text are in units of ${\rm log(O/H)+12}$, 
where ${\rm O/H}$ is the ratio of Oxygen to Hydrogen atoms. This ratio is defined as 
\be 
\frac{\rm O}{\rm H} = \frac{{\rm z}_{\rm SNIa}{\cdot}{\rm f}_{\rm O,SNIa}+{\rm z}_{\rm SNII}{\cdot}{\rm f}_{\rm O,SNII}}{{\rm X}{\cdot}{\rm A}_{\rm O}}
\ee
{\no}where ${\rm f}_{\rm O,SNIa}$ and ${\rm f}_{\rm O,SNII}$ are the mass fraction of Oxygen in 
a typical SNIa and SNII explosion, respectively; X is the primordial abundance of Hydrogen and 
${\rm A}_{\rm O}$ is the atomic weight of Oxygen. Since all our galaxies are star forming galaxies 
with mean stellar ages $\lsim 1\Gyr$, we obtain ${\rm z}_{\rm SNIa}<<{\rm z}_{\rm SNII}$ (as predicted by 
\citet{Ferreras02}). Moreover, since the typical ratio of ${\alpha}$ nucleotides to Fe elements in SNII 
explosions is $\sim 30$ \citep{Ferreras02}, we take ${\rm f}_{\rm O,SNII}$ to be the mass fraction of 
Oxygen relative to ${\alpha}$ nucleotides in SNII explosions. We adopt a value of 
${\rm f}_{\rm O,SNII}=0.5$ based on \citet{Woosley95}. The atomic weight of Oxygen is 
${\rm A}_{\rm O}=16$ and we adopt $X=0.755$. Thus, the final formula is:
\be 
\frac{{\rm O}}{{\rm H}} = 0.5\frac{{\rm z}_{\rm SNII}}{0.755{\cdot}16}
\ee

 
\bsp 
 
\label{lastpage} 
 
\end{document}